\documentclass[aps,prc,showpacs,showkeys,twocolumn,letter,floatfix]{revtex4}

\usepackage{graphicx,epsfig,amsmath,amssymb}
\usepackage[bookmarksnumbered,bookmarksopen]{hyperref}
\usepackage[capitalize]{cleveref}
\usepackage[utf8]{inputenc}
\usepackage{listings}

\AtBeginDocument{\usepackage{booktabs}}
\usepackage{longtable}

\usepackage[usenames]{color}

\begin{document}

\setlength\LTcapwidth{\linewidth}

\title{Strangeness production via resonances in heavy-ion collisions at SIS energies}

\author{V.~Steinberg$^1$, J. Staudenmaier$^{1,3}$, D.~Oliinychenko$^{2}$, F.~Li$^1$, \"O.~Erkiner$^{1,3}$ and H.~Elfner$^{1,3,4}$}

\affiliation{$^1$Frankfurt Institute for Advanced Studies, Ruth-Moufang-Strasse 1, 60438 Frankfurt am Main, Germany}
\affiliation{$^2$Lawrence Berkeley National Lab, 1 Cyclotron Rd, Berkeley, CA 94720, USA}
\affiliation{$^3$Institute for Theoretical Physics, Goethe University, Max-von-Laue-Strasse 1, 60438 Frankfurt am Main, Germany}
\affiliation{$^4$GSI Helmholtzzentrum für Schwerionenforschung, Planckstr. 1, 64291 Darmstadt, Germany}
\date{\today}

\begin{abstract}
Production of strange hadrons in elementary and heavy-ion reactions is studied with the hadronic transport approach SMASH (Simulating Many Accelerated Strongly-interacting Hadrons).
The poorly known branching ratios of the relevant hadronic resonances are constrained from the known elementary hadronic cross sections and from invariant mass spectra of dileptons.
The constrained model is employed as a baseline to compare to heavy-ion-collision experiments at low energies ($E_\text{kin} = 1-2A\,\text{GeV}$) and to predict some of the upcoming pion-beam results by HADES, which are expected to be sensitive to the resonance properties.
The employed vacuum-resonance approach proves to be viable for small systems at these energies, but for large systems additional in-medium effects might be required.
\end{abstract}


\maketitle


\section{Introduction}

Strange quarks produced in heavy-ion collisions are an interesting probe for studying the evolution of the collisions.
Since they do not exist in ordinary nuclear matter, they have to be newly created during the reaction.
Their mass is higher than for up and down quarks and their production mechanism is sensitive to the properties of strongly-interacting matter.
The partonic and hadronic production channels are very different and may serve as a signal of the onset of deconfinement and the quark-gluon plasma, see~\cite{Blume:2017icv} for a recent overview.
Low-energy heavy-ion reactions are usually dominated by hadronic dynamics, but strangeness is enhanced compared to elementary proton-proton collisions~\cite{Blume:2011sb} due to possible secondary reactions.
Recently, the High-Acceptance Di-Electron Spectrometer (HADES) collaboration measured surprisingly high $\phi$ and $\Xi$ multiplicities at energies below the threshold~\cite{Adamczewski-Musch:2017rtf,Agakishiev:2009rr}.
Between the threshold and $\sqrt s = 10A\,\text{GeV}$, the mechanisms of strangeness production in the medium produced in heavy-ion collisions are not well understood, which leads to many open questions:

Do we need kaon-nucleon and antikaon-nucleon potentials?
How important are in-medium cross sections?
What are the production mechanisms in equilibrium and out of equilibrium?
Is the thermal model applicable or are transport models with resonances more appropriate?

Future experiments, such as Facility for Antiproton and Ion Research (FAIR)~\cite{Ablyazimov:2017guv}, Nuclotron-based Ion Collider (NICA)~\cite{Kekelidze:2012zz}, Japan Proton Accelerator Research Complex (J-PARC)~\cite{Sako:2014fha} and the Relativistic Heavy-Ion Collider~(RHIC) Beam Energy Scan~\cite{Aggarwal:2010cw}, will be essential to answer these questions.
In particular, the Compressed Baryonic Matter~(CBM) experiment at~FAIR will provide unique measurements of rare strange particles with high luminosities and precision.

In the past, hadronic transport approaches such as IQMD~\cite{Hartnack:1997ez}, UrQMD~\cite{Bass:1998ca}, HSD~\cite{Cassing:1999es}, JAM~\cite{Nara:1999dz} and GiBUU~\cite{Buss:2011mx} have been successfully employed for modeling the non-equilibrium hadronic phase in heavy-ion collisions at low Schwerionen-Synchrotron (SIS) and high RHIC or Large Hadron Collider (LHC) energies as well as in hybrid approaches.
For a recent, general comparison of such approaches, see~\cite{Zhang:2017esm}.
There are different ideas that describe the data for strangeness production equally well.
On the one hand, GiBUU~\cite{Agakishiev:2014moo}, IQMD and HSD~\cite{Hartnack:2011cn} employ kaon-nucleon and antikaon-nucleon potentials to describe the strangeness production at SIS energies.
On the other hand, the UrQMD approach includes high-mass nucleon resonances for strangeness production~\cite{Steinheimer:2015sha} as well as strangeness exchange reactions~\cite{Graef:2014mra} to do the same.
A third approach has been studied with GiBUU, which was extended with Hagedorn states~\cite{Gallmeister:2017ths}.

In general, it is not clear how the intermediate energy ranges targeted by future experiments (that is, $\sqrt s = 5-20A\,\text{GeV}$) can be described theoretically.
There are attempts to adapt hybrid approaches employed successfully at higher energies to finite baryo-chemical potential~\cite{Karpenko:2015xea,Denicol:2018wdp,Akamatsu:2018olk}.
Alternatively, the recently introduced hadronic transport approach SMASH~\cite{Weil:2016zrk} (Simulating Many Accelerated Strongly-interacting Hadrons) incorporates the newest available experimental data to establish a baseline at low energies that can be extended with additional physics required by intermediate energies.
SMASH has been tested against an analytic solution of the Boltzmann equation~\cite{Tindall:2016try}, utilized to model dilepton production at SIS energies~\cite{Staudenmaier:2017vtq} and to compute the viscosity of a hadron gas~\cite{Rose:2017bjz}.
In this work, a comprehensive study of exclusive elementary cross sections for strangeness production is performed to constrain the resonance properties in SMASH.
This approach is complementary to introducing kaon-nucleon and antikaon-nucleon potentials, which are so far not included in SMASH.
The result is confronted with experimental data from heavy-ion experiments.

As a newly developed hadronic transport approach, SMASH profits from the experiences of the approaches developed during the last three decades and the new experimental data constraining the resonance properties at low energies~\cite{Agakishiev:2014wqa,Patrignani:2016xqp}.

Within SMASH, 106 hadron species (not counting charge and antiparticles) are considered, so there are $\approx 10000$~types of possible 2-body collisions, each of which can have several possible final states.
For most of these reactions, the energy-dependent cross sections have not been measured.
Modeling this multitude of cross sections is one of the challenges a microscopic transport code has to face.
In SMASH, most of them are implemented via resonances:
Using the ansatz for the partial width proposed by Manley and Saleski~\cite{Manley:1992yb} (but with different parameters), and assuming detailed balance, the $1 \leftrightarrow 2$ cross section can be calculated from resonance masses, total decay widths and branching ratios.
Some cross sections are not resonant and have to be parametrized (see~\cite{Weil:2016zrk} and~\cref{sec:elementary}).
The advantage of this approach is that all these vacuum quantities can in principle be measured in experiment.
On the other hand, this results in a model with 1000s of parameters.

First, the model is described in \cref{sec:model_description}.
To minimize the risk of overfitting, we consider the available data on elementary cross sections and branching ratios to constrain the resonance properties within SMASH relevant for strangeness production in \cref{sec:elementary}.
In \cref{sec:hic}, SMASH is compared to independent experimental data on strangeness production in heavy-ion collisions measured by KaoS (Kaon Spectrometer) and HADES.
Finally, the results are summarized and put into the context of future work in \cref{sec:summary}.
A few technical details of the antikaon-nucleon (\cref{sec:KbarN_xs}) and kaon-nucleon cross sections (\cref{sec:KN_xs,sec:KN_amplitude}) are given in the appendices.

\section{Model description}
\label{sec:model_description}

For all calculations in this work, SMASH 1.3 was used.
A detailed model description can be found in~\cite{Weil:2016zrk}.
In the following, we focus on the new features and the properties relevant to strangeness production.

Collisions in SMASH are governed by the geometric collision criterion: Particles interact when their transverse distance~$d_\text{trans}$ is smaller than their interaction radius~$d_\text{int}$,
\begin{equation}
d_\text{trans} < d_\text{int} = \sqrt{\frac{\sigma}{\pi}}
\; .
\end{equation}
Only $2 \leftrightarrow 1$ reactions (resonance formation and decay) and $2 \leftrightarrow 2$ reactions are possible.
In-medium effects (besides the naturally occurring collisional broadening) are neglected and isospin symmetry is assumed.
The cross sections for the different charge states are calculated via the isospin Clebsch-Gordan coefficients.

In SMASH, a test particle ansatz and non-resonant cross section parametrizations similar to GiBUU are employed, while the resonant cross sections are derived from resonance properties akin to UrQMD.
The latter include the major contributions to strangeness production.
Off-shell propagation is not taken into account, unlike in GiBUU and HSD.
A lot of effort is put into providing a flexible, modern Open Source code that can be adapted as a baseline for hadronic systems where densities are low enough that vacuum values can be assumed for the cross sections, pole masses and widths of resonances.
To ensure that ongoing development does not cause regressions in describing experimental data, a large test suite is regularly employed.
This includes for example very extensive tests verifying that detailed balance is maintained for all reactions, which is important for infinite matter calculations.

The branching ratios of the decay channels of the resonances govern most of the cross sections in SMASH, so they are a crucial input for the model.
They can be extracted from experimental data via a partial wave analysis.
A collection of these branching ratios is provided by the Particle Data Group (PDG)~\cite{Patrignani:2016xqp}.
Unfortunately, the experimental data is sometimes rather sparse, especially for heavy resonances above $2\,\text{GeV}$.
The experimental data on exclusive cross sections provides a remedy as shown in \cref{sec:elementary}.

In \cref{sec:new_res}, the new resonances that have been added to SMASH since the previous publication~\cite{Weil:2016zrk} are presented.
This includes the hyperon resonances that are used to model strangeness production in SMASH.
Their and all other branching ratios relevant for strangeness production are discussed in detail in \cref{sec:nucleon_resonances,sec:hyperon_resonances,sec:meson_resonances}, where it is shown how elementary cross section measurements provide a constraint complementary to the PDG data and dilepton spectra.

Employing resonances to model the cross section is limited in energy, because the heaviest known resonances have masses of about $2\,\text{GeV}$.
Due to these limitation, only SIS energies ($E_\text{kin} = 1-2A\,\text{GeV}$) are considered in this work.
For higher energies, a different approach is required.
In SMASH, the high-energy cross sections are implemented via string fragmentation and an additive quark model, but these processes have been switched off for the current work.

\subsection{New resonances in SMASH}
\label{sec:new_res}

Most of the resonance properties in SMASH are based on the data provided by the PDG~\cite{Patrignani:2016xqp}, as has been discussed in great detail in~\cite{Weil:2016zrk}.
In the current version employed for this work, a lot of resonances have been added:
\begin{itemize}
\item Almost all mesons "regarded as established" by the PDG that are made of up, down and strange quarks: $f_0^*$, $f_1^*$, $f_2^*$, $a_0^*$, $a_1^*$, $a_2^*$, $\pi^*$, $\eta^*$, $\rho^*$, $\omega^*$, $K^*$.
\item New hyperon resonances for $\Lambda^*$, $\Sigma^*$, $\Xi^*$ and $\Omega^*$.
\end{itemize}
\cref{tab:particles} lists the individual hadron species that are implemented.
As described in~\cite{Weil:2016zrk}, each resonance contributes to the cross section of its decay products.

The pole masses of $a_0(980)$, $f_0(1500)$ and $K^*_0(1430)$ were increased slightly, well within experimental uncertainty, due to the following technical problem:
In SMASH, the Manley-Saleski ansatz is employed for the mass-dependent width of a resonance~$R$ decaying into children~$a$ and~$b$:
\begin{align}
\label{eq:width}
\Gamma_{R \to ab}(m) &= \Gamma_{R \to ab}(m_0) \frac{\rho_{ab}(m)}{\rho_{ab}(m_0)} \\
\rho_{ab}(m) &= \int dm_a dm_b \mathcal A_a(m_a) \mathcal A_b(m_b) \nonumber\\
&\phantom{= } \times \frac{|\vec p_f|}{m} B_L^2(|\vec p_f| R) \mathcal F_{ab}^2(m)
\end{align}
$m_a$ and $m_b$ are the masses of particles~$a$ and~$b$, $\mathcal A_a$ and $\mathcal A_b$ are their spectral functions.
$B_L$ are the Blatt-Weisskopf functions depending on orbital angular momentum~$L$ and the interaction radius~$R = 1\,\text{fm}$.
$\mathcal F_{ab}$ is a form factor only relevant for unstable children.
$\vec p_f$ is the final-state momentum in the center-of-mass frame.
It is undefined if the (stable) children are heavier than the resonance at its pole ($m_0 < m_a + m_b$).
Normalizing by a mass that is high enough would eliminate this issue, but this is not feasible, because only the particle properties at the pole are given in the experimental data.
Fortunately, this is rarely a problem and only affects the three resonances mentioned above.

\begin{longtable}{cccc}
\caption{
Updated list of hadrons implemented in SMASH~1.3 with their properties and PDG codes (see~\cite{Patrignani:2016xqp} for the definition).
$N^*$ and $\Delta^*$ have been left out, because they did not change compared to the previous publication; see~\cite{Weil:2016zrk}.
The corresponding antiparticles carry a minus sign and have identical properties.
}
\label{tab:particles}
\\
\toprule
\bf{Type} & \bf{Mass} & \bf{Width} & \bf{PDG codes} \\
          & [GeV]     & [GeV]      &                \\
\midrule
$\pi$            & 0.138  & $7.7\cdot10^{-9}$ &      111,     211 \\
$\eta$           & 0.548  & $1.31\cdot10^{-6}$ &      221 \\
$\sigma$         & 0.800  & 0.400   &  9000221 \\
$\rho$           & 0.776  & 0.149   &      113,     213 \\
$\omega$         & 0.783  & $8.49\cdot10^{-3}$ &      223 \\
$\eta'$          & 0.958  & $1.98\cdot10^{-4}$ &      331 \\
$f_0(980)$       & 0.990  & 0.070   &  9010221 \\
$a_0(980)$       & 0.989  & 0.075   &  9000111, 9000211 \\
$\phi$           & 1.019  & $4.27\cdot10^{-3}$ &      333 \\
$h_1(1170)$      & 1.170  & 0.360   &    10223 \\
$b_1(1235)$      & 1.2295 & 0.142   &    10113,   10213 \\
$a_1(1260)$      & 1.23   & 0.42    &    20113,   20213 \\
$f_2$            & 1.275  & 0.185   &      225 \\
$f_1(1285)$      & 1.2819 & 0.024   &    20223 \\
$\eta(1295)$     & 1.294  & 0.05    &   100221 \\
$\pi(1300)$      & 1.30   & 0.4     &   100111,  100211 \\
$a_2(1320)$      & 1.3183 & 0.107   &      115,     215 \\
$f_0(1370)$      & 1.35   & 0.35    &    10221 \\
$\pi_1(1400)$    & 1.354  & 0.33    &  9000113, 9000213 \\
$\eta(1405)$     & 1.409  & 0.051   &  9020221 \\
$f_1(1420)$      & 1.4264 & 0.054   &    20333 \\
$\omega(1420)$   & 1.425  & 0.215   &   100223 \\
$a_0(1450)$      & 1.474  & 0.265   &    10111,   10211 \\
$\rho(1450)$     & 1.465  & 0.400   &   100113,  100213 \\
$\eta(1475)$     & 1.476  & 0.085   &   100331 \\
$f_0(1500)$      & 1.507  & 0.109   &  9030221 \\
$f_2'(1525)$     & 1.525  & 0.0073  &      335 \\
$\pi_1(1600)$    & 1.662  & 0.24    &  9010113, 9010213 \\
$\eta_2(1645)$   & 1.617  & 0.181   &    10225 \\
$\omega(1650)$   & 1.670  & 0.315   &    30223 \\
$\omega_3(1670)$ & 1.667  & 0.168   &      227 \\
$\pi_2(1670)$    & 1.672  & 0.260   &    10115,   10215 \\
$\phi(1680)$     & 1.680  & 0.15    &   100333 \\
$\rho_3(1690)$   & 1.689  & 0.161   &      117,     217 \\
$\rho(1700)$     & 1.720  & 0.25    &    30113,   30213 \\
$f_0(1710)$      & 1.723  & 0.139   &    10331 \\
$\pi(1800)$      & 1.812  & 0.208   &  9010111, 9010211 \\
$\phi_3(1850)$   & 1.854  & 0.087   &      337 \\
$f_2(1950)$      & 1.944  & 0.472   &  9050225 \\
$f_2(2010)$      & 2.010  & 0.20    &  9060225 \\
$a_4(2040)$      & 1.995  & 0.257   &      119,     219 \\
$f_4(2050)$      & 2.018  & 0.237   &      229 \\
$f_2(2300)$      & 2.297  & 0.15    &  9080225 \\
$f_2(2340)$      & 2.350  & 0.32    &  9090225 \\
\midrule
$K$           & 0.494 & 0      & 321, 311 \\
$K^*(892)$    & 0.892 & 0.0508 & 323, 313 \\
$K_1(1270)$   & 1.272 & 0.09   & 10313, 10323 \\
$K_1(1400)$   & 1.403 & 0.174  & 20313, 20323 \\
$K^*(1410)$   & 1.414 & 0.232  & 100323, 100313 \\
$K^*_0(1430)$ & 1.453 & 0.27   & 10311, 10321 \\ 
$K^*_2(1430)$ & 1.429 & 0.104  &   315,   325 \\ 
$K^*(1680)$   & 1.717 & 0.320  & 30313, 30323 \\
$K_2(1770)$   & 1.773 & 0.186  & 10315, 10325 \\
$K^*_3(1780)$ & 1.776 & 0.159  &   317,   327 \\
$K_2(1820)$   & 1.816 & 0.276  & 20315, 20325 \\
$K^*_4(2045)$ & 2.045 & 0.198  &   319,   329 \\
\midrule
$\Lambda$       & 1.116 & 0      & 3122 \\
$\Lambda(1405)$ & 1.405 & 0.0505 & 13122 \\
$\Lambda(1520)$ & 1.520 & 0.0156 & 3124 \\
$\Lambda(1600)$ & 1.600 & 0.1500 & 23122 \\
$\Lambda(1670)$ & 1.670 & 0.0350 & 33122 \\
$\Lambda(1690)$ & 1.690 & 0.0600 & 13124 \\
$\Lambda(1800)$ & 1.800 & 0.3000 & 43122 \\
$\Lambda(1810)$ & 1.810 & 0.1500 & 53122 \\
$\Lambda(1820)$ & 1.820 & 0.0800 & 3126 \\
$\Lambda(1830)$ & 1.830 & 0.0950 & 13126 \\
$\Lambda(1890)$ & 1.890 & 0.1000 & 23124 \\
$\Lambda(2100)$ & 2.100 & 0.2000 & 3128 \\
$\Lambda(2110)$ & 2.110 & 0.2000 & 23126 \\
$\Lambda(2350)$ & 2.350 & 0.1500 & 9903128 \\
\midrule
$\Sigma$       & 1.189 & 0     & 3222, 3212, 3112 \\
$\Sigma(1385)$ & 1.385 & 0.036 & 3224, 3214, 3114 \\
$\Sigma(1660)$ & 1.660 & 0.100 & 13112, 13212, 13222 \\
$\Sigma(1670)$ & 1.670 & 0.060 & 13224, 13214, 13114 \\
$\Sigma(1750)$ & 1.750 & 0.090 & 23112, 23212, 23222 \\
$\Sigma(1775)$ & 1.775 & 0.120 & 3226, 3216, 3116 \\
$\Sigma(1915)$ & 1.915 & 0.120 & 13226, 13216, 13116 \\
$\Sigma(1940)$ & 1.940 & 0.220 & 23114, 23214, 23224 \\
$\Sigma(2030)$ & 2.030 & 0.180 & 3118, 3218, 3228 \\
$\Sigma(2250)$ & 2.250 & 0.100 & 9903118, 9903218, \\
 & & & 9903228 \\
\midrule
$\Xi$       & 1.321 & 0     & 3322, 3312 \\
$\Xi(1530)$ & 1.532 & 0.009 & 3324, 3314 \\
$\Xi(1690)$ & 1.690 & 0.030 & 203312, 203322 \\
$\Xi(1820)$ & 1.820 & 0.024 & 13314, 13324 \\
$\Xi(1950)$ & 1.950 & 0.060 & 103316, 103326 \\
$\Xi(2030)$ & 2.030 & 0.020 & 203316, 203326 \\
\midrule
$\Omega$       & 1.672 & 0     & 3334 \\
$\Omega(2250)$ & 2.252 & 0.055 & 203338 \\
\bottomrule
\end{longtable}

\section{Elementary strangeness production}
\label{sec:elementary}

Let us focus now on the mechanisms to produce strangeness in SMASH at SIS energies.
The goal is to establish a hadronic vacuum baseline calculation that is extendable to larger systems and intermediate energies.

In low-energy heavy-ion collisions, kaons are produced from collisions of nucleons~$N \in \{ p, n \}$ via decays of nucleon resonances~$B^* \in \{ N^*, \Delta^* \}$ into hyperons~$Y \in \{ \Lambda, \Sigma \}$ and kaons~$K \in \{ K^+, K^0 \}$:
\begin{equation}
NN \to NB^* \to NYK
\label{eqn:kaon_prod}
\end{equation}
The decay into hyperons cannot produce antikaons~$\bar K \in \{\bar K^-, \bar K^0\}$, because there are no initial antinucleon collisions producing antihyperons.
A possible reaction chain involves strangeness exchange between pions~$\pi \in \{ \pi^+, \pi^0, \pi^- \}$ and hyperons:
\begin{align}
NN \to NB^* \to NYK
\quad
\pi Y \to Y^* \to \bar K N
\label{eqn:antikaon_prod}
\end{align}
Compared to \cref{eqn:kaon_prod}, this reaction requires an additional pion-hyperon collision that forms a resonance decaying into an antikaon and a nucleon.
This is less likely, resulting in a significantly lower antikaon than kaon production in nucleus-nucleus collisions in the resonance picture.
Indeed, measurements by KaoS and HADES show that there are two orders of magnitudes less $\bar K^-$ than $K^+$ in heavy-ion collisions at low energies~\cite{Forster:2007qk,Agakishiev:2009ar,Adamczewski-Musch:2017rtf}.

Another important channel for antikaon production proceeds via $\phi$ decays:
\begin{align}
NN \to NN^*
\quad
N^* \to \phi N
\quad
\phi \to \bar KK
\label{eqn:phi_prod}
\end{align}
However, $\phi$ production from $N^*$~decays has not been measured in experiment, suggesting the branching ratio is small.

In the following subsections, we look at each of these three contributions (\cref{eqn:kaon_prod,eqn:antikaon_prod,eqn:phi_prod}) in detail and show how the properties of the relevant resonances are constrained by the available experimental data.

\subsection{Nucleon resonances}
\label{sec:nucleon_resonances}

In heavy-ion collisions in SMASH, nucleon resonances are responsible for the hyperon production.
Therefore, let us take a close look at the experimental data constraining $\Lambda$ and $\Sigma$ production.

When simulating heavy-ion collisions with SMASH, $\Lambda$~baryons are mostly produced via the formation and decay of $N^*$ resonances ($NN \to NN^* \to N \Lambda K$).
The $N^* \to \Lambda K$ branching ratios are constrained by PDG data~\cite{Patrignani:2016xqp}and a recent HADES partial wave analysis~\cite{Munzer:2017hbl}, see \cref{tab:lambda_res}.
However, the data still leaves a lot of leeway to choose branching ratios.
It is helpful to consider measurements of elementary cross sections, because they are very sensitive to branching ratios and there exists a wealth of experimental data~\cite{Balewski:1998pd,Sewerin:1998ky,Kowina:2004kr,AbdElSamad:2010tz,AbdelBary:2010pc,Bilger:1998jf,AbdelSamad:2006qu,LaBoer}.

By comparing the contributions of the different resonances to the different (exclusive) cross sections, the branching ratios are tuned to fit the experimental data better.
Increasing the $N^* \to \Lambda K$ branching ratio increases the $pp \to \Lambda p K^+$ and $p \pi^- \to \Lambda K^0$ cross section.
Describing both cross sections simultaneously is challenging:
A good fit to the $pp \to \Lambda p K^+$ cross section can lead to underestimating the $p \pi^- \to \Lambda K^0$ cross section.
However, varying the $N^* \to \pi N$ branching ratios within the experimental errors margins only affects the latter cross section, resulting in a good simultaneous fit that is compatible to the other observables on pion production.

\begin{figure}
\centering
\includegraphics[width=\linewidth]{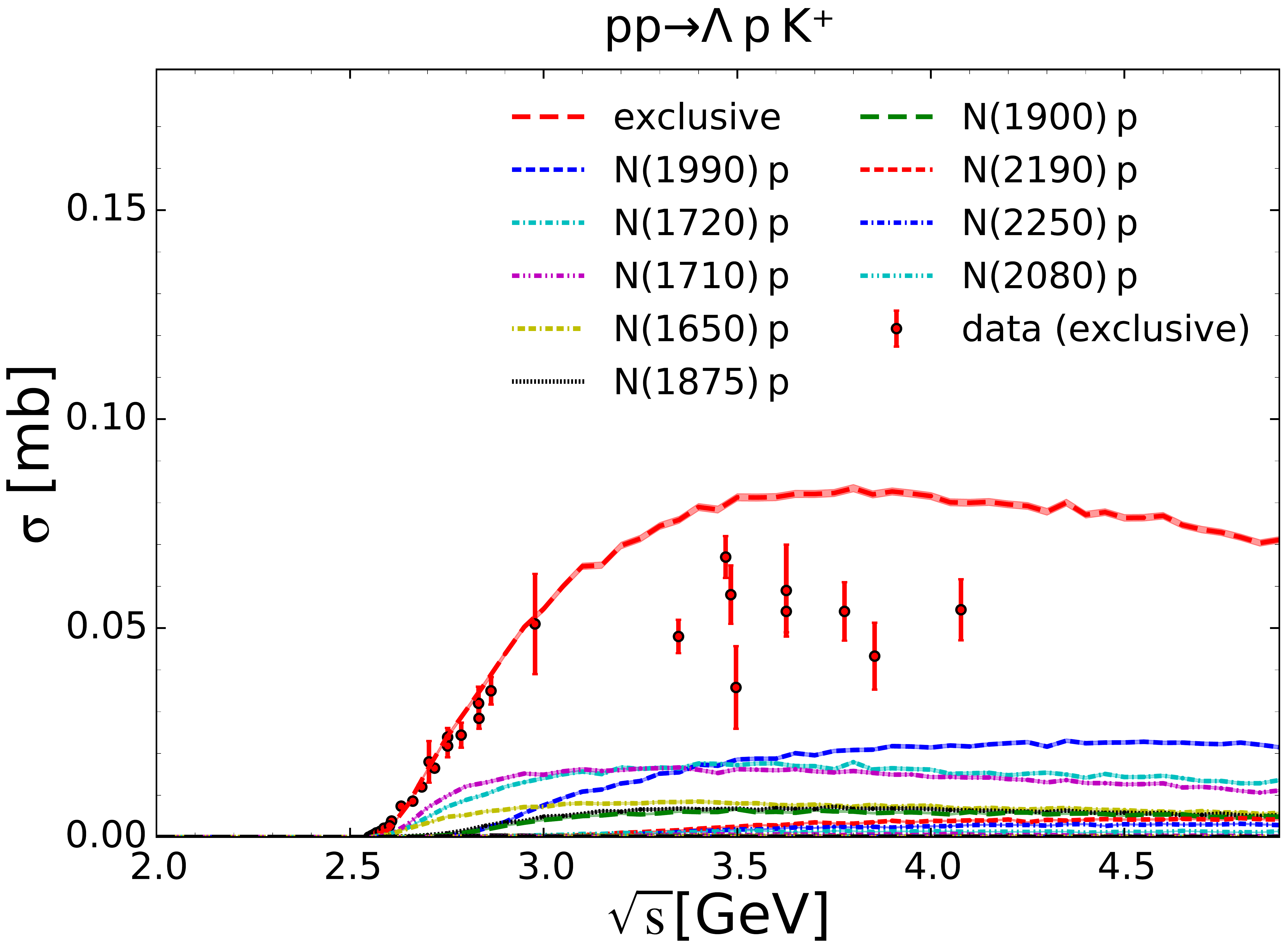}
\caption{$p p \to \Lambda p K^+$ cross section from SMASH compared to experimental data~\cite{Balewski:1998pd,Sewerin:1998ky,Kowina:2004kr,AbdElSamad:2010tz,AbdelBary:2010pc,Bilger:1998jf,AbdelSamad:2006qu,LaBoer}.}
\label{fig:xs_PP_LambdaPK+}
\end{figure}

\begin{figure}
\centering
\includegraphics[width=\linewidth]{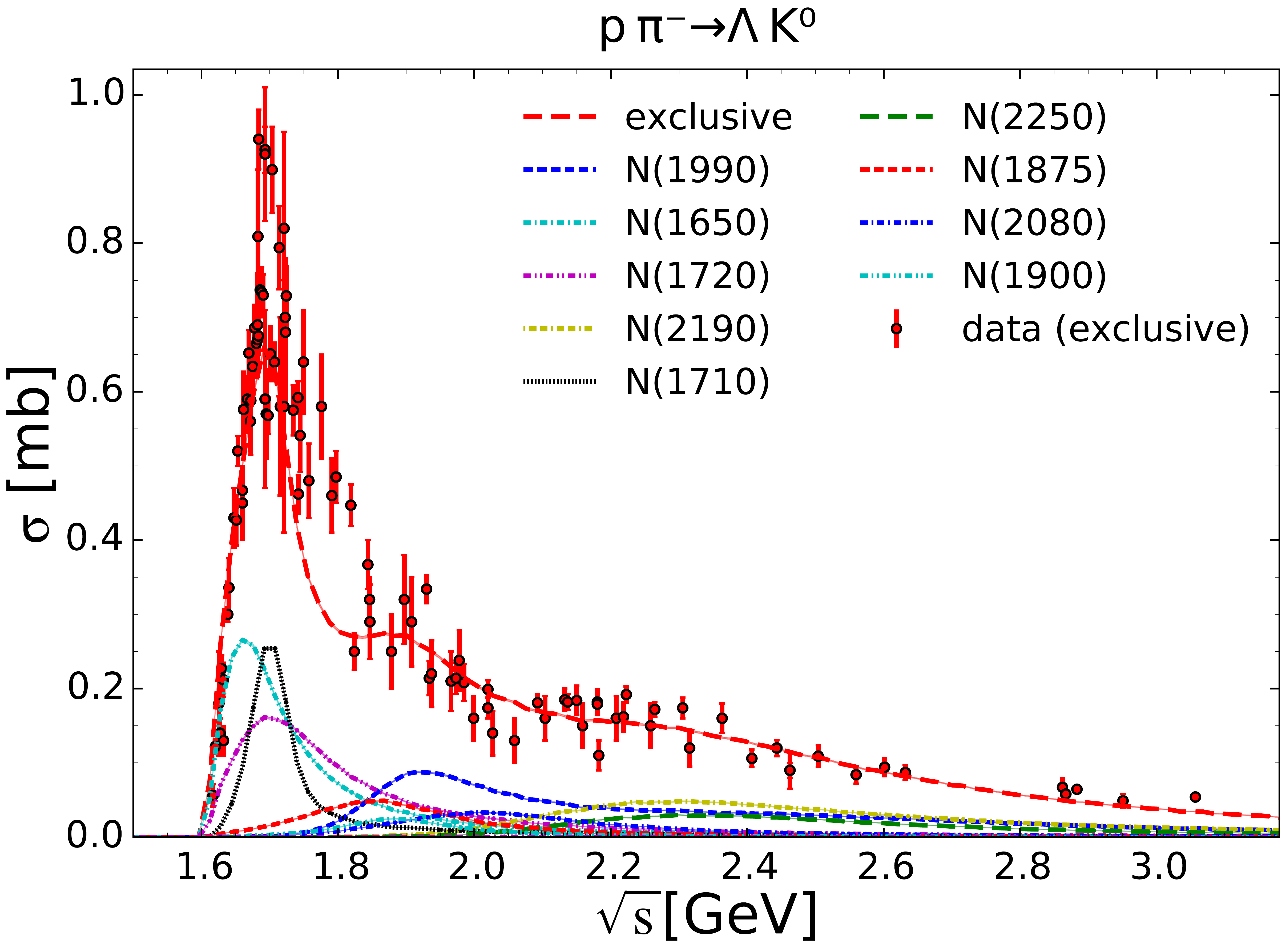}
\caption{$p \pi^- \to \Lambda K^0$ cross section from SMASH compared to experimental data~\cite{LaBoer}.}
\label{fig:xs_pi-P_LambdaK0}
\end{figure}

To reconstruct the elementary cross section from SMASH output, the elementary collisions are simulated many times with a random impact parameter.
The results are utilized to determine the maximal impact parameter~$b_\text{max}$, which directly gives the corresponding geometrical cross section~$\sigma = \pi b_\text{max}^2$.
Exclusive cross sections~$\sigma_\text{excl}$ are computed from the ratio of the number of exclusive reactions~$N_\text{reac,excl}$ and inclusive reactions~$N_\text{reac,incl}$:
\begin{equation}
\sigma_\text{excl} = \frac{N_\text{reac,excl}}{N_\text{reac,incl}} \sigma
\; .
\end{equation}
The cross sections of both the $\Lambda$ production channels $pp \to \Lambda p K^+$ and $p\pi^- \to \Lambda K^0$ are compared to experimental data from~\cite{Balewski:1998pd,Sewerin:1998ky,Kowina:2004kr,AbdElSamad:2010tz,AbdelBary:2010pc,Bilger:1998jf,AbdelSamad:2006qu,LaBoer} in \cref{fig:xs_pi-P_LambdaK0,fig:xs_PP_LambdaPK+}, respectively.
Additionally, the contributions of the different resonances are shown.
It can be seen how the intermediate $N^*$ states add up to the $\Lambda$ production cross section.
The threshold is well reproduced in both figures, but at $\sqrt s > 3.3\,\text{GeV}$ the $pp \to \Lambda p K^+$ cross section is slightly overestimated while the $p \pi^- \to \Lambda K^0$ cross section is slightly underestimated at $\sqrt s \approx 1.75\,\text{GeV}$.
In the present resonance approach, this cannot be alleviated without deviating significantly from the PDG branching ratios.
Assuming non-resonant contributions to the $pp \to \Lambda p K^+$ cross section would relax the tension, but also introduce additional model parameters.

\begin{table}
\caption{
$N^* \to \Lambda K$ branching ratios given by the PDG~\cite{Patrignani:2016xqp} and a HADES partial wave analysis~\cite{Munzer:2017hbl} compared to the values in SMASH 1.3.
$N(1880)$ and $N(1895)$ do not exist in SMASH and are not listed by the PDG~\cite{Patrignani:2016xqp}.
The error of including them in the partial wave analysis or not is reflected in the errors provided by HADES.
$N(1990)$, $N(2080)$, $N(2220)$ and $N(2250)$ were introduced to better reproduce the elementary nucleon-nucleon cross sections.
They are similar to the ones in UrQMD~\cite{Bass:1998ca}.
}
\label{tab:lambda_res}
\begin{tabular}{cccc}
\toprule
& \multicolumn{3}{c}{branching ratio $N^* \to \Lambda K$} \\
resonance & PDG & HADES & SMASH \\
\midrule
$N(1650)$ & $5-15\%$ & $7\pm4\%$ & $4\%$ \\
$N(1710)$ & $5-25\%$ & $15\pm10\%$ & $13\%$ \\
$N(1720)$ & $4-5\%$ & $8\pm7\%$ & $5\%$ \\
$N(1875)$ & $>0$ & $4\pm2\%$ & $2\%$ \\
$N(1880)$ & & $2\pm1\%$ & \\
$N(1895)$ & & $18\pm5\%$ & \\
$N(1900)$ & $2-20\%$ & $5\pm5\%$ & $2\%$ \\
$N(1990)$ & & & $2\%$ \\
$N(2080)$ & & & $0.5\%$ \\
$N(2190)$ & $0.2-0.8\%$ & & $0.8\%$ \\
$N(2220)$ & & & $0$ \\
$N(2250)$ & & & $0.5\%$ \\
\bottomrule
\end{tabular}
\end{table}

\begin{figure}
\centering
\includegraphics[width=\linewidth]{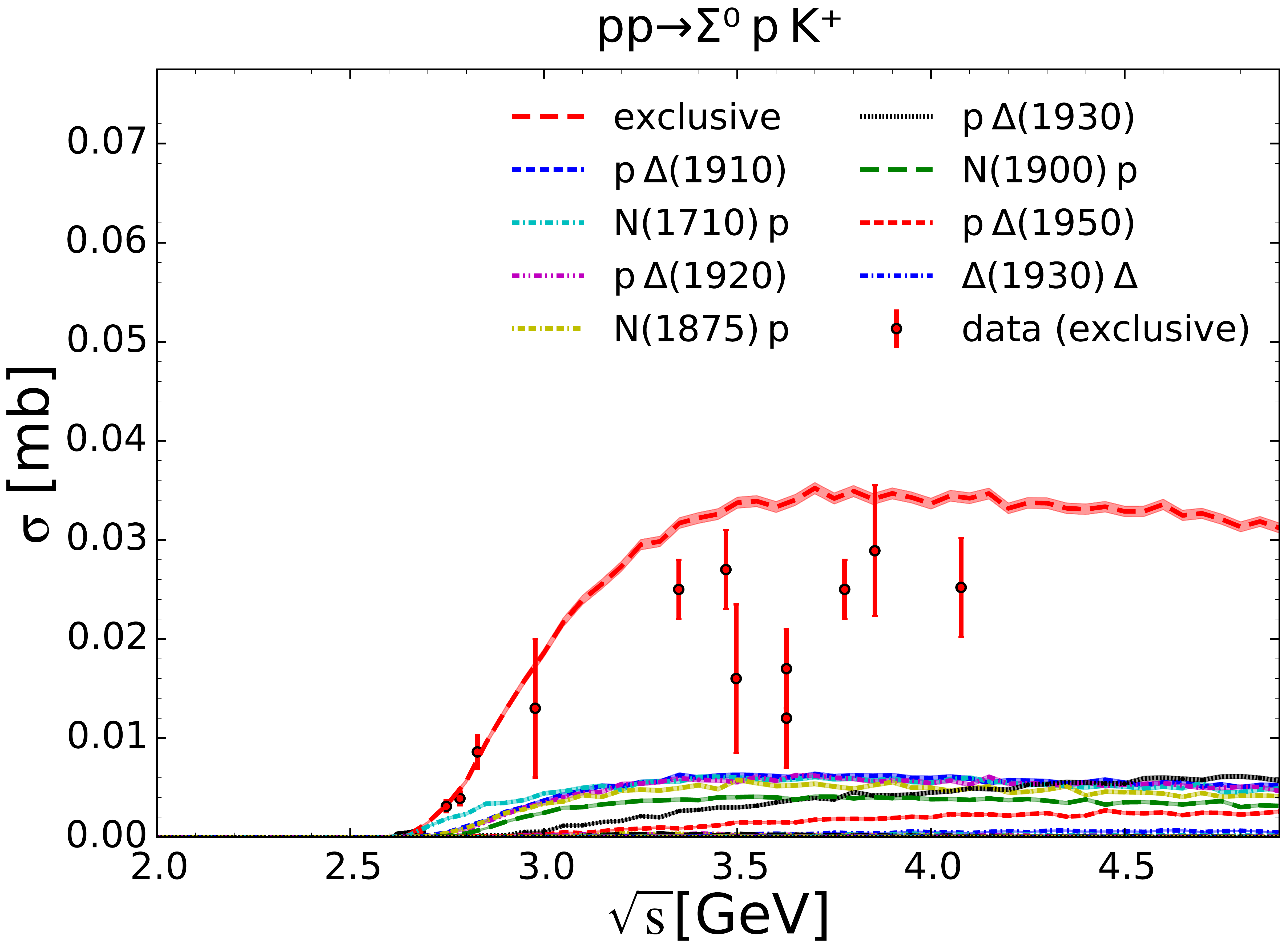}
\caption{$p p \to \Sigma^0 p K^+$ cross section from SMASH compared to experimental data~\cite{Sewerin:1998ky,Kowina:2004kr,AbdelBary:2010pc,LaBoer}.}
\label{fig:PP_Sigma0PK+}
\end{figure}

\begin{figure}
\centering
\includegraphics[width=\linewidth]{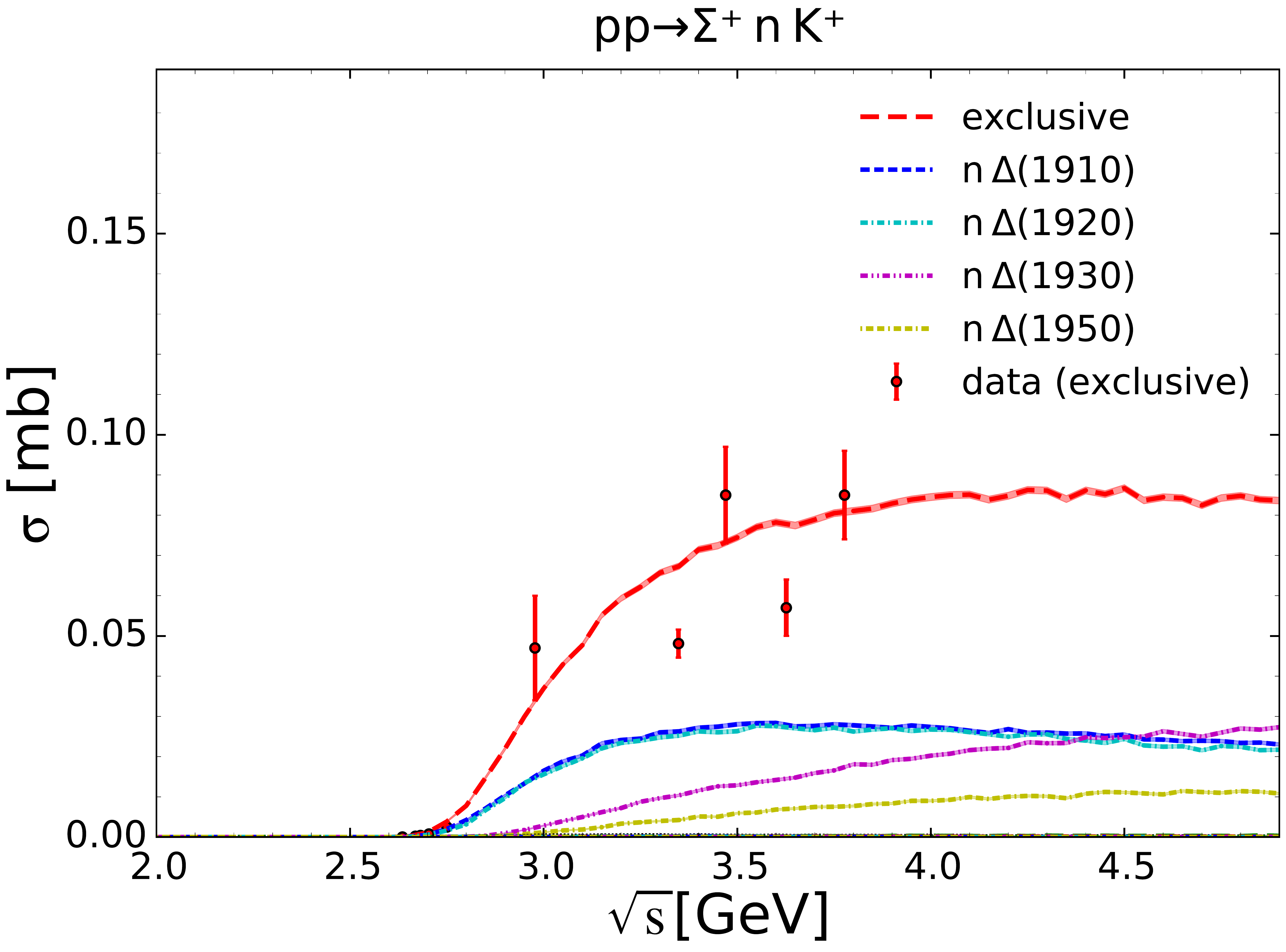}
\caption{$p p \to \Sigma^+ n K^+$ cross section from SMASH compared to experimental data~\cite{Valdau:2010kw,Valdau:2007re,LaBoer}.}
\label{fig:PP_Sigma+NK+}
\end{figure}

\begin{figure}
\centering
\includegraphics[width=\linewidth]{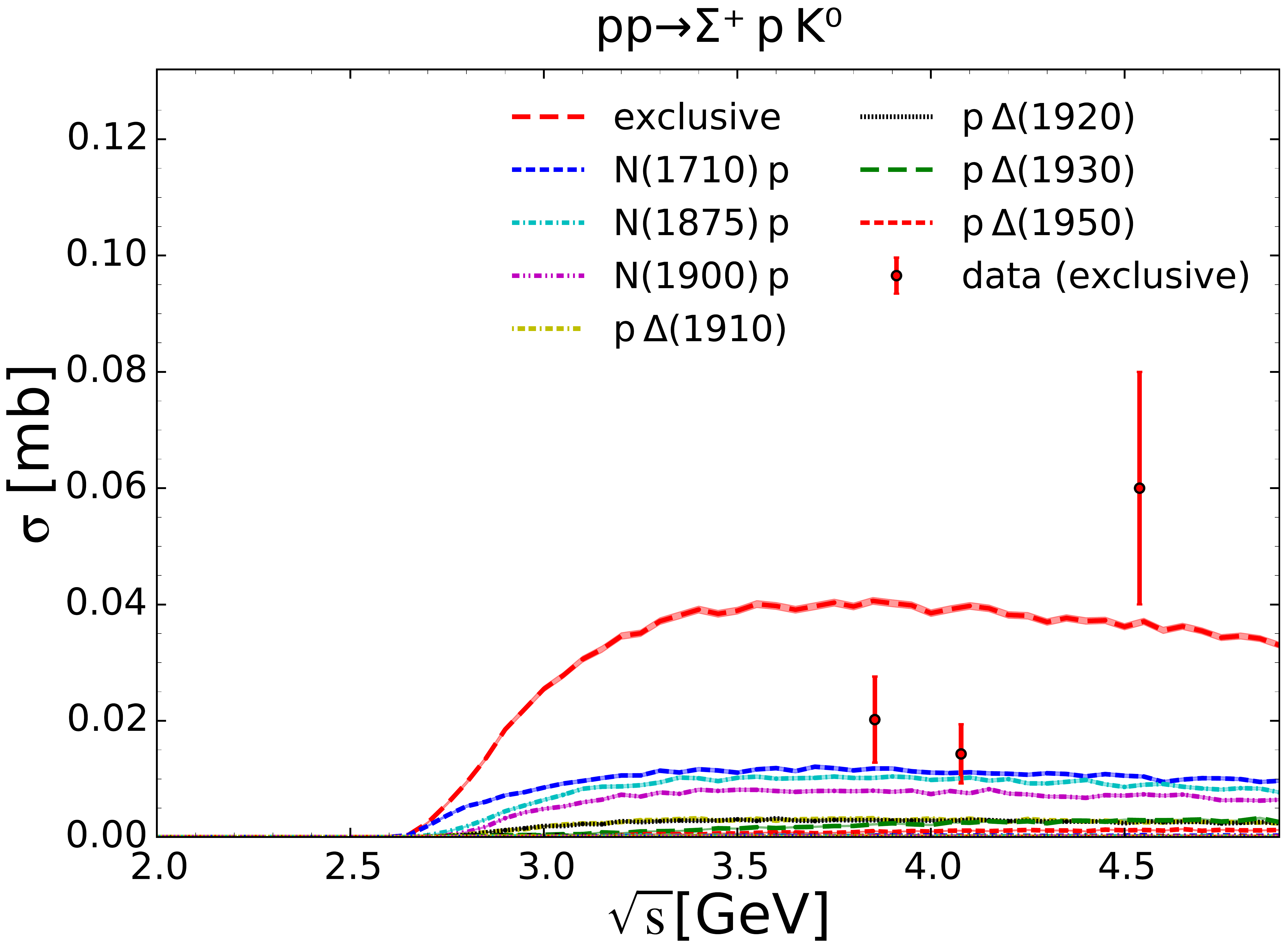}
\caption{$p p \to \Sigma^+ p K^0$ cross section from SMASH compared to experimental data~\cite{LaBoer}.}
\label{fig:PP_Sigma+PK0}
\end{figure}

\begin{figure}
\centering
\includegraphics[width=\linewidth]{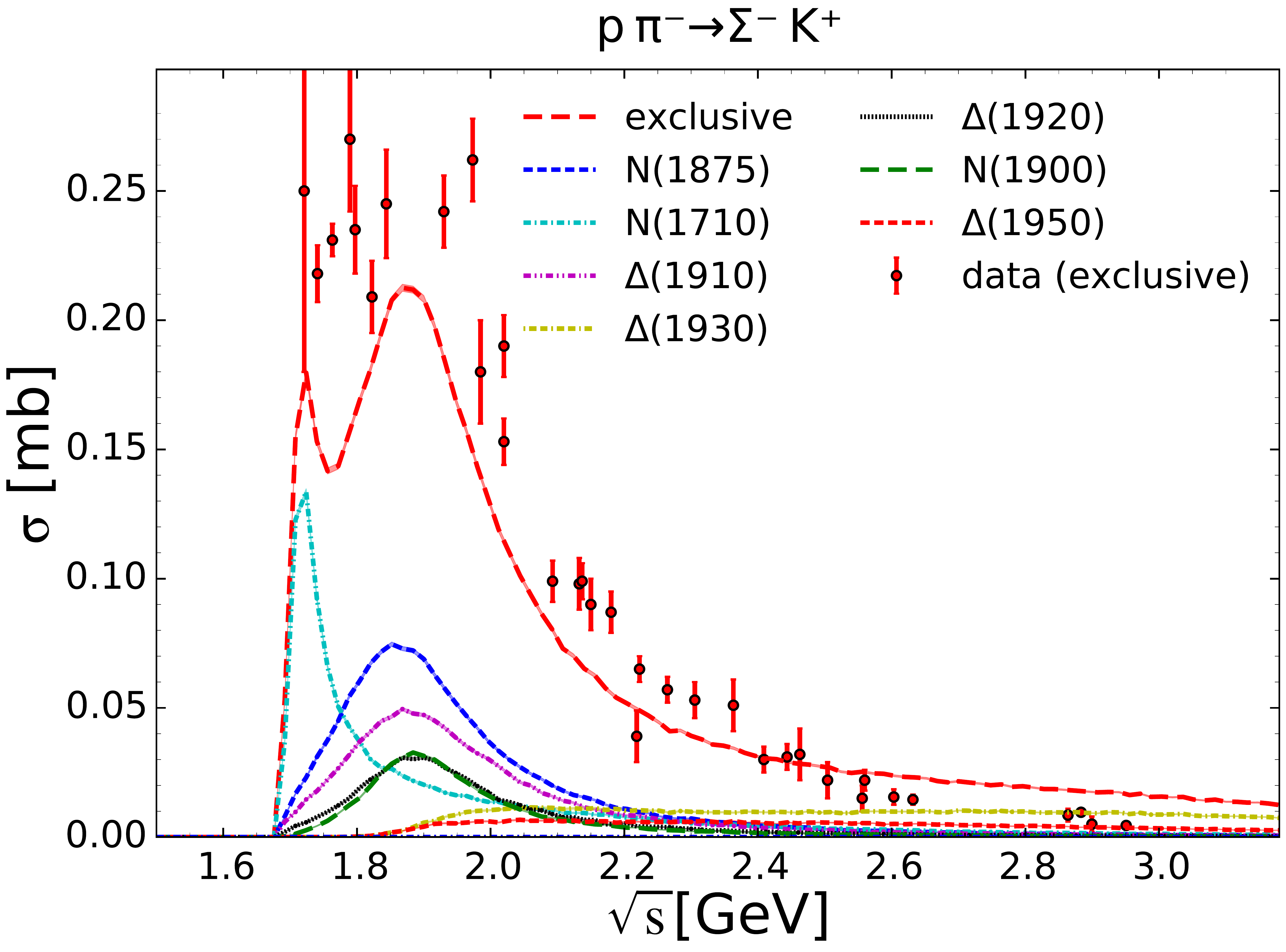}
\caption{$p \pi^- \to \Sigma^- K^+$ cross section from SMASH compared to experimental data~\cite{LaBoer}.}
\label{fig:xs_pi-P_Sigma-K+}
\end{figure}

Analogously to $\Lambda$ production, the branching ratios for $\Sigma$ production are constrained by the PDG data.
Again, there is a lot of leeway, but because of the different possible charges there are more measurements of elementary cross sections~\cite{Valdau:2010kw,Valdau:2007re,Sewerin:1998ky,Kowina:2004kr,AbdelBary:2010pc,LaBoer} constraining the branching ratios:
\begin{align}
p p &\to \Sigma^+ n K^+, \Sigma^+ p K^0, \Sigma^0 p K^+
\\
\pi^+ p &\to \Sigma^+ K^+
\\
\pi^- p &\to \Sigma^- K^+
\end{align}
As before, there is some tension since too many $\Sigma$ are produced in $pp$ (see \cref{fig:PP_Sigma0PK+,fig:PP_Sigma+NK+,fig:PP_Sigma+PK0}) but too few in $p\pi^-$ (see \cref{fig:xs_pi-P_Sigma-K+}), where the cross section is underestimated at $\sqrt s < 2.05\,\text{GeV}$.
This can be somewhat compensated by increasing $N^* \to N\pi$, until the upper limit given by the PDG branching ratios is reached.

\begin{figure}
\centering
\includegraphics[width=\linewidth]{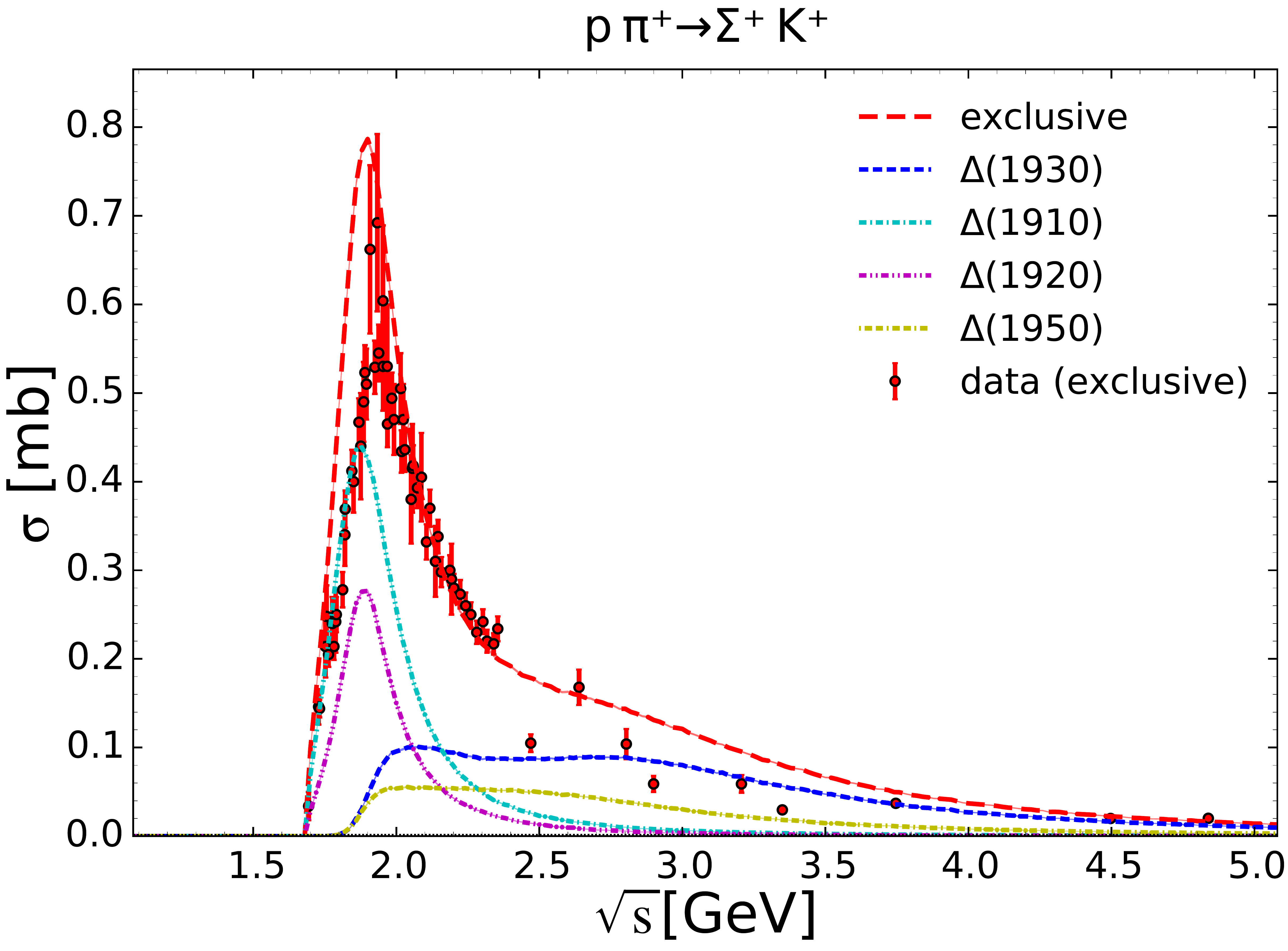}
\caption{$p \pi^+ \to \Sigma^+ K^+$ cross section from SMASH compared to experimental data~\cite{LaBoer}.}
\label{fig:xs_pi+P_Sigma+K+}
\end{figure}

In contrast to $\Lambda$ production, the $\Delta^*$ resonances are an important contribution to the $\Sigma$ production:
Reactions like $\Delta^{*++} \to \Sigma^+ K^+$ are not possible with $N^*$ resonances, so $\Delta^*$ resonances are necessary to describe $\pi^+ p \to \Sigma^+ K^+$ (\cref{fig:xs_pi+P_Sigma+K+}).
Contrary to the $p \pi^- \to \Sigma^- K^+$ cross section, the $p \pi^+ \to \Sigma^+ K^+$ cross section is overestimated at $\sqrt s = 1.75 - 1.95\,\text{GeV}$.
This discrepancy is hard to reconcile, because the $N^*$ contributions are already maximized within the limits of the PDG branching ratios, while the $\Delta^*$ contribution cannot be reduced without decreasing the already underestimated $p \pi^- \to \Sigma^- K^+$ cross section.



\subsection{Hyperon resonances}
\label{sec:hyperon_resonances}

In this work, the most important reason to look at the antikaon-nucleon cross section is to constrain the relevant hyperon branching ratios ($Y^* \to \bar K N, \pi Y$).
Antikaon-nucleon scatterings happen rarely in low-energy heavy-ion collisions, because only few antikaons are produced.
However, the corresponding backward reaction is crucial: the exchange of a strange quark between a pion and a hyperon was proposed 35~years ago as a dominant antikaon production mechanism, based on kinetic theory and thermodynamics~\cite{Ko:1983zp}:
\begin{equation}
\pi Y \leftrightarrow \bar K N
\label{eqn:s_exchange}
\end{equation}
In SMASH, this strangeness exchange is mostly modeled via hyperon resonances~$Y^* \in \{ \Lambda^*, \Sigma^* \}$:
\begin{equation}
\pi Y \leftrightarrow Y^* \leftrightarrow \bar K N
\end{equation}
As known from scattering theory, interactions via resonances correspond to scattering on an attractive potential.
In this regard, hyperon resonances have an effect qualitatively similar to attractive antikaon-nucleon potentials.
(In contrast, kaon-nucleon scattering does not involve intermediate resonances in SMASH.)

In IQMD, the dominant channel for antikaon production is $BY \to NN \bar K$ where $B \in \{ N, \Delta \}$.~\cite{Hartnack:2011cn}
This would correspond to $BY \to BY^*$ in SMASH and is currently not implemented.
UrQMD and GiBUU do not have a $BY$ channel either and it has not been measured experimentally.
Introducing a $BY \leftrightarrow BY^*$ channel would reduce the lifetime of the hyperon resonances in the medium.
It is not clear whether such a channel would increase or decrease antikaon multiplicities.

To further constrain the branching ratios given by the PDG~\cite{Patrignani:2016xqp}, the following cross sections measured in experiments~\cite{Patrignani:2016xqp,LaBoer} are considered:
\begin{itemize}
\item $\bar K^- p \to X,\, \bar K^- p,\, \Lambda \pi^0,\, \Sigma^- \pi^+,\, \Sigma^+ \pi^-,\, \Sigma^0 \pi^0$,
\item $\bar K^- n \to X,\, \bar K^- n,\, \Lambda \pi^-,\, \Sigma^- \pi^0,\, \Sigma^0 \pi^-$,
\end{itemize}
where $X$ means "anything", denoting total cross sections.
Because resonances are not sufficient to describe the $\bar K N$ cross sections, additional contributions have to be parametrized to constrain the branching ratios at $\sqrt s < 2\,\text{GeV}$:
\begin{enumerate}
\item an inelastic background diverging towards the threshold,
\item an elastic background,
\item charge exchange.
\end{enumerate}
The parametrizations employed in SMASH for these contributions are discussed in \cref{sec:KbarN_xs}.

\begin{figure}
\centering
\includegraphics[width=\linewidth]{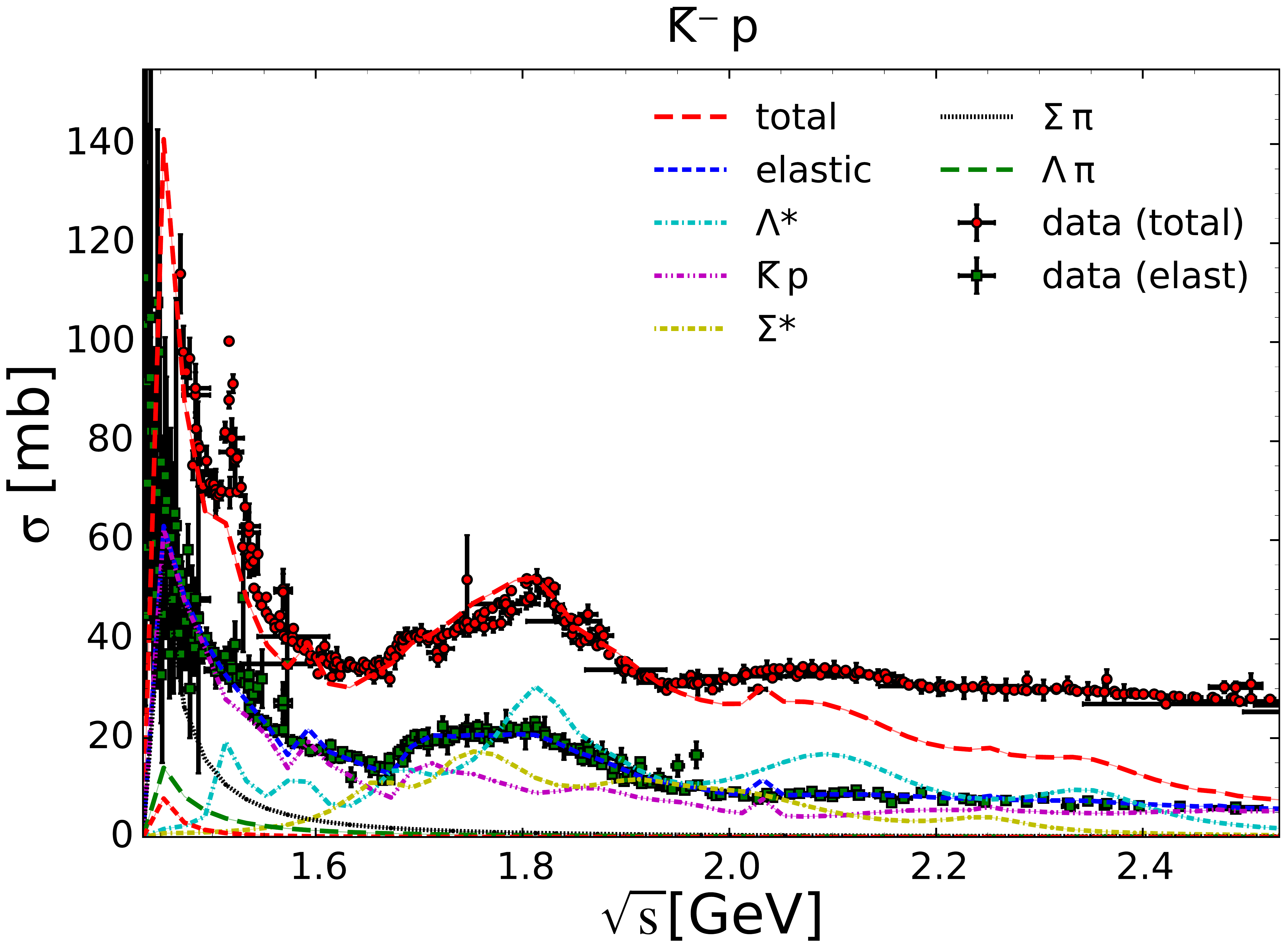}
\caption{$\bar K^- p$ cross section from SMASH compared to experimental data~\cite{Patrignani:2016xqp}.}
\label{fig:xs_K-P}
\end{figure}

In \cref{fig:xs_K-P}, the total and elastic $\bar K^- p$ cross section and the contributions by the intermediate states after the first collision are shown.
For instance, \lq$\bar K p$\rq\, corresponds to the elastic parametrization that is given by the difference of the experimental data and the elastic contribution of the resonances, and \lq$\Lambda\pi$\rq, \lq$\Sigma\pi$\rq\, correspond to the parametrized strangeness exchange.
The elastic and total cross section are mostly well reproduced, until about $\sqrt s = 2\,\text{GeV}$, where the total cross section falls off due to a lack of resonances.
By default, this cross section is reproduced in SMASH with an additive quark model, however, for simplicity this contribution is not included in the present resonance study.
The total $\bar K^- p$ cross section has clear peaks from the $\Lambda^*$~and $\Sigma^*$~resonances, which are sensitive to the parametrizations.
The $\Lambda(1520)$ peak in the $\bar KN$ cross section at $1.52\,\text{GeV}$ is underestimated.
Increasing the $\Lambda(1520) \to \bar KN$ branching ratio is unfortunately not possible without significantly deviating from the PDG values.
On the other hand, the error bounds and discrepancy in the experimental cross section data at that energy leave room for adjustment.

\begin{figure}
\centering
\includegraphics[width=\linewidth]{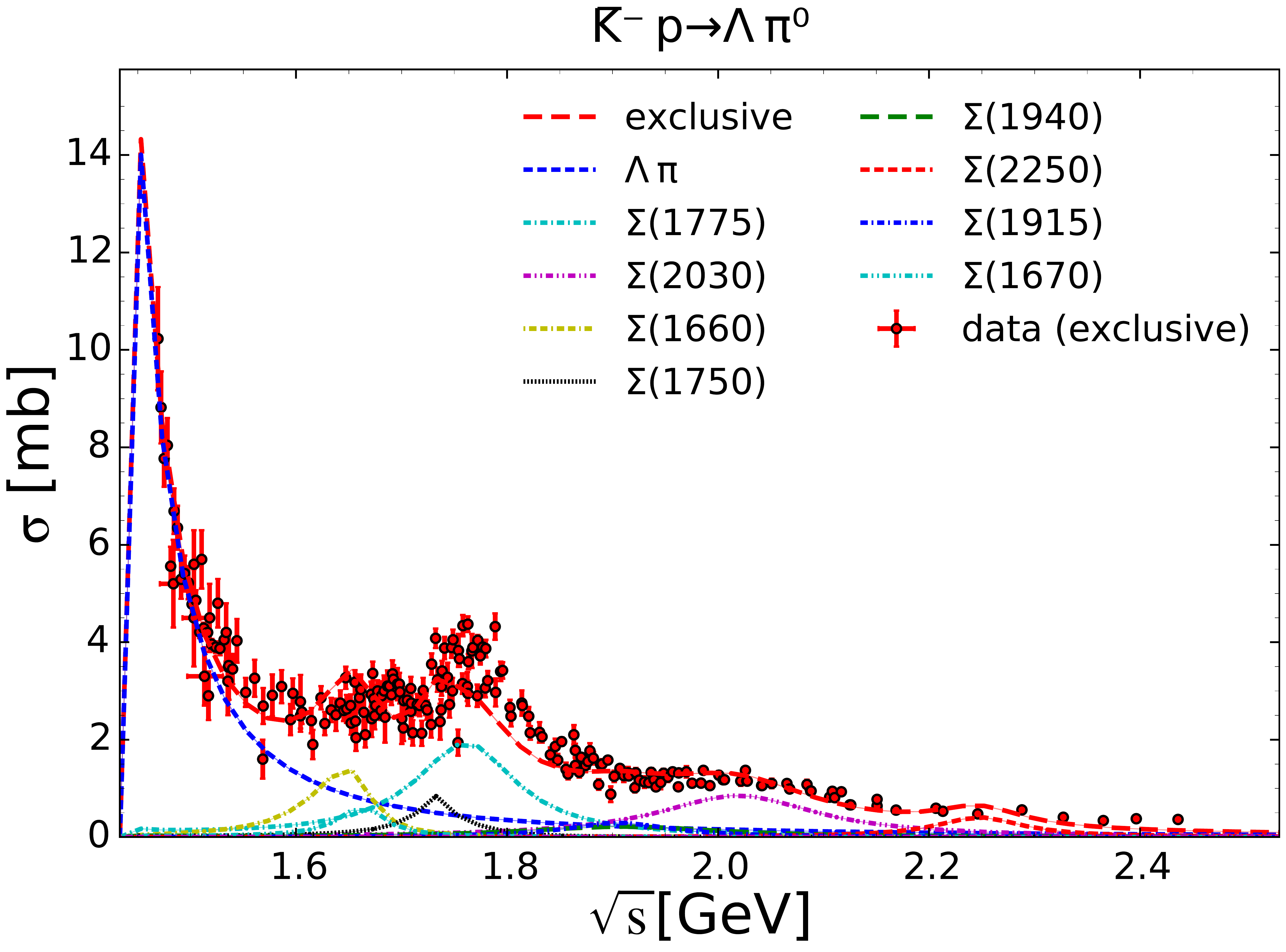}
\caption{$\bar K^- p \to \Lambda \pi^0$ cross section from SMASH compared to experimental data~\cite{LaBoer}.}
\label{fig:xs_K-P_Lambdapi0}
\end{figure}

\begin{figure}
\centering
\includegraphics[width=\linewidth]{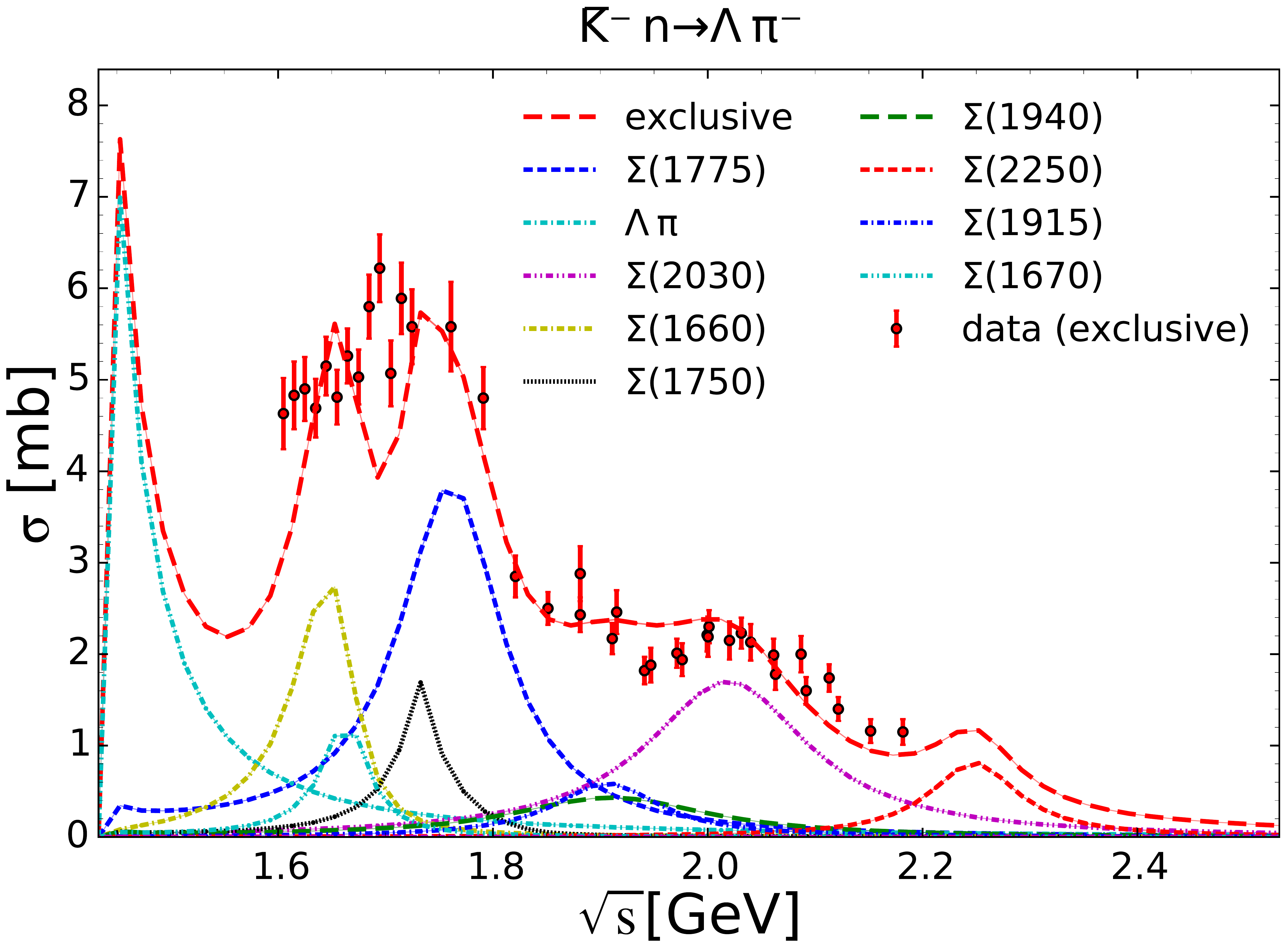}
\caption{$\bar K^- n \to \Lambda \pi^-$ cross section from SMASH compared to experimental data~\cite{LaBoer}.}
\label{fig:xs_K-N_Lambdapi-}
\end{figure}

For the $\bar K^- p \to \Lambda\pi^0$ cross section shown in \cref{fig:xs_K-P_Lambdapi0}, the intermediate $\Lambda^*$ state is forbidden by isospin.
Therefore, this cross section is useful for constraining the $\Sigma^*$ branching ratios, without being influenced by $\Lambda^*$.
The background parametrization of the strangeness exchange reproduces the experimental data well, and the contribution of the different resonances sum up reasonably well to the total cross section given by the experimental data.
The $\Sigma(1660)$ peak is a bit too high and the $\Sigma(1775)$ peak may be a bit too low; a compromise with the $\bar K^- n \to \Lambda\pi^-$ data in \cref{fig:xs_K-N_Lambdapi-} has been chosen.

\begin{figure}
\centering
\includegraphics[width=\linewidth]{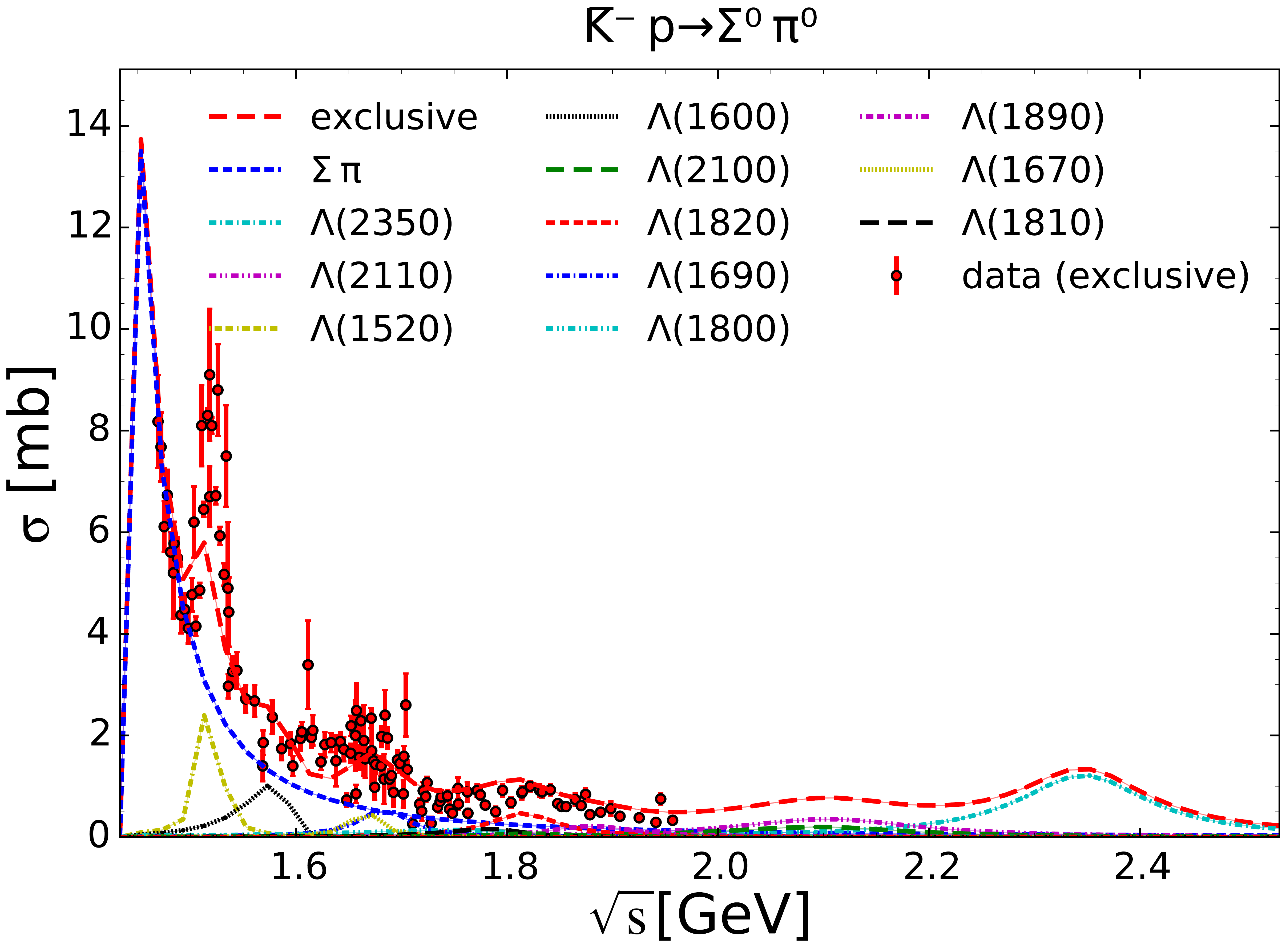}
\caption{$\bar K^- p \to \Sigma^0 \pi^0$ cross section from SMASH compared to experimental data~\cite{LaBoer}.}
\label{fig:xs_K-P_Sigma0pi0}
\end{figure}

Similarly, the $\bar K^- p \to \Sigma^0\pi^0$ cross section in \cref{fig:xs_K-P_Sigma0pi0} exclusively constrains the $\Lambda^*$ branching ratios, because the intermediate $\Sigma^{*0}$ states are forbidden.
Again, the strangeness exchange background and the resonance contributions are well reproduced.
As for the total $\bar K^- p$ cross section, the $\Lambda(1520)$ peak is too low, but the branching ratios of this particular resonance are tightly constrained by the PDG data.

\begin{figure}
\centering
\includegraphics[width=\linewidth]{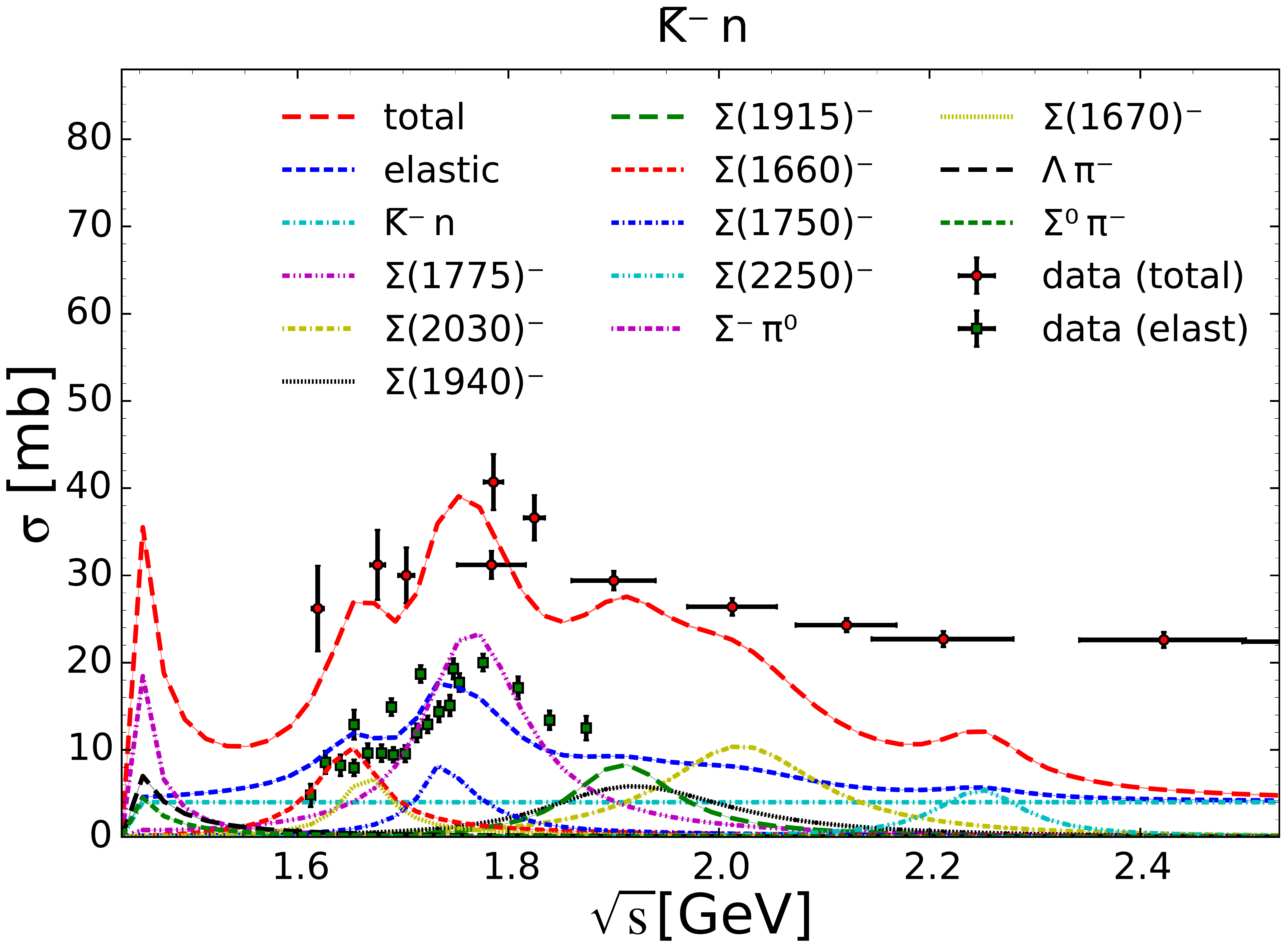}
\caption{$\bar K^- n$ cross section from SMASH compared to experimental data~\cite{Patrignani:2016xqp}.}
\label{fig:xs_K-N}
\end{figure}

\begin{figure}
\centering
\includegraphics[width=\linewidth]{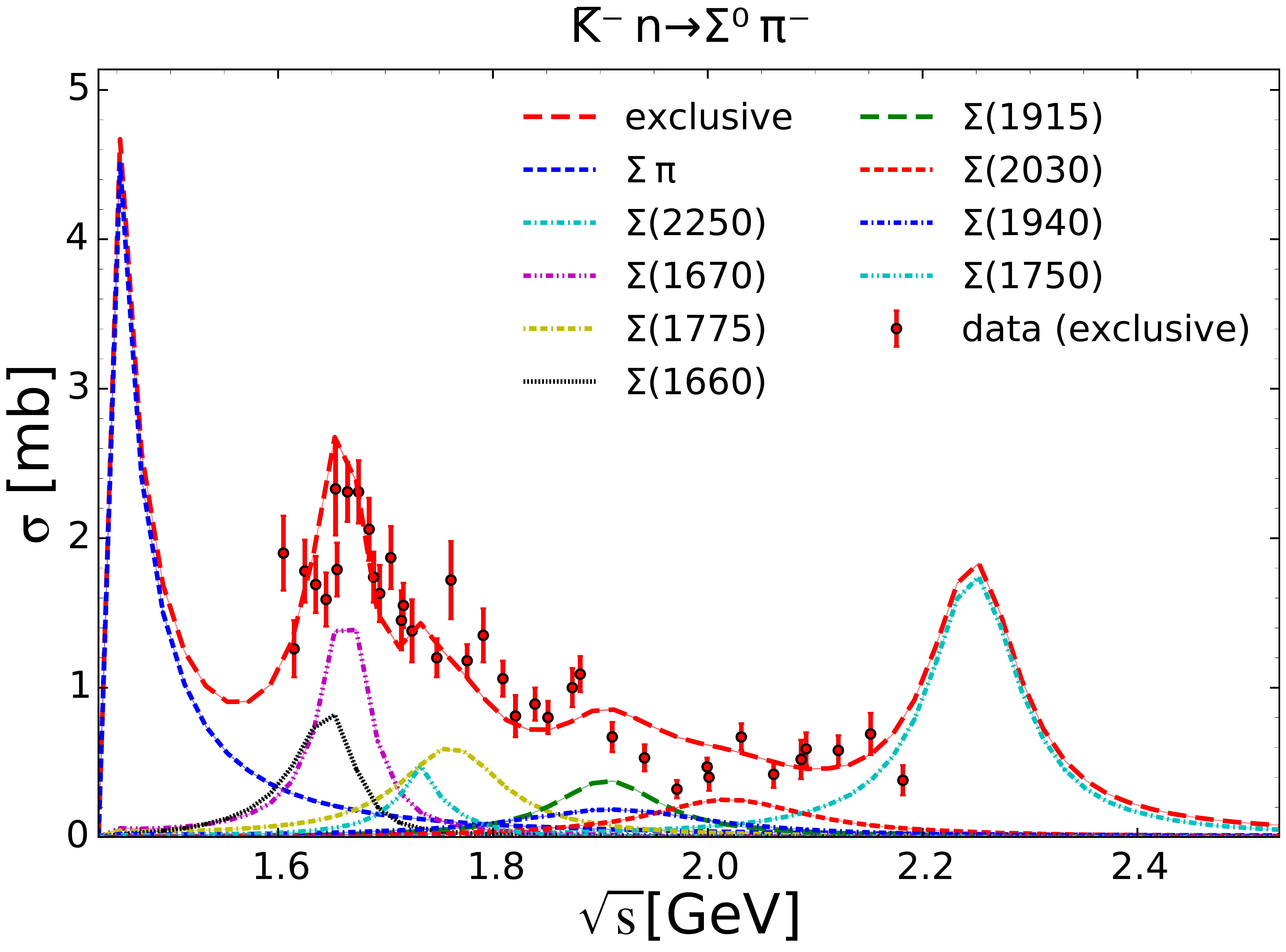}
\caption{$\bar K^- n \to \Sigma^0 \pi^-$ cross section from SMASH compared to experimental data~\cite{LaBoer}.}
\label{fig:xs_K-N_Sigma0pi-}
\end{figure}

The corresponding $\bar K^- n$ cross sections shown in \cref{fig:xs_K-N,fig:xs_K-N_Lambdapi-,fig:xs_K-N_Sigma0pi-} only differ from $\bar K^- p$ in the isospin Clebsch-Gordan coefficients, which only allow intermediate $\Sigma^*$ resonances.
The available data are much sparser and do not constrain the strangeness exchange background parametrization.
The total and elastic $\bar K^- n$ cross section in \cref{fig:xs_K-N} is resonably well reproduced up to $2\,\text{GeV}$, as are the $\bar K^- n \to \Lambda\pi^-$ cross section in \cref{fig:xs_K-N_Lambdapi-} and the $\bar K^- n \to \Sigma^0 \pi^-$ cross section in \cref{fig:xs_K-N_Sigma0pi-}.
The former has a small gap at $1.7\,\text{GeV}$, where there are no resonances.
Introducing one at that energy improves the agreement with experimental data, but there is no evidence for such a resonance, therefore we refrain from doing so.

The $\bar K^- p \to \Sigma^\mp \pi^\pm$ and $\bar K^- n \to \Sigma^- \pi^0$ cross sections (\cref{fig:xs_K-P_Sigma-pi+,fig:xs_K-P_Sigma+pi-,fig:xs_K-N_Sigma-pi0}) are dominated by the strangeness exchange background and are discussed in \cref{sec:KN_xs}.
They show a $\Lambda(1520)$ peak that is again too low.

Taking the cross sections discussed in this section and the PDG data into account results in the $\Lambda^* \to \bar KN$ branching ratios listed in \cref{tab:antikaon_res}.

Unlike the $\bar KN$~cross sections, the modeled $KN$~cross sections do not absorb any kaons and do not involve the hyperon resonances.
They do not constrain any branching ratios, but they do affect kinematics: For example, in IQMD they are responsible for depleting the yield in forward direction in heavy-ion collisions and changing the momentum spectra~\cite{Hartnack:2011cn}.
For the details of the implementation of $KN$ reactions in SMASH, see \cref{sec:KN_xs}.

\begin{table}
\caption{
$\Lambda^* \to \bar KN$ branching ratios given by the PDG~\cite{Patrignani:2016xqp} compared to the values employed in UrQMD 3.4 and in SMASH 1.3.
For $\Lambda(2350)$, the sum of the branching ratios listed by the PDG was rescaled to one.
}
\label{tab:antikaon_res}
\begin{tabular}{cccc}
\toprule
& \multicolumn{3}{c}{branching ratio $\Lambda^* \to \bar K N$} \\
resonance & PDG & UrQMD & SMASH \\
\midrule
$\Lambda(1405)$ & 0 & 0 & 0 \\
$\Lambda(1520)$ & $45 \pm 1\%$ & $45\%$ & $46.2\%$ \\
$\Lambda(1600)$ & $15-30\%$ & $35\%$ & $15\%$ \\
$\Lambda(1670)$ & $20-30\%$ & $20\%$ & $29.2\%$ \\
$\Lambda(1690)$ & $20-30\%$ & $25\%$ & $25\%$ \\
$\Lambda(1800)$ & $25-40\%$ & $40\%$ & $40\%$ \\
$\Lambda(1810)$ & $20-50\%$ & $35\%$ & $34\%$ \\
$\Lambda(1820)$ & $55-65\%$ & $65\%$ & $65\%$ \\
$\Lambda(1830)$ & $3-10\%$ & $10\%$ & $3\%$ \\
$\Lambda(1890)$ & $20-35\%$ & $35\%$ & $40\%$ \\
$\Lambda(2100)$ & $25-35\%$ & $35\%$ & $45\%$ \\
$\Lambda(2110)$ & $5-25\%$ & $25\%$ & $30\%$ \\
$\Lambda(2350)$ & $55\%$ & & $50\%$ \\
\bottomrule
\end{tabular}
\end{table}

\subsection{Meson resonances}
\label{sec:meson_resonances}

\begin{figure}
\centering
\includegraphics[width=\linewidth]{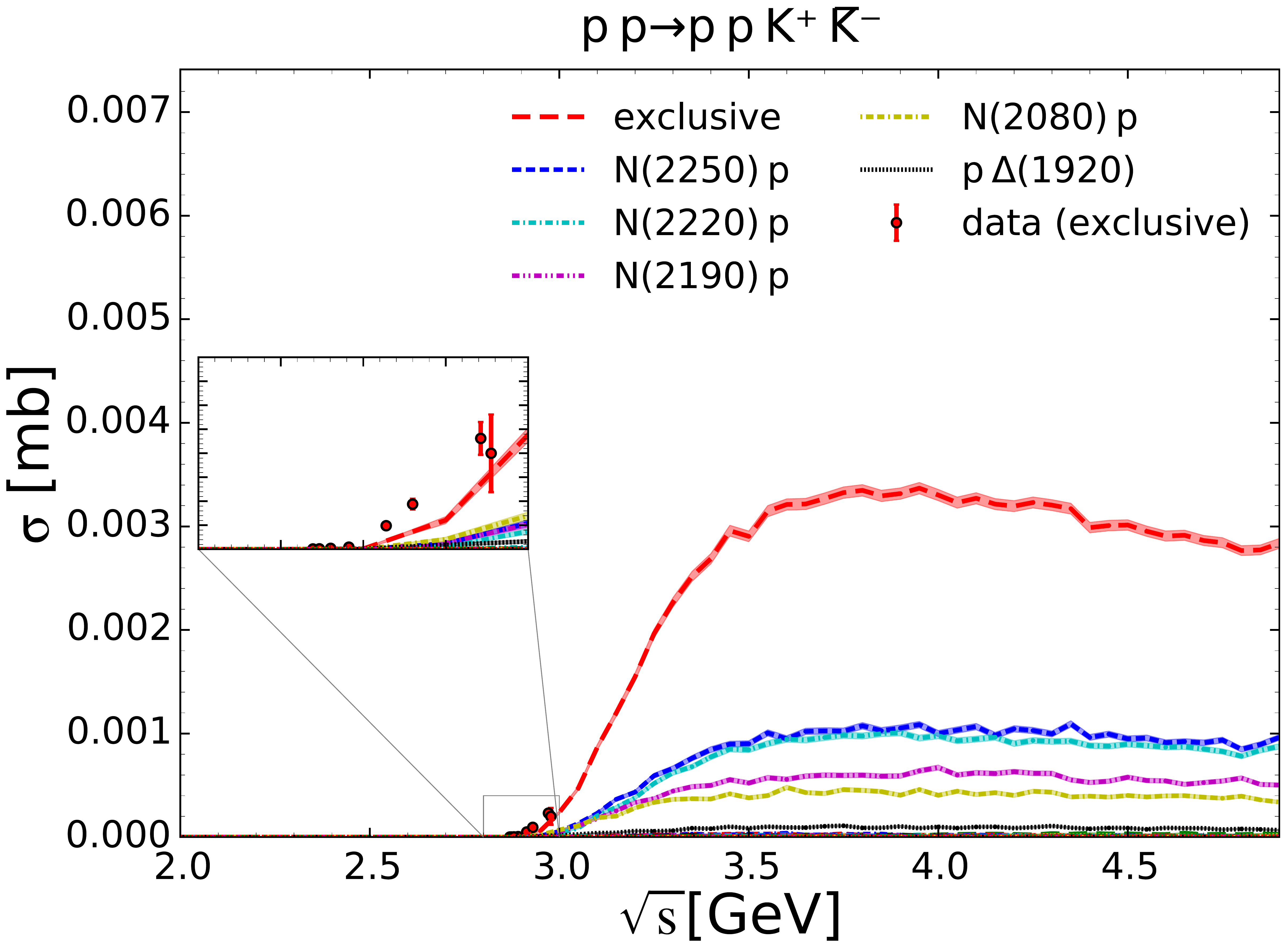}
\caption{$p p \to p p K^+ \bar K^-$ cross section from SMASH compared to experimental data~\cite{Wolke_thesis,Winter:2006vd,Quentmeier:2001ec,Maeda:2007cy,Balestra:2000ex}.
}
\label{fig:xs_PP_PPK+K-}
\end{figure}

The HADES collaboration measured a high $\phi/\bar K^-$ ratio of about~0.5 in gold-gold collisions at $1.23A\,\text{GeV}$~\cite{Adamczewski-Musch:2017rtf}.
Since the branching ratio for the $\phi \to K^+ \bar K^-$ decay is about 0.5~\cite{Patrignani:2016xqp}, this indicates that ca.~25\% of the antikaons are produced via $\phi$ decays.
None of the $N^*$ and $\Delta^*$ resonances have a decay into a $\phi$ meson listed by the PDG.
To produce $\phi$ within SMASH, the experimental uncertainty is exploited to introduce a speculative $N^* \to \phi N$ decay with small, constant branching ratio for all $N^*$ beyond $2\,\text{GeV}$, as proposed in~\cite{Steinheimer:2015sha}.
The branching ratio is constrained by measurements of the $pp \to pp K^+ \bar K^-,\, pp \phi$ cross sections.
According to the $pp \to K^+ \bar K^-$ cross section computed in SMASH (\cref{fig:xs_PP_PPK+K-}), the main contribution to $\phi \to K^+ \bar K^-$ production via $N^*$~resonances is at energies around $\sqrt s = 3.5-4.0\,\text{GeV}$ .
Unfortunately, there is no experimental data for $\sqrt s > 3.0\,\text{GeV}$: The cross sections have mostly been measured close to the threshold~\cite{Wolke_thesis,Winter:2006vd,Quentmeier:2001ec,Maeda:2007cy,Balestra:2000ex}.

\begin{figure}
\centering
\includegraphics[width=\linewidth]{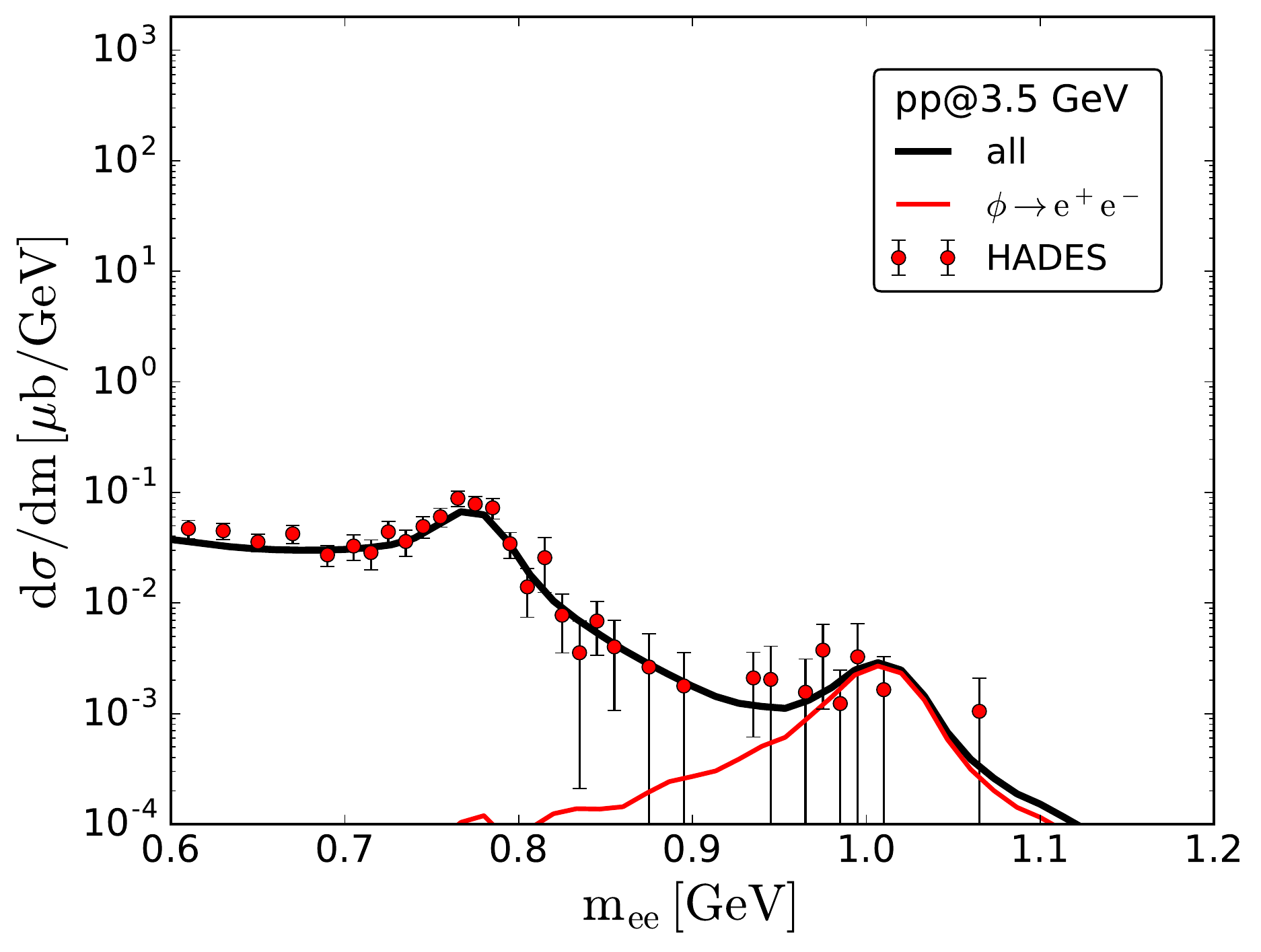}
\caption{
SMASH dilepton spectrum (lines) in proton-proton collisions at $E_\text{kin} = 3.5\,\text{GeV}$ compared to HADES data~\cite{HADES:2011ab} (points).
The $\phi$ contribution (red line) and the total (black line) are shown.
}
\label{fig:dileptons_pp}
\end{figure}

\begin{figure}
\centering
\includegraphics[width=\linewidth]{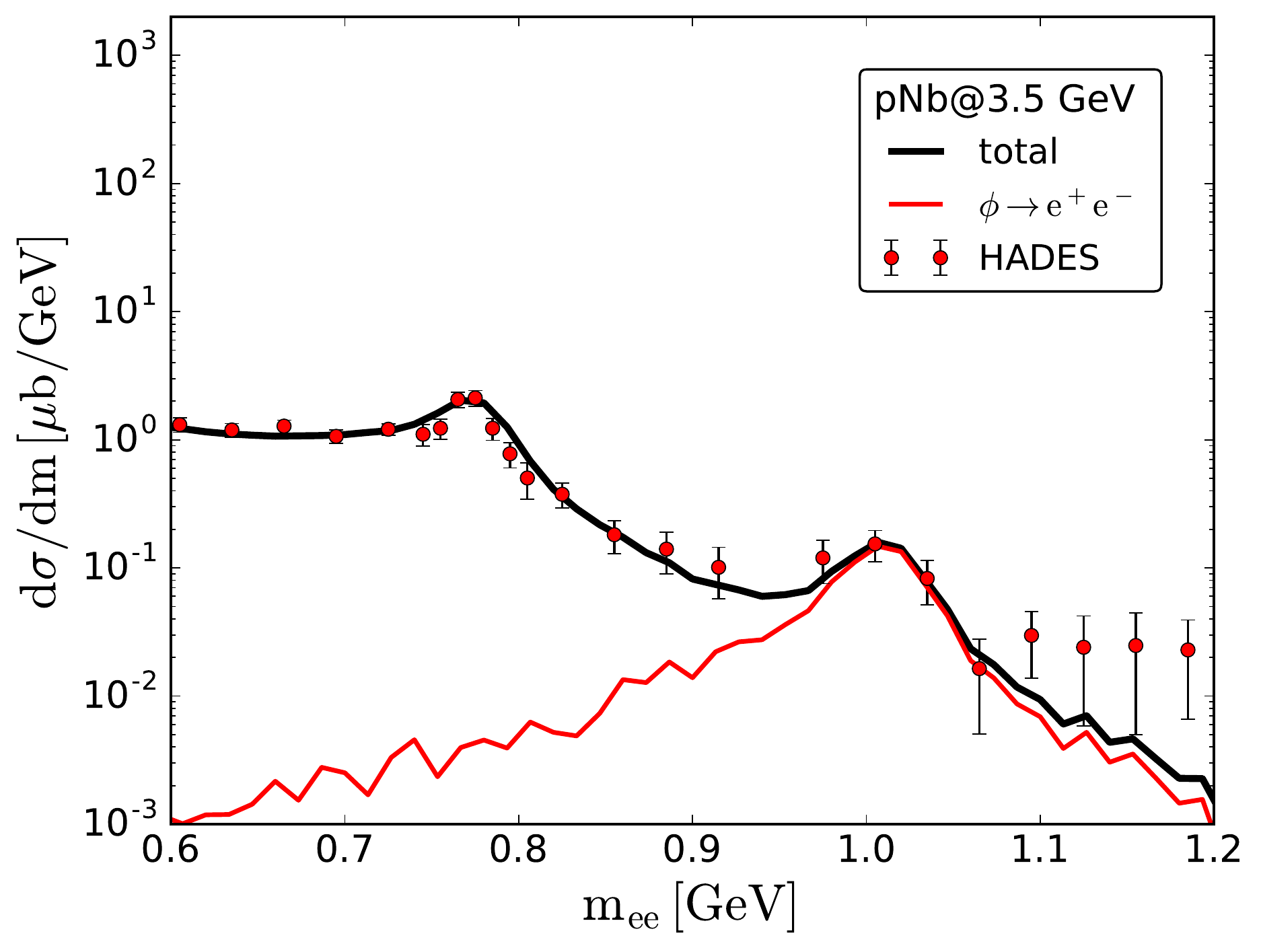}
\caption{
SMASH dilepton spectrum (lines) in proton-niobium collisions at $E_\text{kin} = 3.5\,\text{GeV}$ compared to HADES data~\cite{Agakishiev:2012vj} (points).
The $\phi$ contribution (red line) and the total (black line) are shown.
}
\label{fig:dileptons_pnb}
\end{figure}

Additional constraints of $\phi$-resonant production are imposed by dielectron measurements of the HADES collaboration, specifically the dielectron mass distribution in proton-proton and proton-niobium collisions at $E_\text{kin} = 3.5\,\text{GeV}$~\cite{HADES:2011ab,Agakishiev:2012vj}, which resolves the peaks of many resonances.
While the $\phi$ peak in $pp$ is poorly resolved and only provides a rough upper limit (\cref{fig:dileptons_pp}), the peak in $p \mathit{Nb}$ constrains the $\phi$ production rather well (\cref{fig:dileptons_pnb}).
It is expected that the upcoming HADES gold-gold dilepton spectra will resolve the $\phi$ peak even more precisely.
In SMASH, dileptons are produced during collisions via the shining method~\cite{Staudenmaier:2016hmh,Weil:2016fxr} and the HADES $p \mathit{Nb}$ data has been applied to scale the $N(>\! 2000) \to \phi N$ branching ratios:
\begin{equation}
\frac{\Gamma_{N^* \to \phi N}}{\Gamma_{N^* \to X}} = 0.5\%
\end{equation}
This result for the $\phi$ is larger than the value employed by UrQMD~\cite{Steinheimer:2015sha}.
It should be noted that the only in-medium effect exerted on the $\phi$ in SMASH is collisional broadening.
There may be significant additional in-medium effects on the cross sections, effectively changing the $\phi$ production branching ratio.
%
Applying the dilepton constraints, the exclusive cross section shown in \cref{fig:xs_PP_PPK+K-} is decently described by SMASH except for two underestimated data points close to the threshold.


It is not clear how the non-$\phi$ contribution to the $pp \to pp K^+ \bar K^-$ cross section should be described.
Previous studies suggested that final state interactions or a mixture of~$a_0(980)$ and~$f_0(980)$ resonances may play a role~\cite{Maeda:2007cy}.
However, introducing an $N^* \to f_0(980) N$ branching ratio of 0.1\% did not change the results in this work.


\subsection{Momentum spectra in proton-proton collisions by HADES}
\label{sec:HADES_pp}

\begin{figure}
\centering
\includegraphics[width=0.45\linewidth]{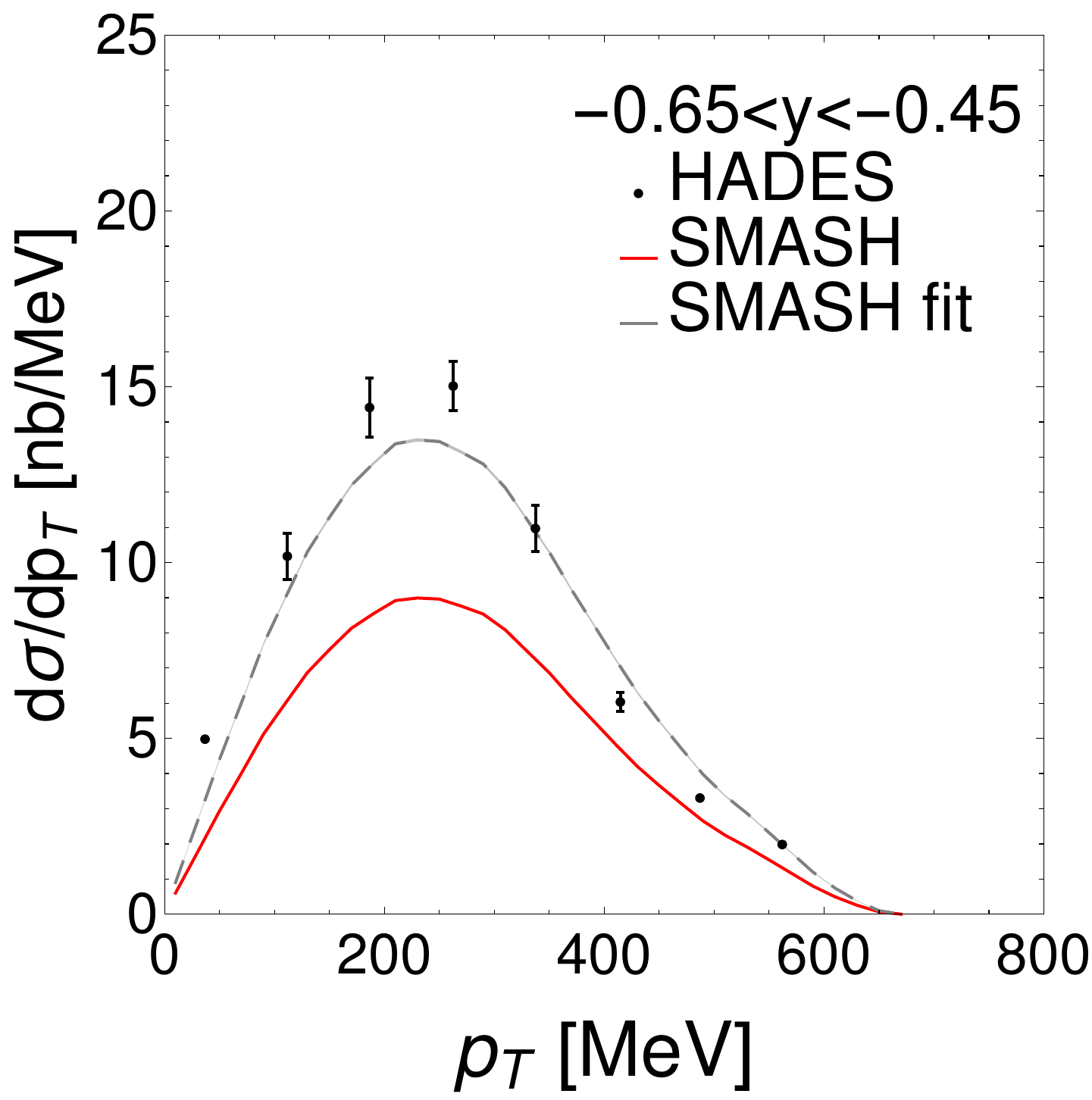}
\includegraphics[width=0.45\linewidth]{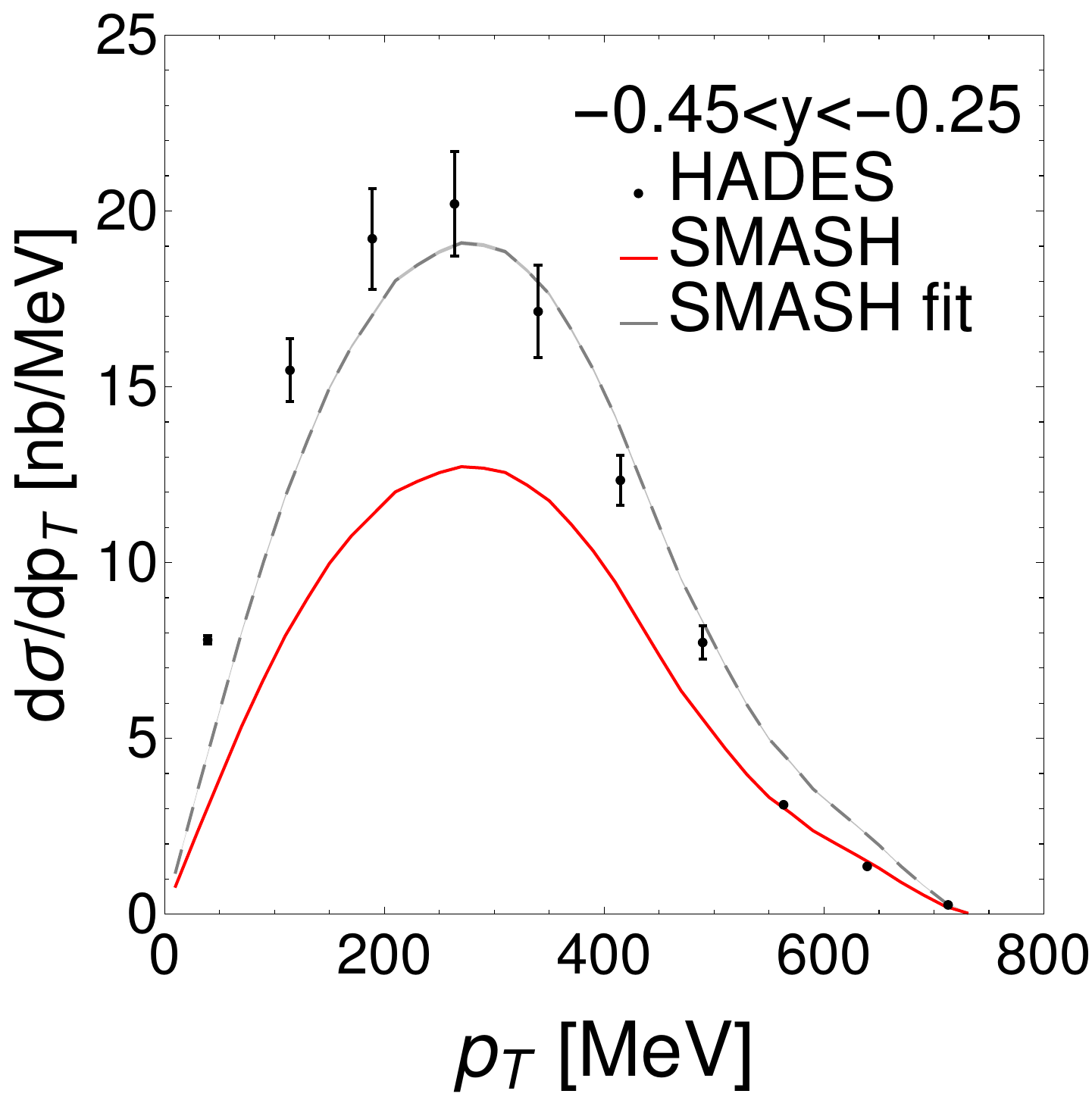}
\includegraphics[width=0.45\linewidth]{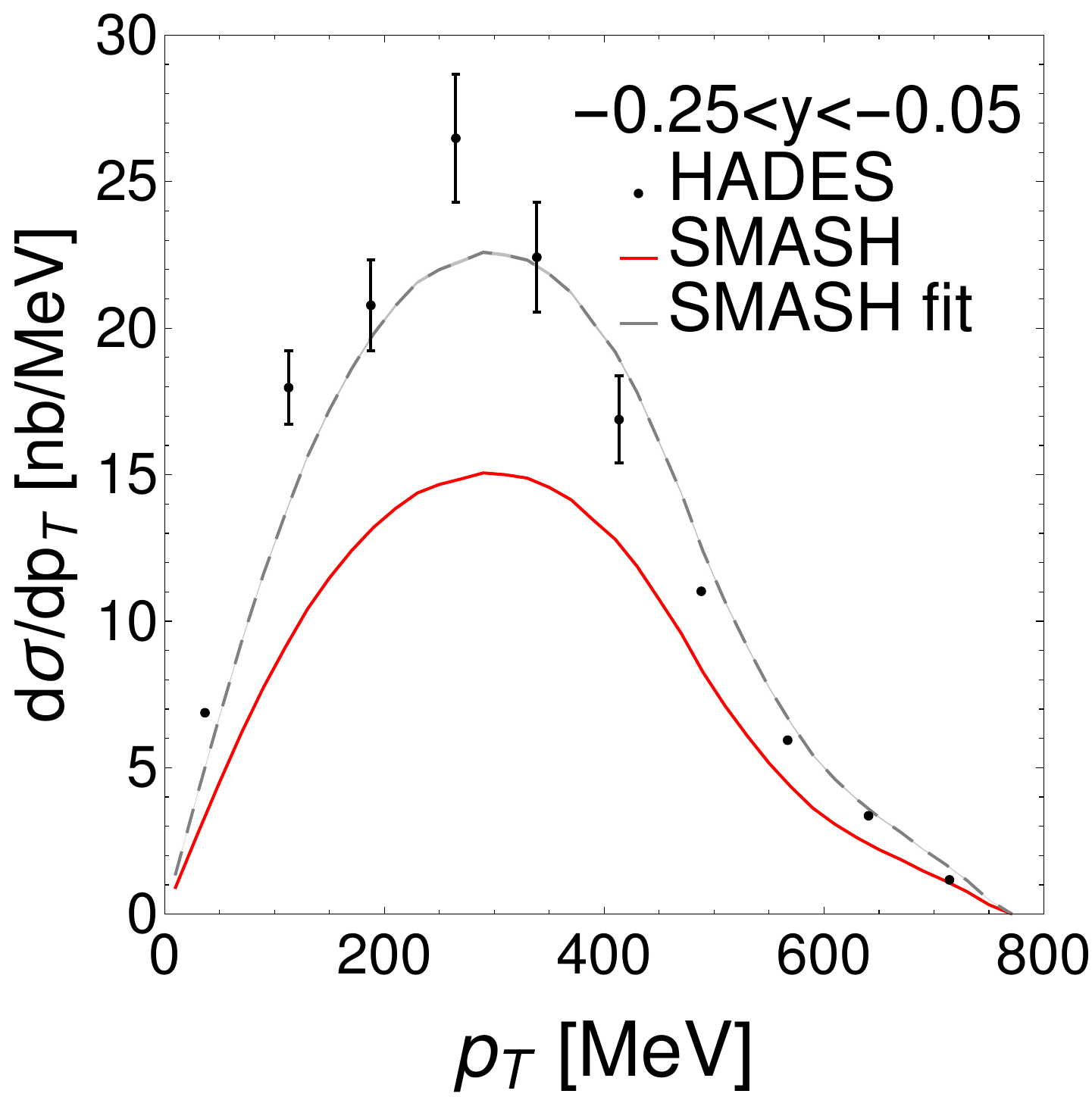}
\includegraphics[width=0.45\linewidth]{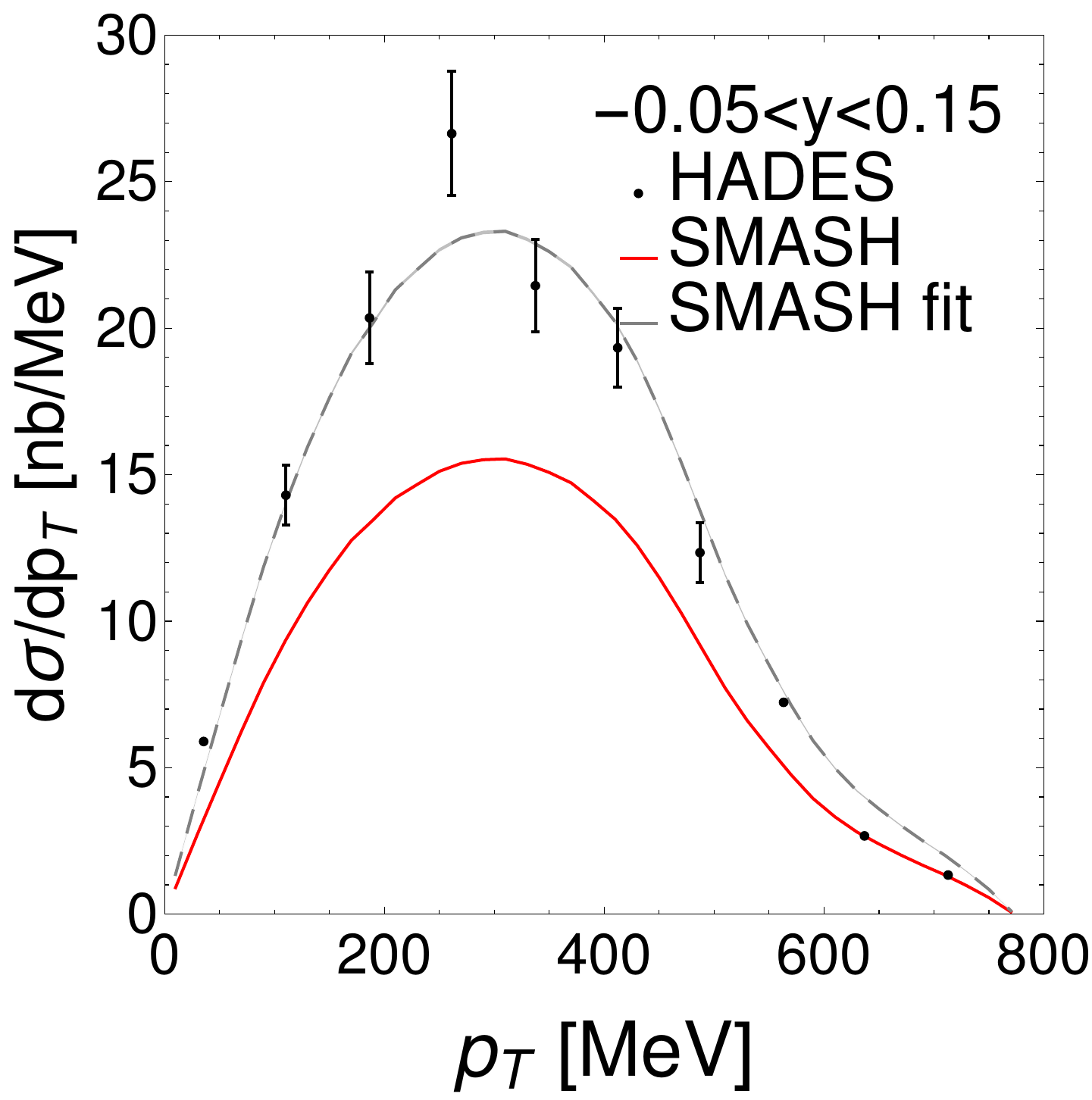}
\includegraphics[width=0.45\linewidth]{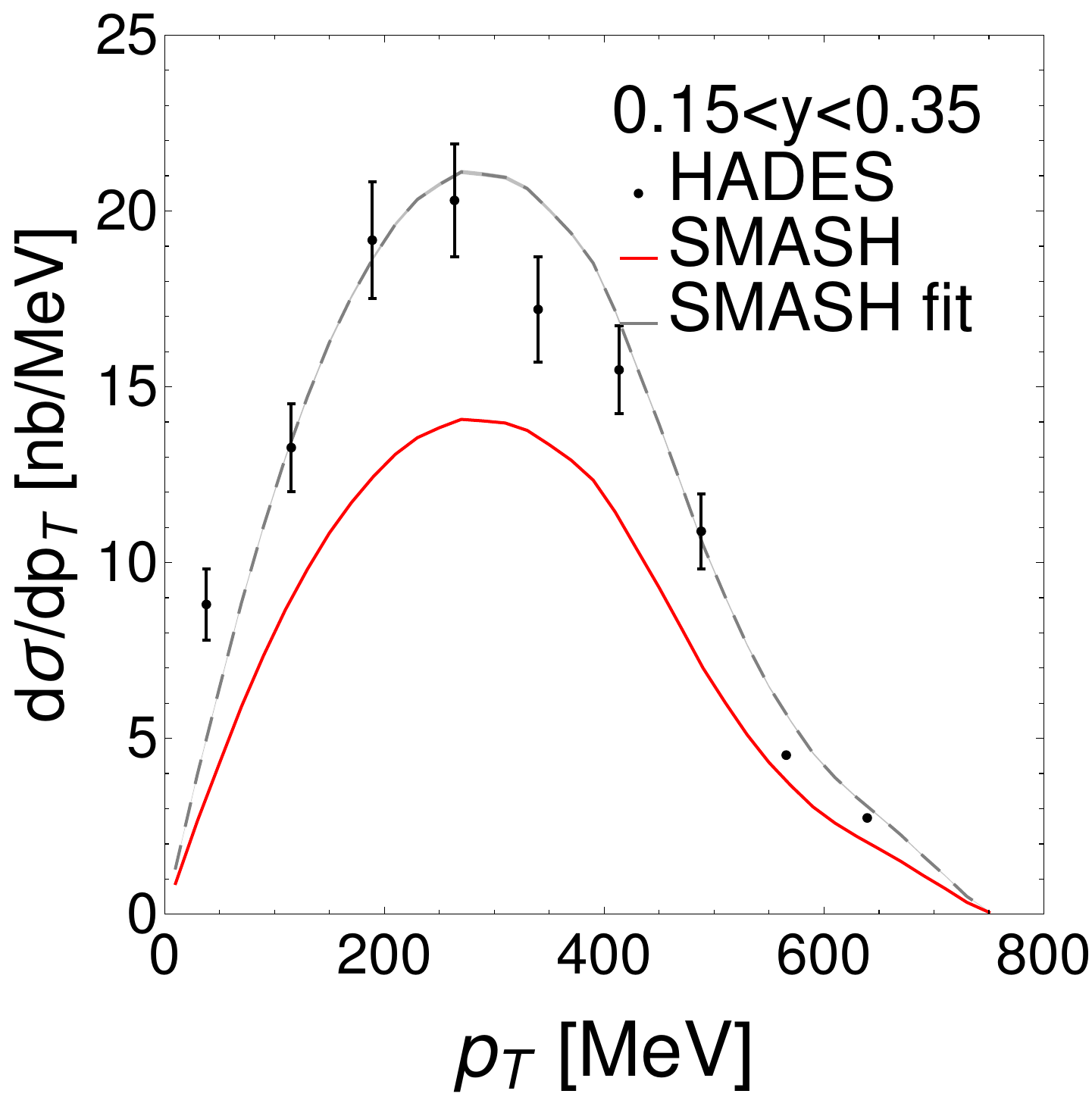}
\includegraphics[width=0.45\linewidth]{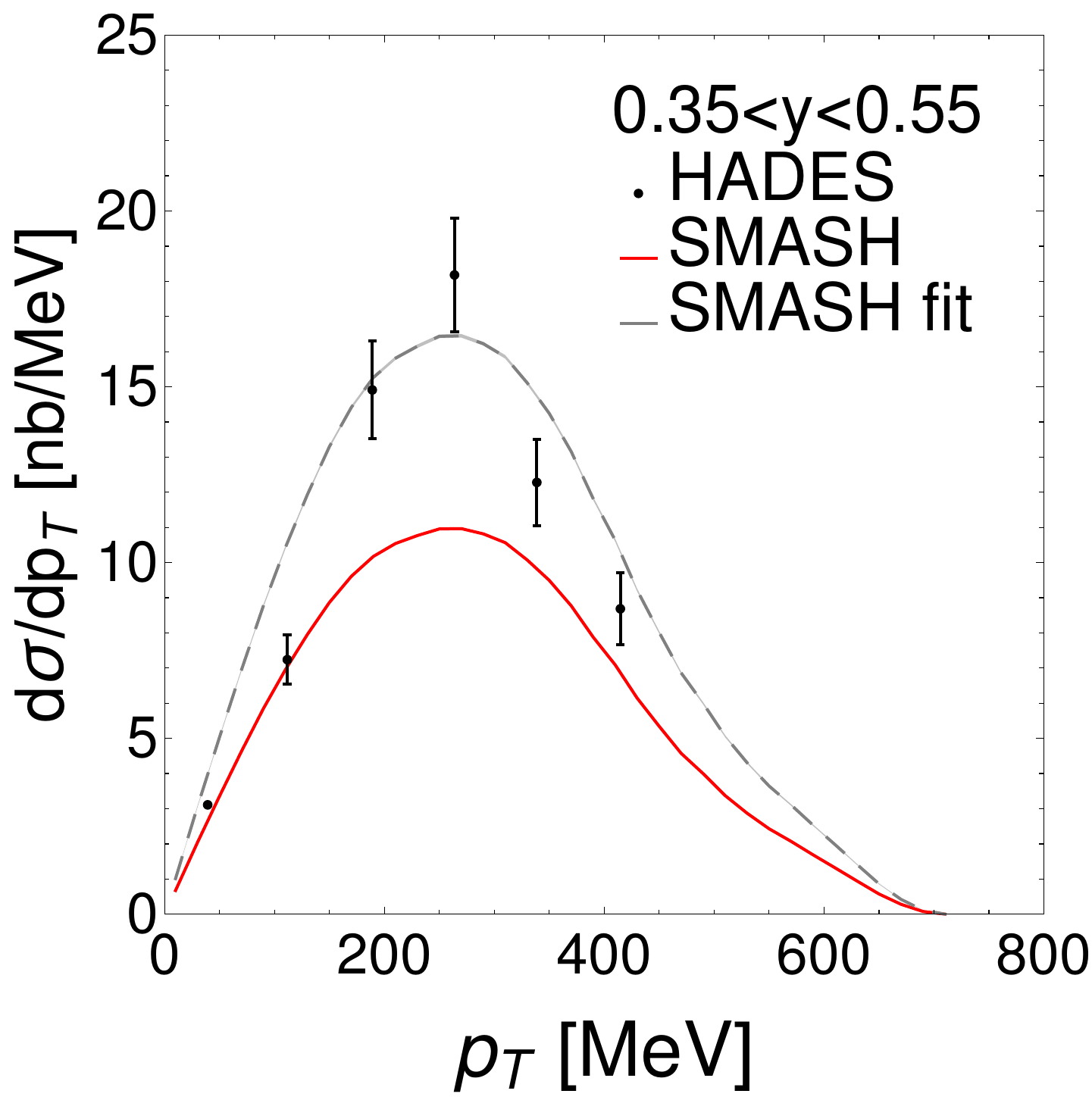}
\caption{ $pp\to K^0_SX$ differential cross sections as functions of the transverse momenta $p_T$ of $K^0_S$ at $E_{\rm kin} = 3.5$ GeV from SMASH output (red band) and the scaled SMASH output (gray band) compared to experimental data~\cite{Agakishiev:2014moo} within different rapidity bins.
}
\label{fig:xs_pT_pp}
\end{figure}

\begin{figure}
\centering
\includegraphics[width=\linewidth]{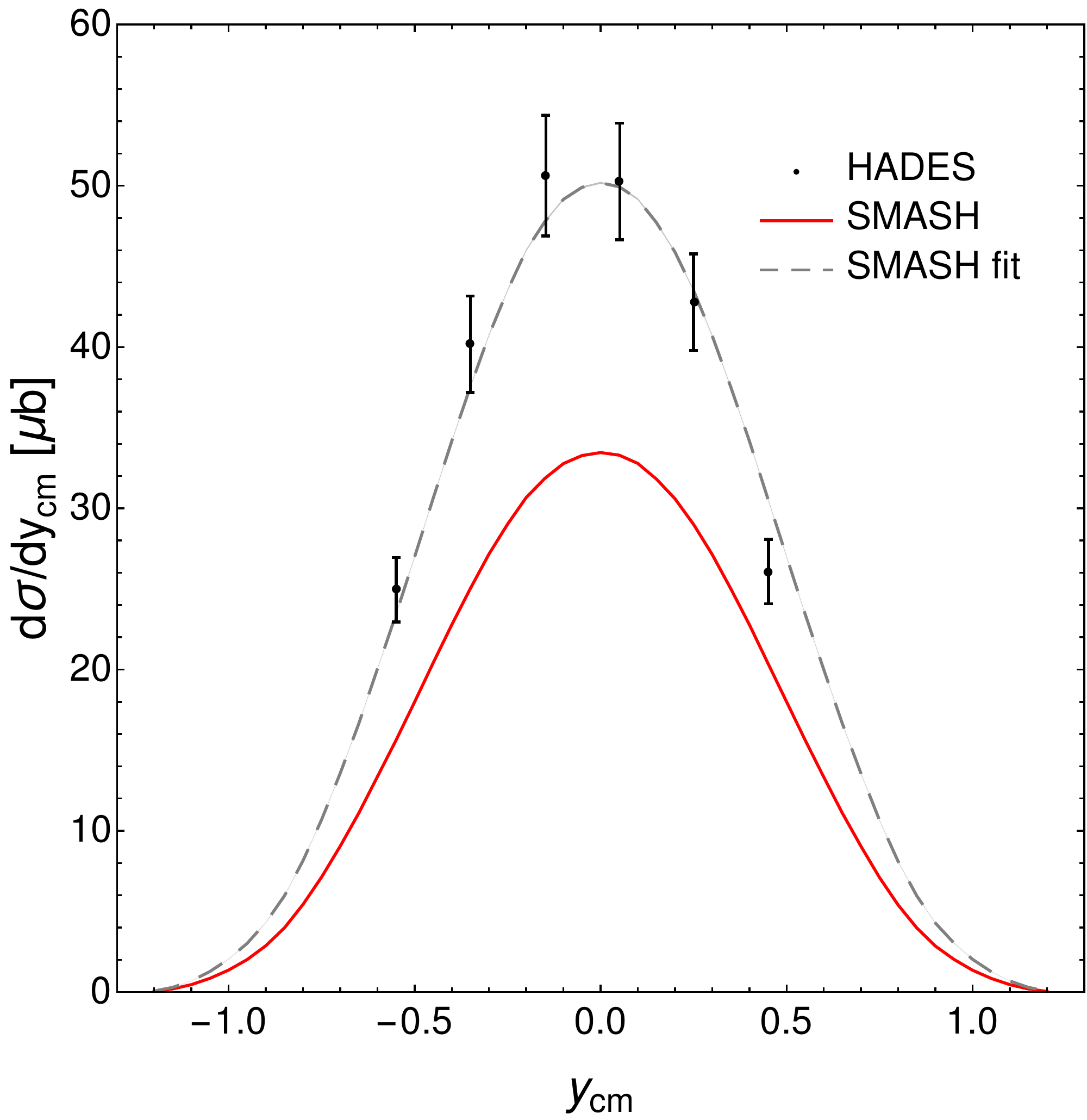}
\caption{ $pp\to K^0_SX$ differential cross sections as a function of the rapidity $y_{\rm cm}$ of $K^0_S$ at $E_{\rm kin} = 3.5$ GeV from SMASH (red band) and rescaled SMASH (gray band) compared to experimental data~\cite{Agakishiev:2014moo}.
}
\label{fig:xs_y_pp}
\end{figure}

The $K^0_S$ production cross section in proton-proton collisions at $E_\text{kin} = 3.5\,\text{GeV}$ as a function of transverse momentum and rapidity was measured by the HADES collaboration and compared to the GiBUU transport model~\cite{Agakishiev:2014moo}.
The influence of the implemented $KN$ potential was found to be negligible.
To reproduce the spectra, the individual $K^0$ production cross sections had to be rescaled in GiBUU.
This is not easily possible in SMASH, because the relevant cross sections are not directly parametrized but rather derived from the resonance properties.

The differential cross sections for $K^0_S$ production can be reconstructed from SMASH output.
Because SMASH only has $K^0$ and $\bar K^0$ as degrees of freedom, it is assumed that half of the $K^0$ and $\bar K^0$ in SMASH correspond to a $K^0_S$.
When confronting the SMASH results with the measured differential cross sections as a function of transverse momentum for different rapidity bins (\cref{fig:xs_pT_pp}) and as a function of rapidity (\cref{fig:xs_y_pp}), the differential cross section is underestimated for all rapidities.
Scaling the SMASH cross section up by a factor~1.5 improves the agreement with the data for all rapidities and transverse momenta, showing that the shape is reproduced.
Within our production model, where strangeness in $pp$ is produced via $pp \to NB^* \to YNK$, this underestimation is in tension with the $pp \to \Sigma^+ p K^0$ cross section shown in~\cref{fig:PP_Sigma+PK0}, which was somewhat overestimated at the energy corresponding to the HADES measurement.
On the other hand, there may be other $K^0$~production mechanisms missing in SMASH.
Furthermore, the experimental data in~\cref{fig:PP_Sigma+PK0} is much less extensive and was measured before 1988~\cite{LaBoer}, while the more recent HADES data is from 2014~\cite{Agakishiev:2014moo}.
However, the newer data cannot be applied to improve the branching ratios in SMASH, because the $K^0$ production via decays into $\Sigma^+ K^0$ can only be scaled via the $N^*, \Delta^* \to K\Sigma$ branching ratios, which also affect $K^+$ production.
The latter is already large enough, as seen in \cref{fig:PP_Sigma0PK+,fig:PP_Sigma+NK+,fig:xs_pi-P_Sigma-K+,fig:xs_pi+P_Sigma+K+}.
An increase by~50\% would result in a much worse agreement with the data.

\section{Strangeness production in heavy-ion collisions}
\label{sec:hic}

In the previous section, our hadron-resonance approach was tuned to describe elementary reactions.
Now we can compare it to the experimental data obtained in heavy-ion reactions, where secondary reactions and Fermi motion provide additional energy, allowing for sub-threshold production of strange particles.
Furthermore, resonances can act as an energy storage and can be formed in secondary collisions with the medium.
They are affected by in-medium effects such as collisional broadening.
In this section, the different systems studied at SIS are discussed in historical order:
nickel-nickel and gold-gold by KaoS (\cref{sec:NiNi_AuAu}), and Ar-KCl (\cref{sec:ArKCl}), gold-gold (\cref{sec:AuAu}), pion-carbon (\cref{sec:pion_beam}) by HADES.
The different sizes of the systems allow conclusions about possible in-medium effects.

In SMASH, the nuclei are initialized with a Wood-Saxon distribution as described in~\cite{Weil:2016zrk}.
For this work, nucleon-nucleon potentials and Pauli blocking are not employed.
In our preliminary studies, we found that they do not affect the multiplicities of strange particles, while consuming a lot of computational resources.
To allow for Fermi motion without the potentials holding the nuclei together, the Fermi momentum of each nucleon is ignored for propagation until its first interaction.
This treatment is refered to as the frozen Fermi approximation.


\subsection{Ni-Ni and Au-Au collisions by KaoS}
\label{sec:NiNi_AuAu}

\begin{figure}
\centering
\includegraphics[height=0.29\textheight]{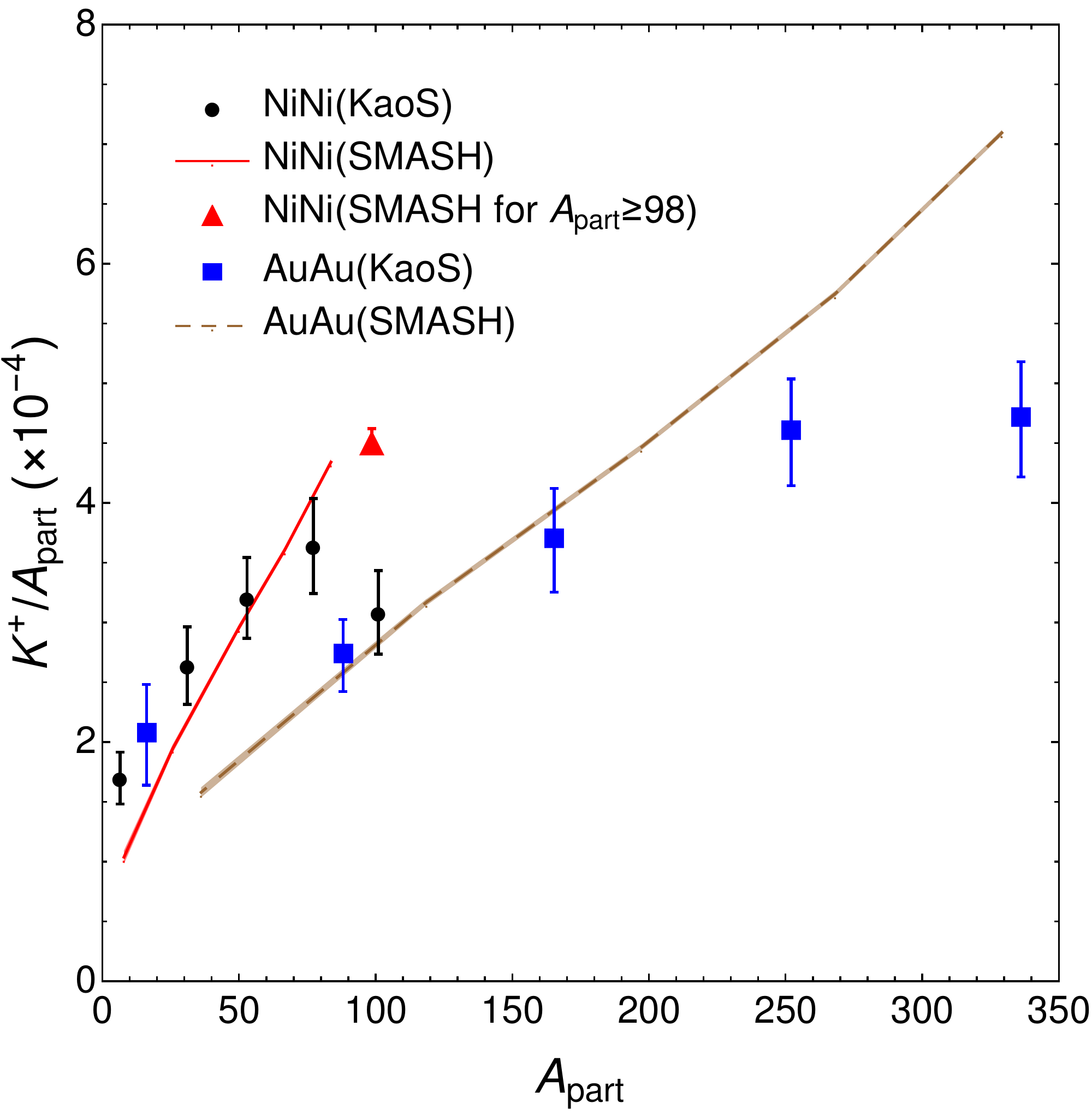}
\includegraphics[height=0.29\textheight]{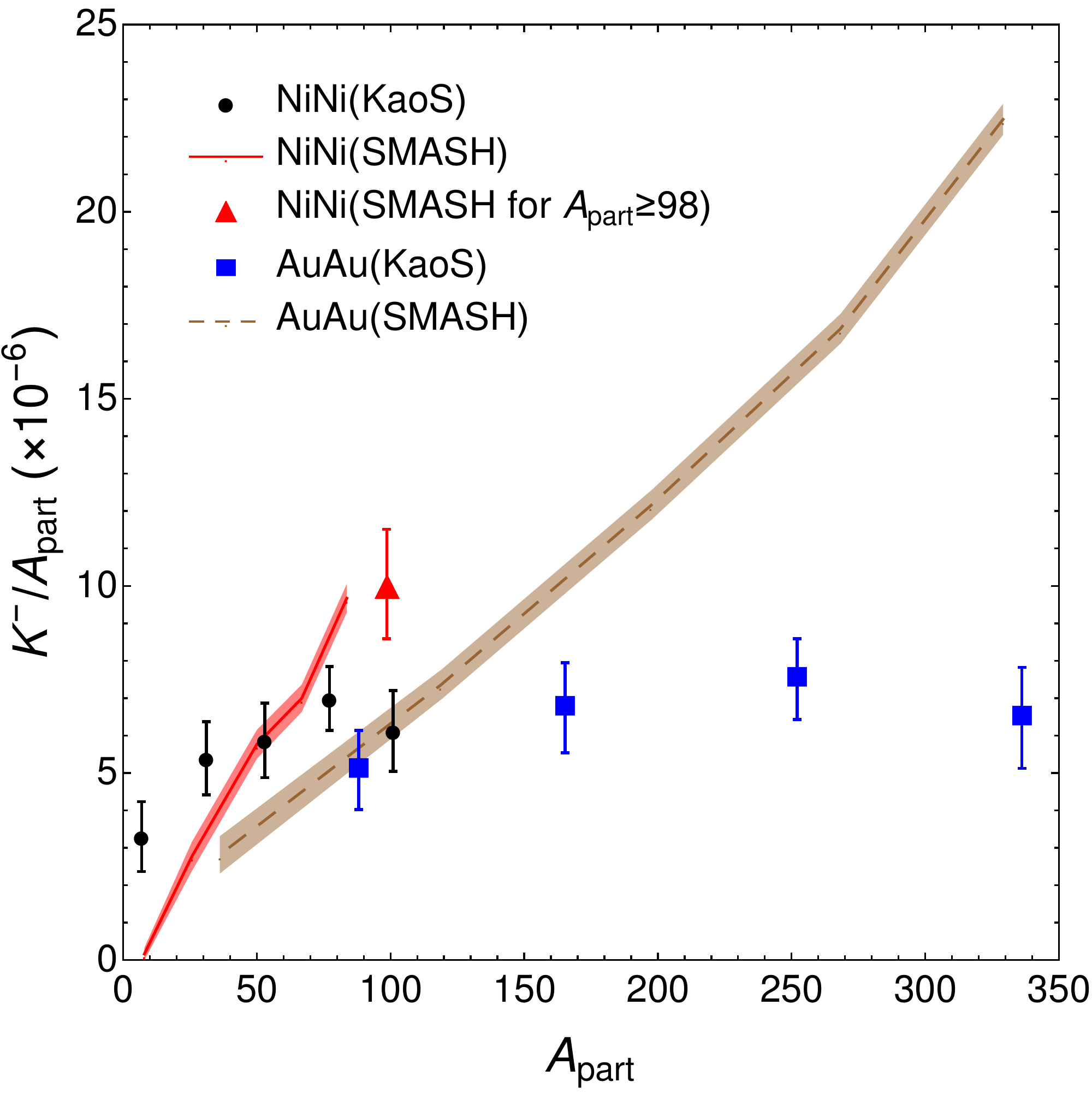}
\includegraphics[height=0.29\textheight]{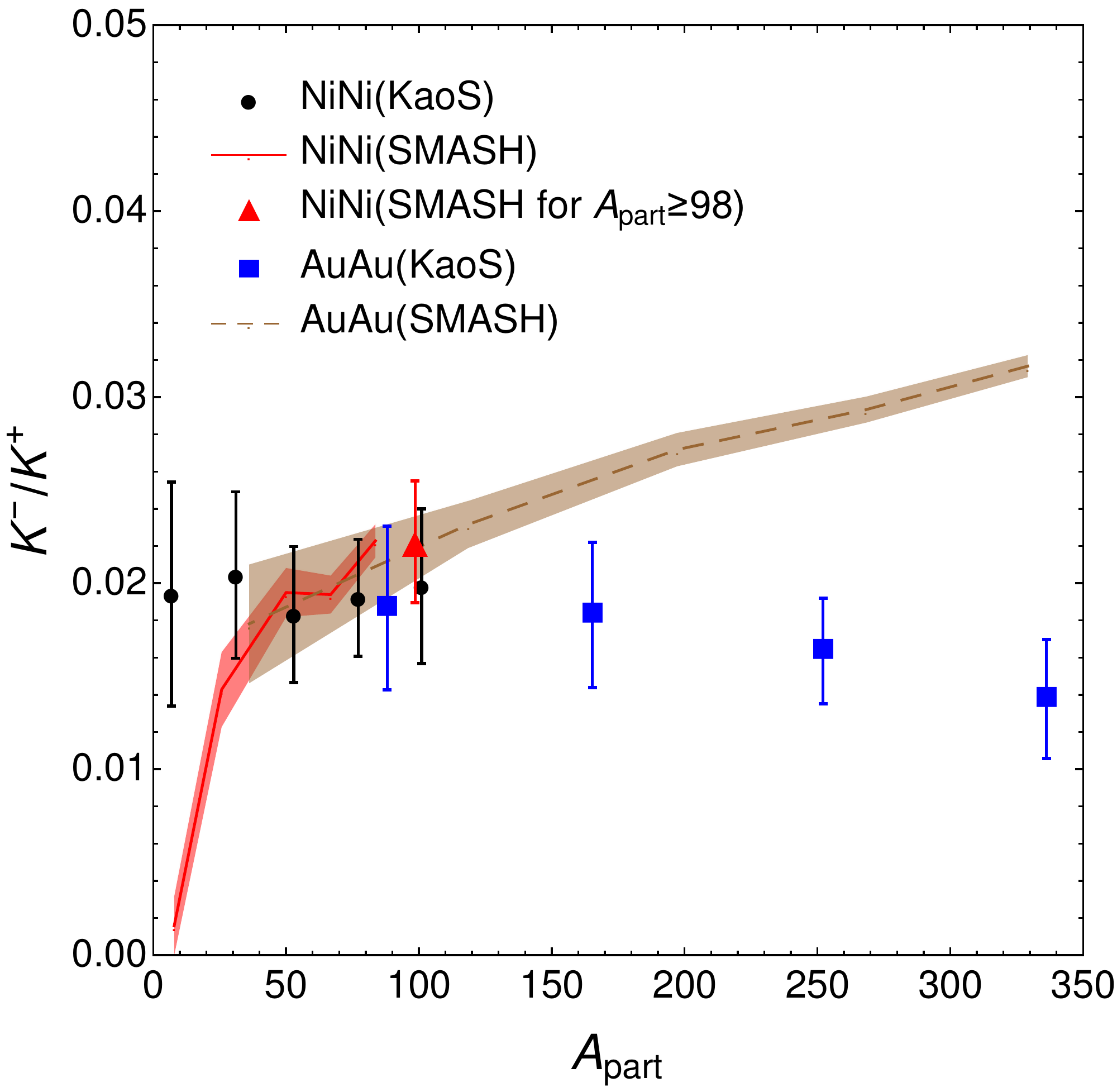}
\caption{
The multiplicities of~$K^+$ (upper panel) and~$\bar K^-$ (middle panel) per number of participants, as well as their ratio (lower panel), produced in both the Ni-Ni and Au-Au collisions at $E_\text{kin} = 1.5A\, \text{GeV}$ with different centralities as a function of the participant number~$A_\text{part}$~\cite{Forster:2007qk}.
Additional red triangular points are shown for the events among the head-on Ni-Ni collisions which have at least 98 participants.
}
\label{fig:K_Apart}
\end{figure}

The KaoS collaboration measured the multiplicities of $K^+$ and $\bar K^-$ in nickel-nickel and gold-gold collisions as a function of the number of participants~$A_\text{part}$~\cite{Forster:2007qk}.
The multiplicities for these systems with SMASH are compared to the experimental data in \cref{fig:K_Apart}.
In the experiment, the number of participants is estimated with a Glauber model.
In SMASH, it is determined microscopically: Any initial particle that scatters inelastically is assumed to be a participant.
(It is important to exclude elastic scatterings, because in our simulation they are common among spectators.)
The number of participants are adjusted by choosing different ranges of the impact parameter.
However, the expectation value of the participant number in SMASH is below~100 even for head-on nickel-nickel collisions.
To get the multiplicities of $K^+$ and $\bar K^-$ at a participant number of about~$100$, the head-on central collisions which have at least 98 participants were utilized.

For nickel-nickel (red bands in \cref{fig:K_Apart}), the multiplicities of $K^+$ and $\bar K^-$ from SMASH are similar to the experimental ones.
There is some underestimation at low~$A_\text{part}$ and an overestimation at high~$A_\text{part}$.
The underestimation might be due to systematic differences in the centrality determination, which could be more prominent for small numbers of participants.
The ratio agrees with the data, except for the bin with the lowest number of participants.
For gold-gold (brown bands in \cref{fig:K_Apart}), the agreement is good for low~$A_\text{part}$, but for higher~$A_\text{part}$ the multiplicities are increasingly overestimated.
This happens more drastically for $\bar K^-$ than for $K^+$.
As a consequence, the ratio is well described for low participant numbers, but increasingly overestimated for higher numbers.
The IQMD transport model, which includes repulsive kaon-nucleon and attractive antikaon-nucleon potentials, showed a similar linear rise and a similar overestimation when comparing to the Au-Au collisions measured by KaoS~\cite{Hartnack:2011cn}.

Even when the magnitudes of the multiplicities~$N$ are similar, their slope is different: With SMASH, $N / A_\text{part}$ increases linearly, while the experimental data shows a plateau at large $A_\text{part}$.
In RHIC gold-gold collisions at higher energies, a stronger saturation was observed~\cite{Adamczyk:2017iwn} which can be understood in terms of a core-corona model~\cite{Becattini:2008ya}:

At its core, the colliding system behaves like a hadron gas in chemical equilibrium.
The produced multiplicities are proportional to the volume of the core, that is proportional to~$A_\text{part}$.
In the corona surrounding the core, the system behaves like many independent collisions.
The multiplicities are proportional to number of interactions, which scales as some function of~$A_\text{part}$ between $A_\text{part}$ (participants interact once) and $A_\text{part}^2$ (participants interact with every other participant).

Fits of the core-corona model to the centrality-dependence of the experimental data suggest that for low~$A_\text{part}$, particle production from the corona dominates, while for large~$A_\text{part}$ production in the core becomes more important.
Qualitatively, the saturation in the KaoS data would be expected from a core-dominated production, but there is less than one kaon per collision, rendering a chemical equilibrium in the core implausible.
In this regard, the ballistic production expected from the corona at low centralities is captured by SMASH, but even in central collisions the kaons are produced ballistically, which differs from the data.
A similar behavior was observed with the IQMD model~\cite{Hartnack:2011cn}.

The dependence of the multiplicities of $K^+$ and $\bar K^-$ on the number of participants~$A_\text{part}$ can be fitted by a power function proportional to $A_{\rm part}^\alpha$, where the power index $\alpha\in[1,2]$ has been determined by least square regression to the SMASH results: $\alpha_{K^+}({\rm Ni})=1.61\pm 0.073$, $\alpha_{\bar K^-}({\rm Ni})=1.85\pm 0.307$, $\alpha_{K^+}({\rm Au})=1.87\pm 0.053$, and $\alpha_{\bar K^-}({\rm Au})=1.93\pm 0.148$, which are all larger than the experimental values listed in~\cite{Forster:2007qk}.

\begin{figure}
\centering
\includegraphics[height=0.3\textheight]{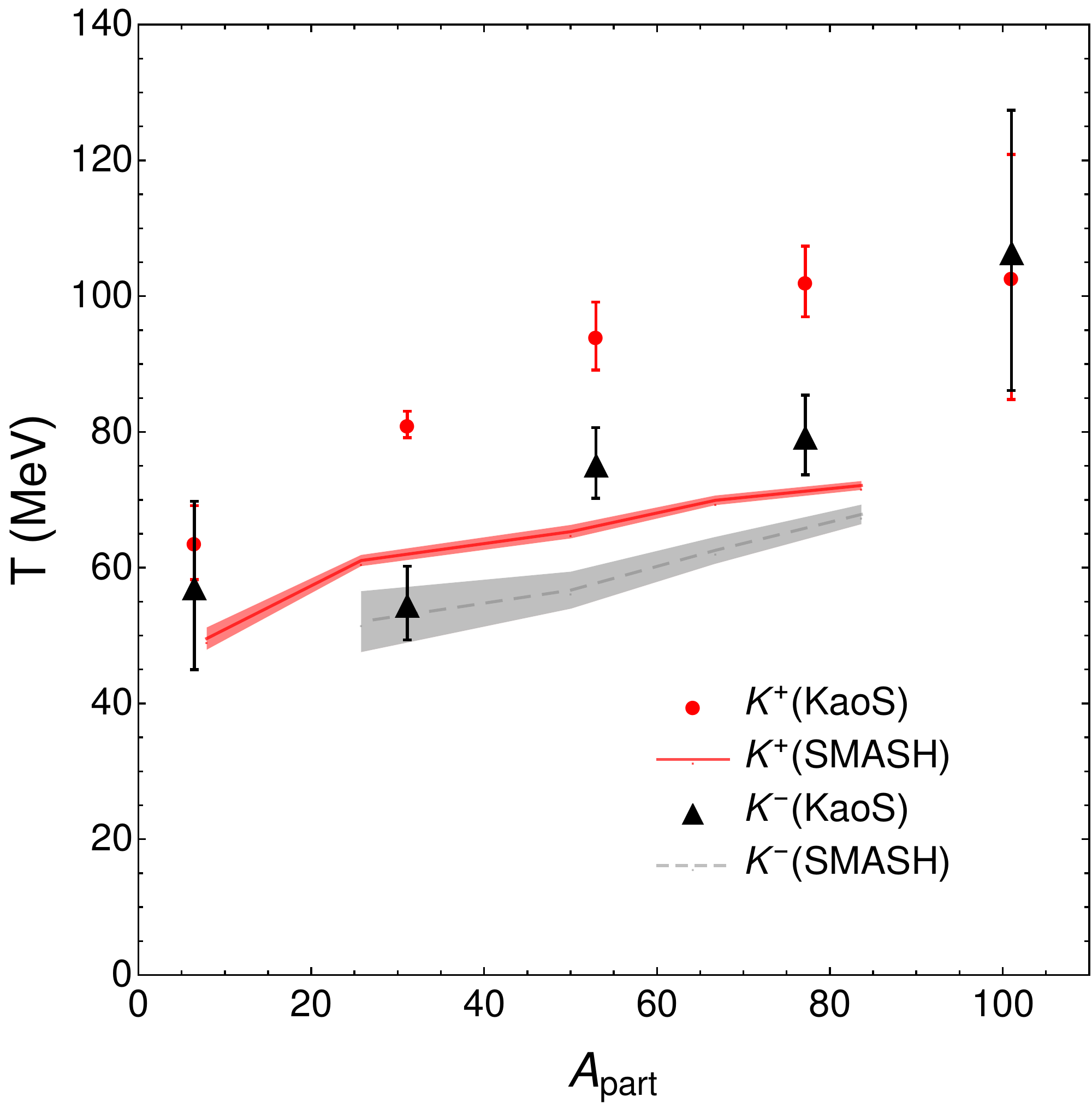}
\caption{
The inverse slope parameter~$T$ of $K^+$ and $\bar K^-$ produced in $4\,000\,000$ Ni-Ni collisions at $E_\text{kin} = 1.5A\, \text{GeV}$ with different centralities as the functions of the participant number~$A_\text{part}$~\cite{Forster:2007qk}.
}
\label{fig:T_NiNi_KaoS}
\end{figure}

\begin{figure}
\centering
\includegraphics[height=0.3\textheight]{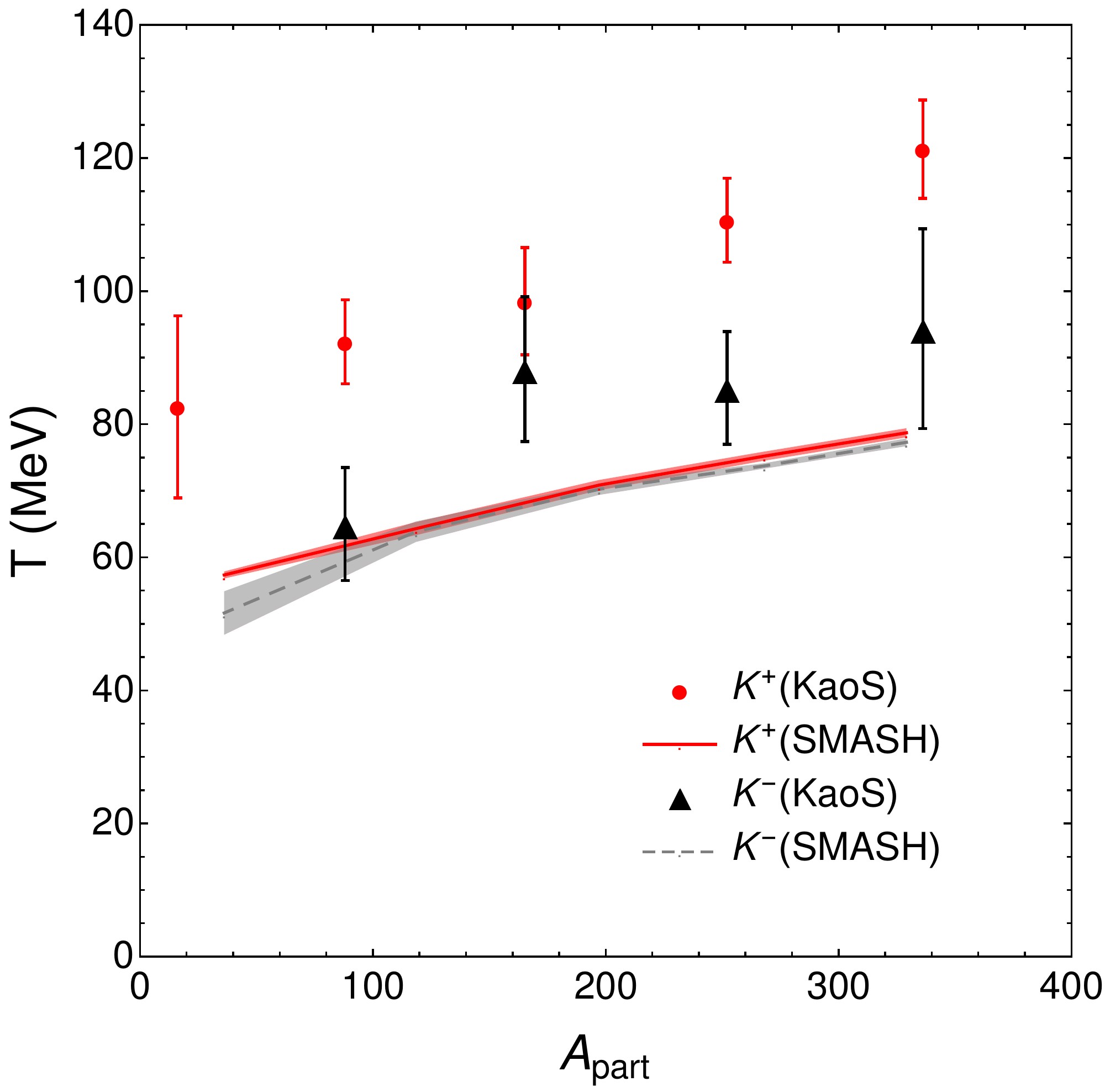}
\caption{
The inverse slope parameter~$T$ of $K^+$ and $\bar K^-$ produced in $2\,000\,000$ Au-Au collisions at $E_\text{kin} = 1.5A\, \text{GeV}$ with different centralities as the functions of the participant number~$A_\text{part}$~\cite{Forster:2007qk}.
}
\label{fig:T_AuAu_KaoS}
\end{figure}

From a Boltzmann fit to the transverse mass spectra, the inverse slope parameter~$T$ is calculated with SMASH for the~$K^+$s and the~$\bar K^-$s produced in both the nickel-nickel and gold-gold collisions and compared to the KaoS data~\cite{Forster:2007qk} in \cref{fig:T_NiNi_KaoS} and \cref{fig:T_AuAu_KaoS}, respectively.
Both the simulation results and the experimental data show that the inverse slopes increase with the participant number.
However, it is shown in the experimental data that the inverse slopes of the $\bar K^-$s are lower than those of the $K^+$s for all the centralities and both systems, which could be interpreted as a later freeze-out of $\bar K^-$ than $K^+$.
Meanwhile, the SMASH results show inverse slopes that are only slightly higher for $K^+$ than for $\bar K^-$.
For $\bar K^-$, the SMASH slopes are more similar to the data than for $K^+$.
Within IQMD, the kaon-nucleon potential was shown to be the main reason why the inverse slopes of the $K^+$s are higher than those of the $\bar K^-$s, since the repulsive forces between the $K^+$s and the nucleons will enhance the $K^+$s' transverse momenta, while the attractive forces between the $\bar K^-$s and the nucleons will reduce the $\bar K^-$s' transverse momenta~\cite{Hartnack:2011cn}.
Compared to IQMD (where only the $\Delta$ is implemented as a resonance and $\bar KN \leftrightarrow \pi Y$ is parametrized), SMASH employs 22~hyperon resonances to model $\bar KN$ scattering, which corresponds to an effectively attractive interaction.
There is no repulsive kaon-nucleon interaction in SMASH, which might explain the underestimation of the $K^+$ inverse slope.

In summary, the results from SMASH are in good agreement with the data for small systems and at low numbers of participants, but overestimate the kaon and antikaon multiplicities when the number of participants increases, which is similar to previous studies.
This could be due to in-medium effects that are not included in SMASH.
As seen in the IQMD calculations~\cite{Hartnack:2011cn}, a repulsive kaon-nucleon potential can reduce the multiplicity and increase the inverse slope of kaons, while an attractive antikaon-nucleon potential would do the opposite to the antikaons.


\subsection{Ar-KCl collisions by HADES}
\label{sec:ArKCl}

The dynamics of strangeness production have been investigated by the HADES collaboration~\cite{Agakishiev:2010rs} by measuring the transverse mass spectra of $K^+$ and $\bar K^-$ and $\Lambda$~hyperons in Ar-KCl collisions.
For the simulations with SMASH, KCl was approximated by averaging the number of neutrons and protons, which corresponds to Ar-37.

\begin{figure}
\centering
\includegraphics[width=\linewidth]{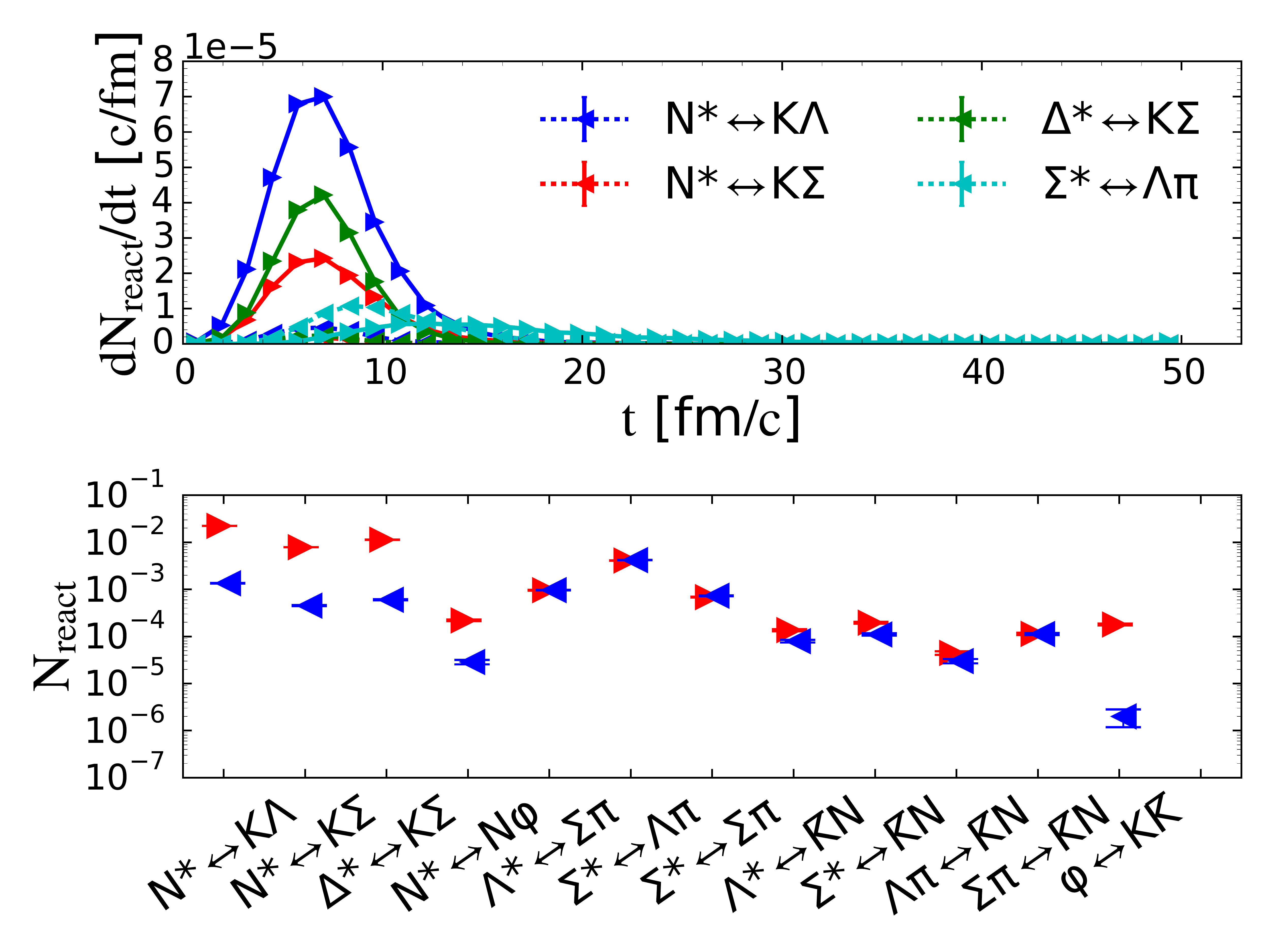}
\caption{
Average strangeness production in Ar-KCl collisions at $E_\text{kin} = 1.76A\,\text{GeV}$.
The upper plot shows the most important production and absorption rates as a function of time.
The lower one shows the total number of forward (red) and backward (blue) reactions involving strange particles.
}
\label{fig:rates_ArKCl}
\end{figure}

To illustrate how strangeness production in this system proceeds in SMASH, \cref{fig:rates_ArKCl} shows the reaction rates for the different strangeness production channels, averaged over $58\,800\,000$~events.
In the beginning of the collision, strangeness production via $N^*$ and $\Delta^*$ decays into hyperons and kaons dominates.
At about $5\,\text{fm}/c$, the backward reactions, indicated by the left pointing triangles, kick in.
For $N^*$ and $\Delta^*$, they are dominated by the forward reactions, but for $\Sigma^*$ decaying into $\Lambda$s and pions, the backward reactions are dominant.
This changes at $12\,\text{fm}/c$, where the forward reactions are more numerous.
They persist until about $30\,\text{fm}/c$, while the strangeness production via $N^*$ and $\Delta^*$ ends after approximately $16\,\text{fm}/c$.

Looking at the absolute rates for each channel over the whole evolution of the heavy-ion collision reveals that $N^*$ and $\Delta^*$ are responsible for producing kaons, while the meson decays do not have a significant net contribution.
As discussed in \cref{sec:meson_resonances}, $\phi$ mesons are produced via the decay of (heavy) $N^*$ resonances.
The $\phi$ decays are an important source of antikaons, while their contribution to kaon production is insignificant compared to the dominating channels.
Hyperon decays also produce a significant amount of antikaons, while the parametrized strangeness exchange channels effectively absorb them.

\begin{figure}
\centering
\includegraphics[width=\linewidth]{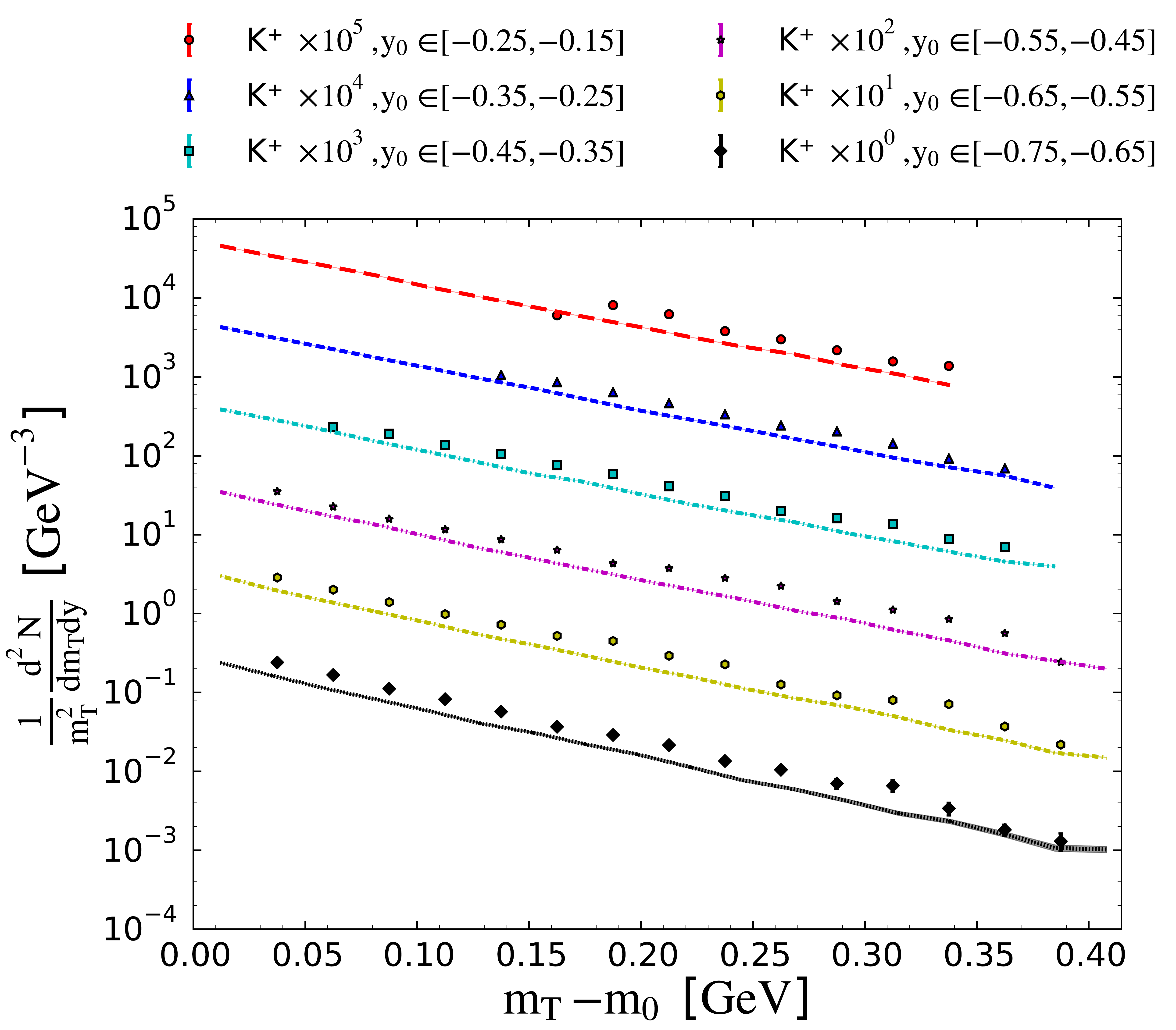}
\includegraphics[width=\linewidth]{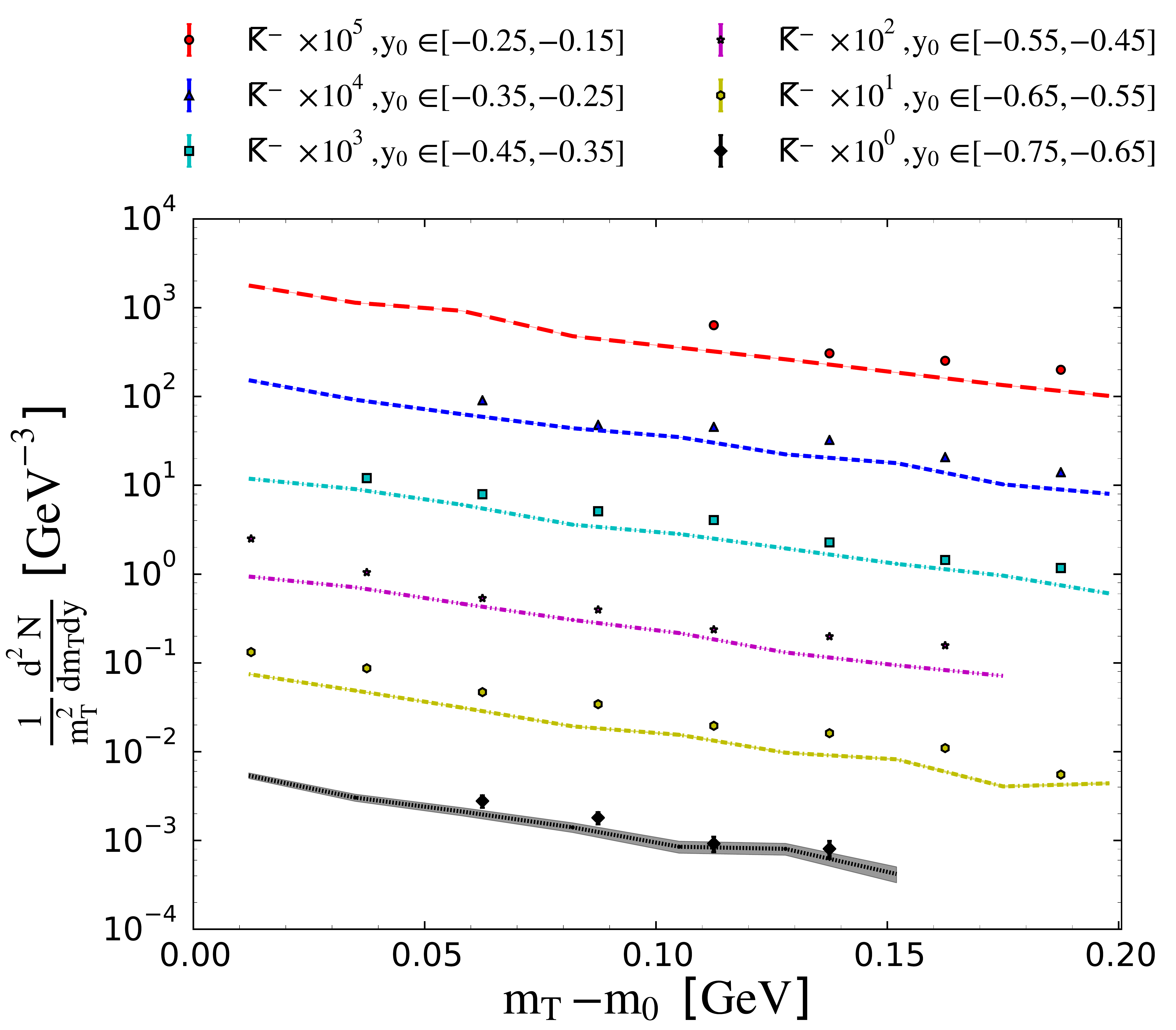}
\includegraphics[width=\linewidth]{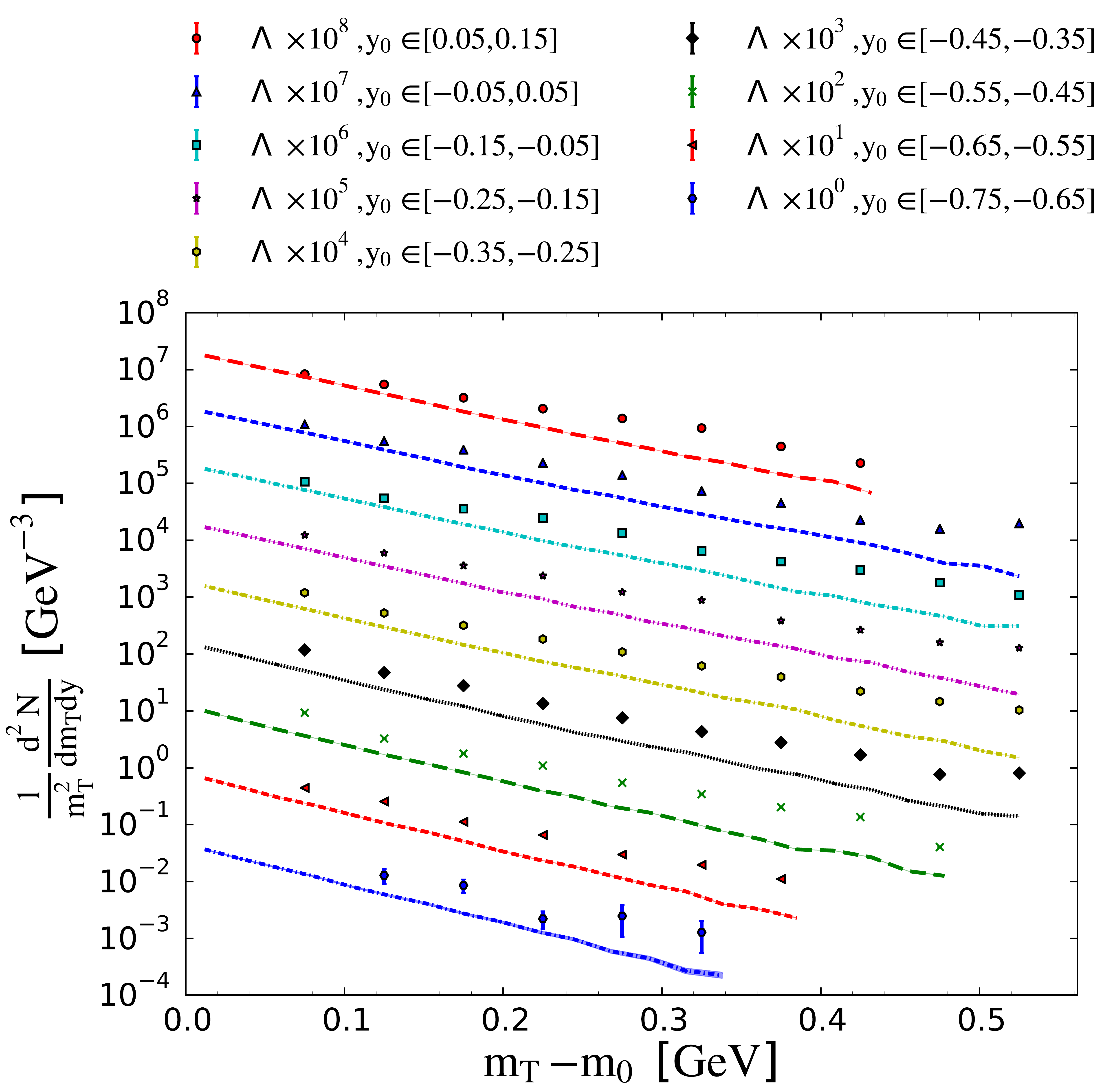}
\caption{
$m_T$ spectra of $K^+$, $\bar K^-$, $\Lambda$ produced in Ar-KCl collisions at $E_\text{kin} = 1.76A\,\text{GeV}$ within different rapidity bins.
Data measured by HADES~\cite{Agakishiev:2010rs} (points) is compared to SMASH simulations (lines).
}
\label{fig:mt_ArKCl}
\end{figure}

In \cref{fig:mt_ArKCl}, the HADES $m_T$ spectra for strange particles in different rapidity windows are compared to SMASH simulations.
For the kaons and antikaons, the slopes are similar to the experimental data, but the production is underestimated.
For the $\Lambda$ hyperons, the underestimation is worse and the slope is steeper in SMASH compared to the data.

\begin{figure}
\centering
\includegraphics[width=\linewidth]{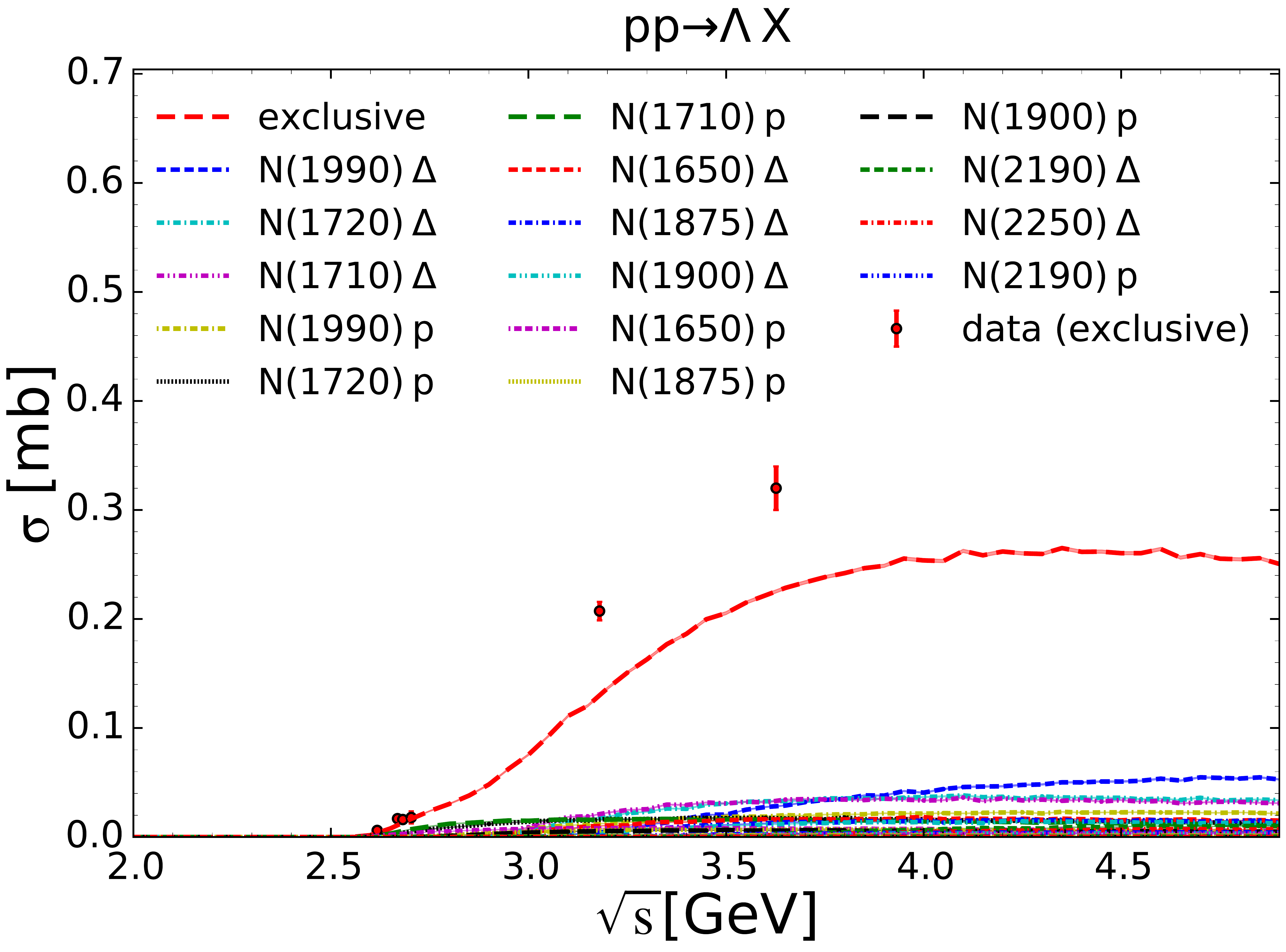}
\caption{$p p \to \Lambda \,\text{anything}$ cross section from SMASH compared to experimental data~\cite{Valdau:2010kw,Adamczewski-Musch:2016vrc,LaBoer}.}
\label{fig:xs_PP_LambdaX}
\end{figure}

There are at least the following possible reasons for this underestimation:
\begin{enumerate}
\item As shown in \cref{sec:nucleon_resonances}, the elementary exclusive $\Lambda$ production is reproduced. On the other hand, the $pp \to \Lambda\, \text{anything}$ cross section (see \cref{fig:xs_PP_LambdaX}) is too low, because $\Lambda$ production channels with more than three particles in the final state are missing.
\item SMASH does not have any $\pi N \to \Lambda \pi K,\, \Lambda \pi \pi K$ cross sections.
Within the resonance approach taken here, these could be emulated by introducing $N^* \to \Lambda K^*, \Lambda^* K$ decays, but such decays have not been observed.
\item Neglected in-medium effects, such as in-medium cross sections, kaon/antikaon-nucleon potentials and kaon/antikaon self-energies, may affect strangeness production~\cite{Hartnack:2011cn}.
\end{enumerate}


\subsection{Au-Au collisions by HADES}
\label{sec:AuAu}

\begin{figure}
\centering
\includegraphics[width=\linewidth]{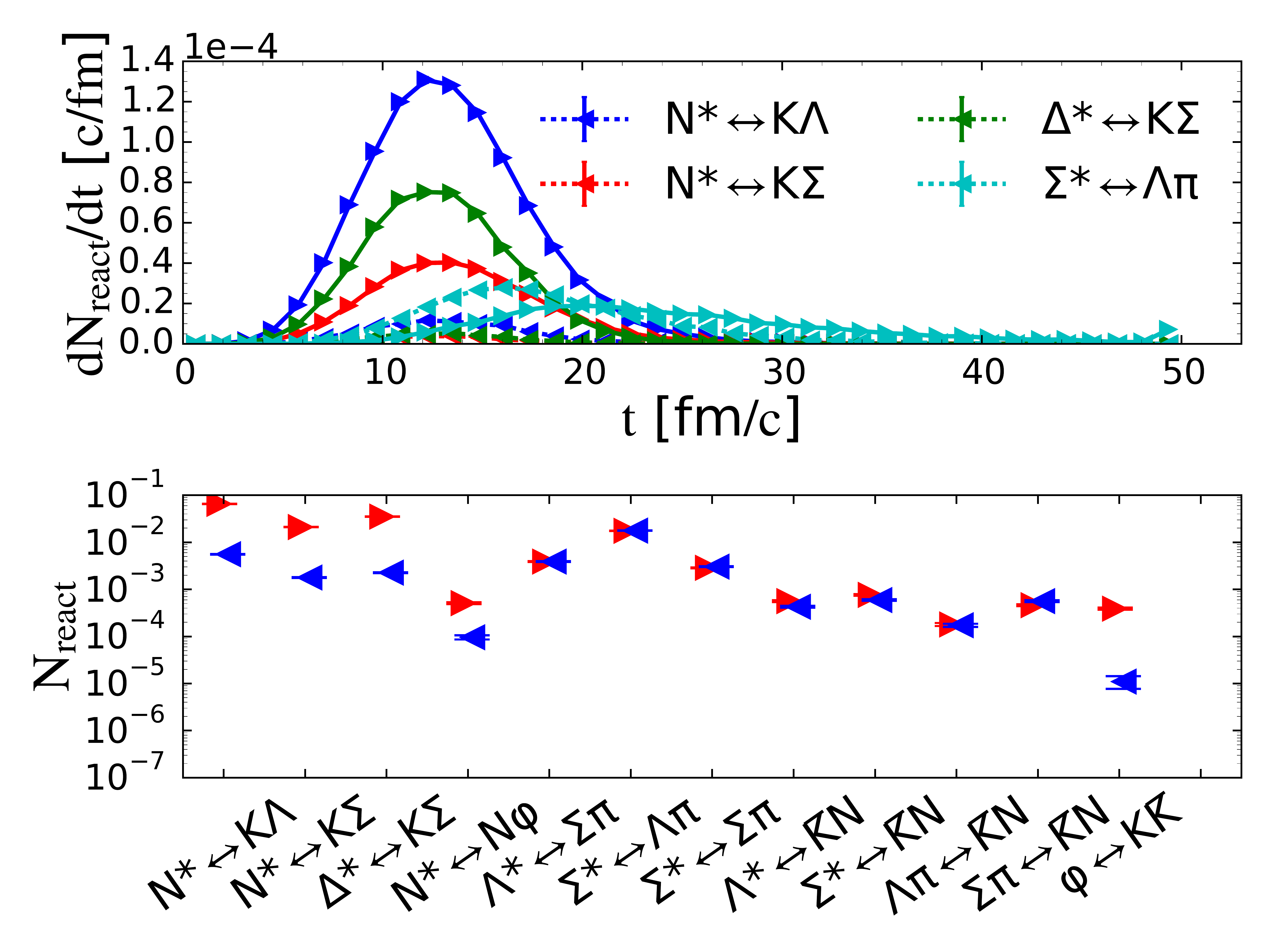}
\caption{
Average strangeness production in gold-gold collisions at $E_\text{kin} = 1.23A\,\text{GeV}$.
The upper plot shows the most important production and absorption rates as a function of time.
The lower one shows the total number of forward (red) and backward (blue) reactions involving strange particles.
}
\label{fig:rates_AuAu}
\end{figure}

After looking at intermediate-sized systems, it is of interest to consider larger systems, because they are more strongly affected by secondary reactions and possibly other in-medium effects, as listed at the end of the previous section.
Such a system has been investigated by the HADES collaboration by measuring the transverse mass spectra in Au-Au collisions at $E_\text{kin} = 1.23A\,\text{GeV}$ for kaons, antikaons and $\phi$ mesons~\cite{Adamczewski-Musch:2017rtf}.
As before, it is instructive to take a look at the reaction rates shown in \cref{fig:rates_AuAu}, which are averaged over ca. $20\,000\,000$~events.
They are similar to the ones observed in the smaller Ar-KCl system (\cref{sec:ArKCl}):
$N^*$ and $\Delta^*$ decays dominate the kaon production, while hyperon and $\phi$ decays are responsible for the antikaon production.
The backward reactions start at a similar time of ca.~$6\,\text{fm}/c$, but the production via $N^*$ and $\Delta^*$ stops later at about $25\,\text{fm}/c$.
The break-even point for $\Sigma^* \leftrightarrow \Lambda\pi$ is at about $21\,\text{fm}/c$, which is significantly later than for the smaller system.
Production via $\Sigma^*$ persists until much later times for ca.~$45\,\text{fm}/c$.
There is a jump at the last time step, because all unstable particles are forced to decay at the end of the simulation.
As before, the non-resonant strangeness exchange (see \cref{fig:rates_AuAu}) absorbs antikaons.

When comparing $\phi$ multiplicities to experimental data, it has to be taken into account that only $\phi$s decaying into $K^+ K^-$ can be reconstructed and only if the decay products do not rescatter afterwards such that they decorrelate.
As an approximation when comparing results from SMASH to the data, only $\phi$s decaying into $K^+ K^-$ which do not rescatter are considered.
The $\phi/K^-$ ratio given by HADES is rescaled by the $\phi \to K^+ K^-$ branching ratio.
This rescaling is also applied to the $\phi$ multiplicity reconstructed from SMASH.

\begin{figure}
\centering
\includegraphics[width=\linewidth]{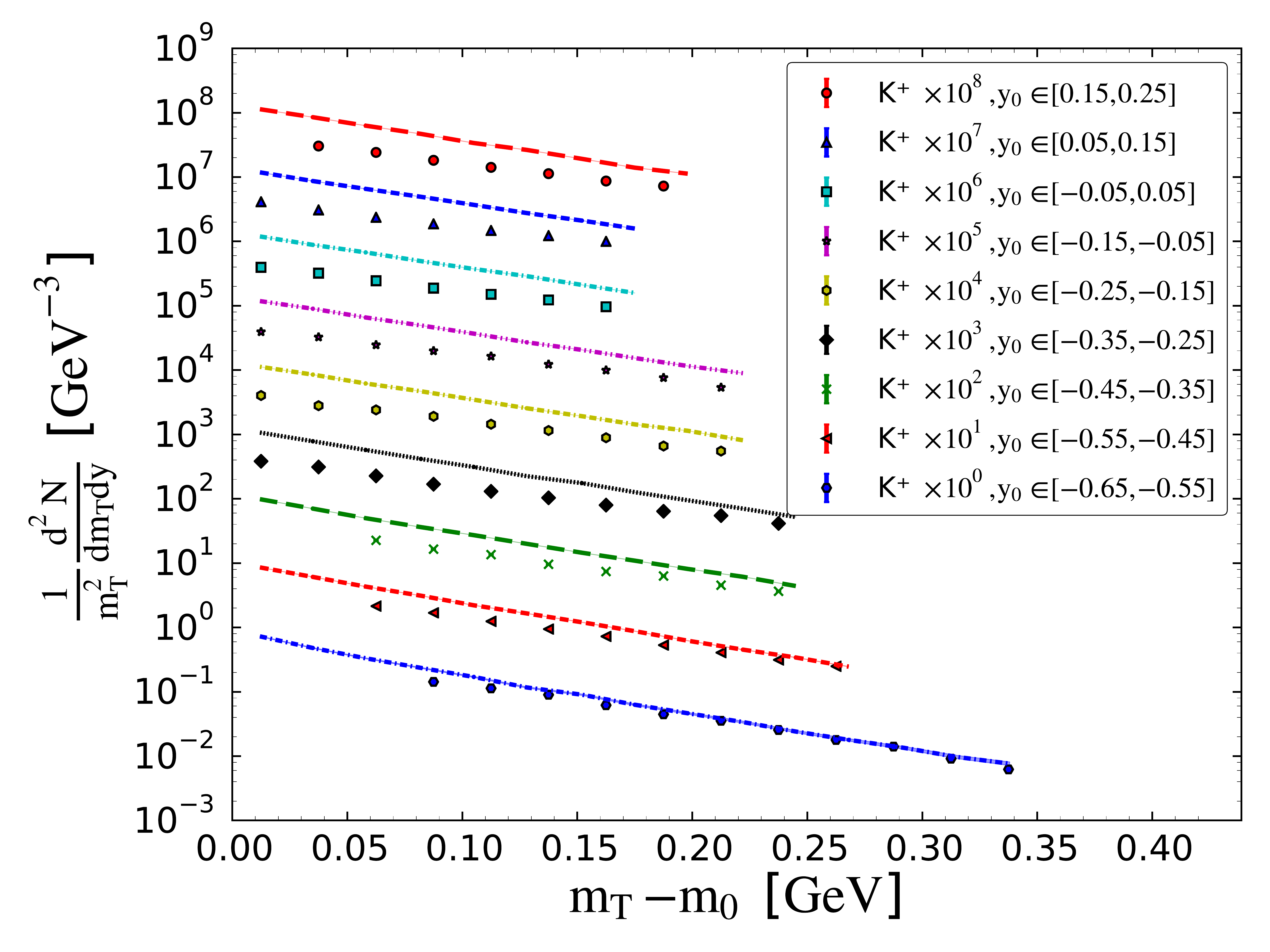}
\includegraphics[width=\linewidth]{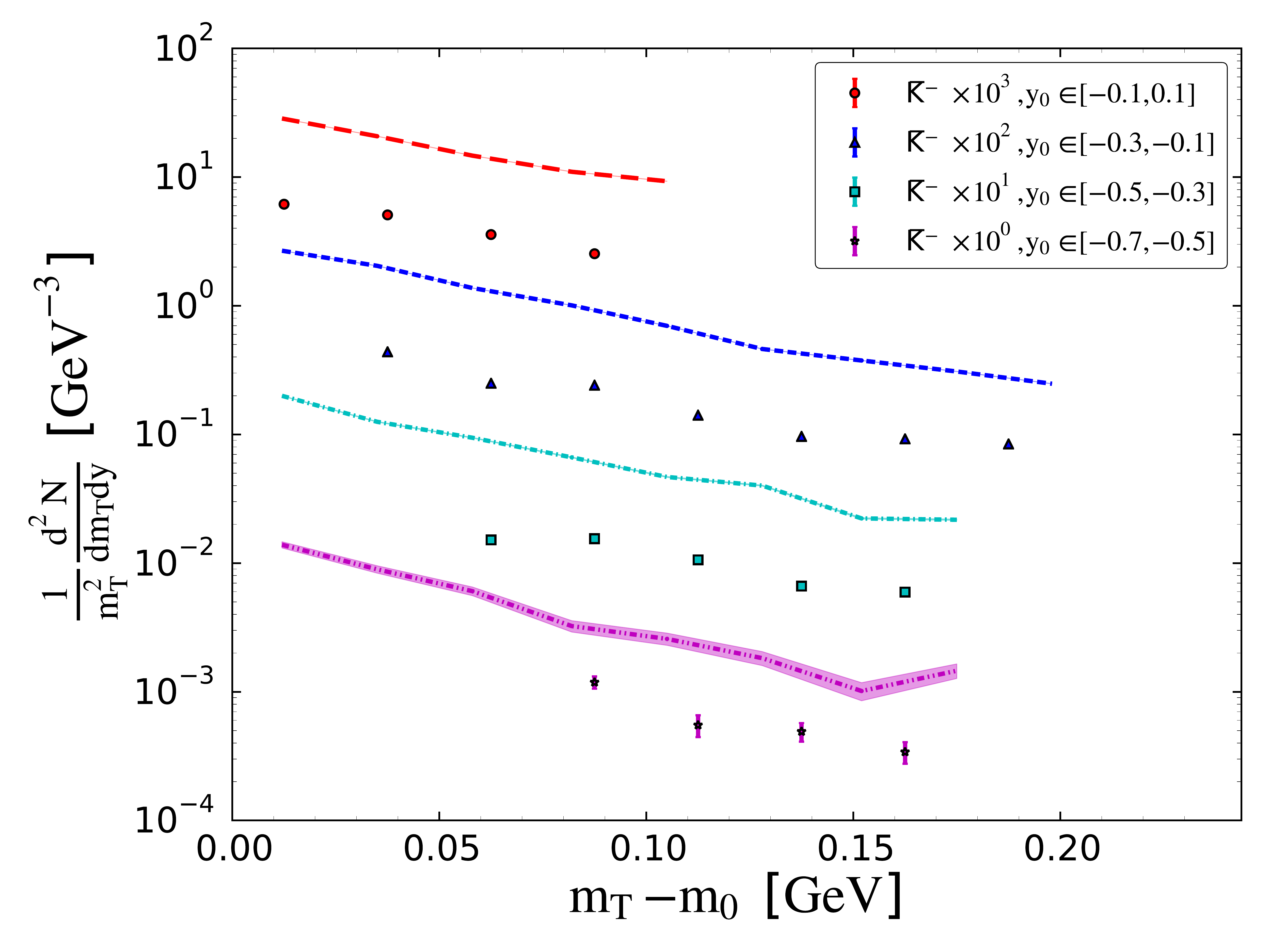}
\includegraphics[width=\linewidth]{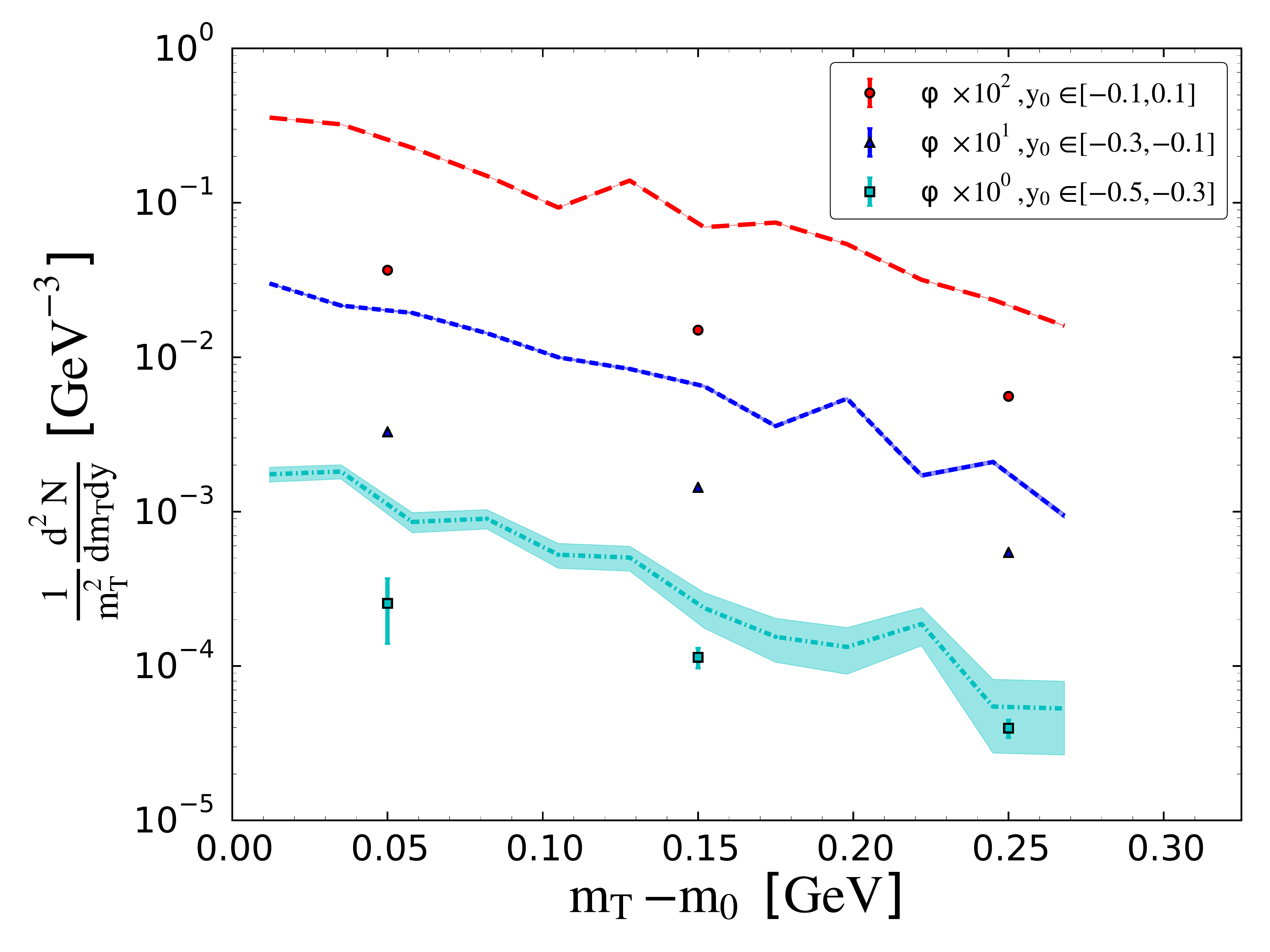}
\caption{
$m_T$ spectra of $K^+$, $\Lambda$, $\bar K^-$ produced in gold-gold collisions at $E_\text{kin} = 1.23A\,\text{GeV}$ within different rapidity bins.
Data measured by HADES~\cite{Adamczewski-Musch:2017rtf} (points) is compared to SMASH simulations (lines).
}
\label{fig:mt_AuAu}
\end{figure}

Comparing transverse mass spectra in SMASH to the HADES data in \cref{fig:mt_AuAu} shows a dependency on the rapidity window: For large rapidities SMASH is in good agreement with the experimental data for $K^+$ and to a limited extent $\phi$, but the agreement gets worse for midrapidity, where strangeness production is overestimated by SMASH.
This effect was not visible when comparing to the smaller Ar-KCl system (where $K^+$ was underestimated), suggesting that SMASH is missing a strangeness-suppressing in-medium effect important in larger systems such as Au-Au.
The $\bar K^-$ production is strongly overestimated for all rapidities, in stark contrast to the Ar-KCl results where it is slightly underestimated, again hinting at a strangeness-suppression mechanism missing in SMASH.

\begin{figure}
\centering
\includegraphics[width=\linewidth]{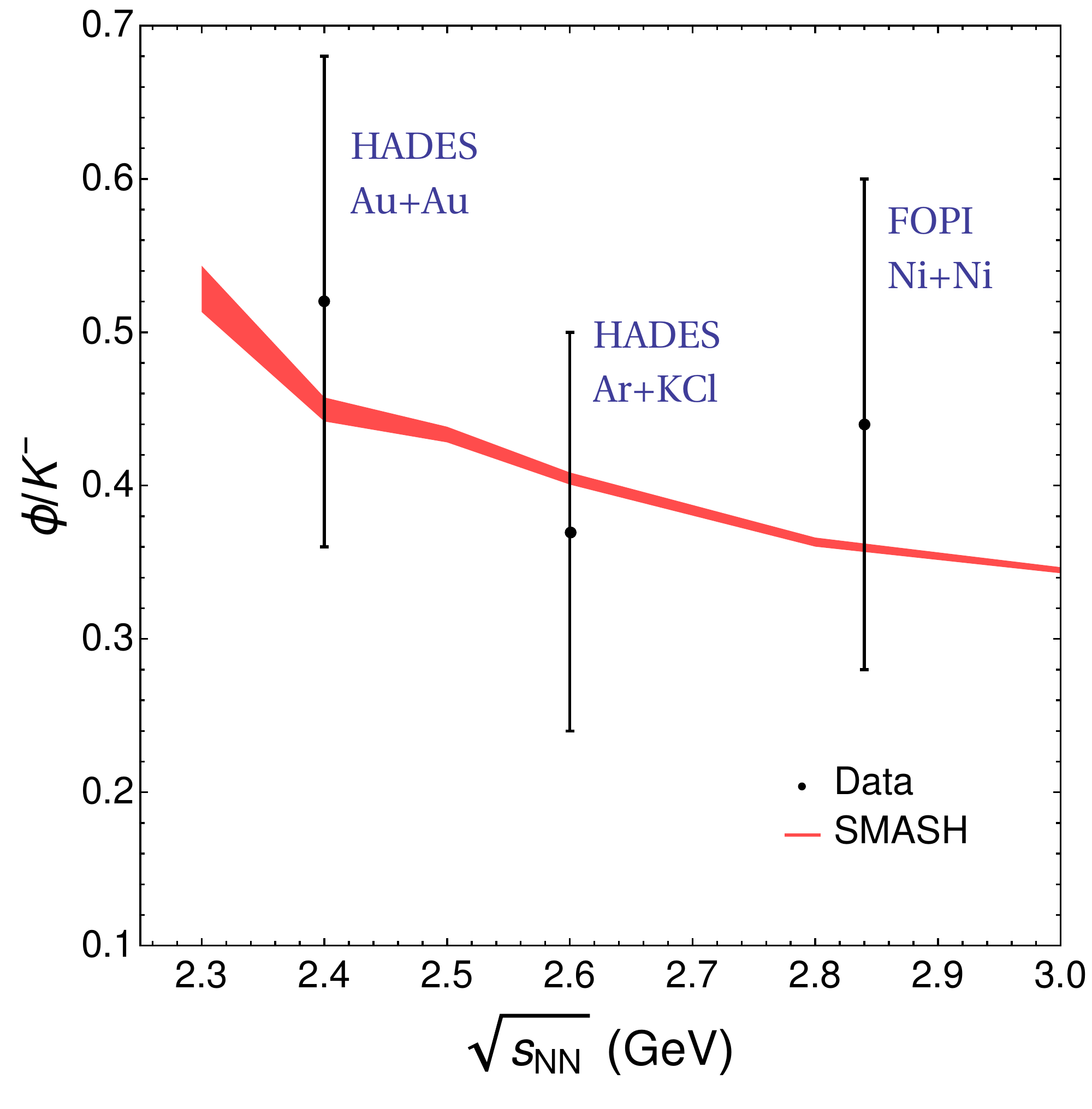}
\caption{
The ratio of the multiplicity of the mid-rapidity~$\phi$ to the multiplicity of the mid-rapidity~$\bar K^-$ obtained in gold-gold collisions with the impact parameter~$b < 3.4~{\rm fm}$ at different collision energies compared to the experimental data obtained from different beam energies and systems~\cite{Mangiarotti:2002mw, Agakishiev:2009ar, Adamczewski-Musch:2017rtf}.
}
\label{fig:ratio_phi_kminus}
\end{figure}

However, the $\phi/K^-$ ratio in SMASH is very similar to the one measured by HADES~\cite{Lorenz:2014eja} (ca.~0.5) and the trend for higher energies per nucleon agrees with other experiments measuring smaller systems, see \cref{fig:ratio_phi_kminus}.

The same data has been studied by other transport models, with comparable results:
A calculation employing GiBUU extended with Hagedorn resonances and a strangeness suppression factor was able to obtain a good agreement with HADES for the $m_t$ spectra integrated over rapidity, with a slightly worse $\phi/K^-$ ratio of $\approx 0.8$~\cite{Gallmeister:2017ths}.
The UrQMD approach (where the $N^* \to N\phi$ channel was first introduced) managed to predict the $\phi/K^-$ ratio~\cite{Steinheimer:2015sha}.

\subsection{Pion beam by HADES}
\label{sec:pion_beam}

The HADES collaboration has measured transverse momentum spectra of kaons and $\Lambda$~baryons in $\pi^-$-C and $\pi^-$-W collisions.
This is a very interesting system for a resonance approach as exercised in SMASH, because it is more sensitive to the $\pi N$ branching ratios than the usual $NN$ collisions.
As the HADES results have not been published so far, we only show predictions.

\begin{figure}
\centering
\includegraphics[width=\linewidth]{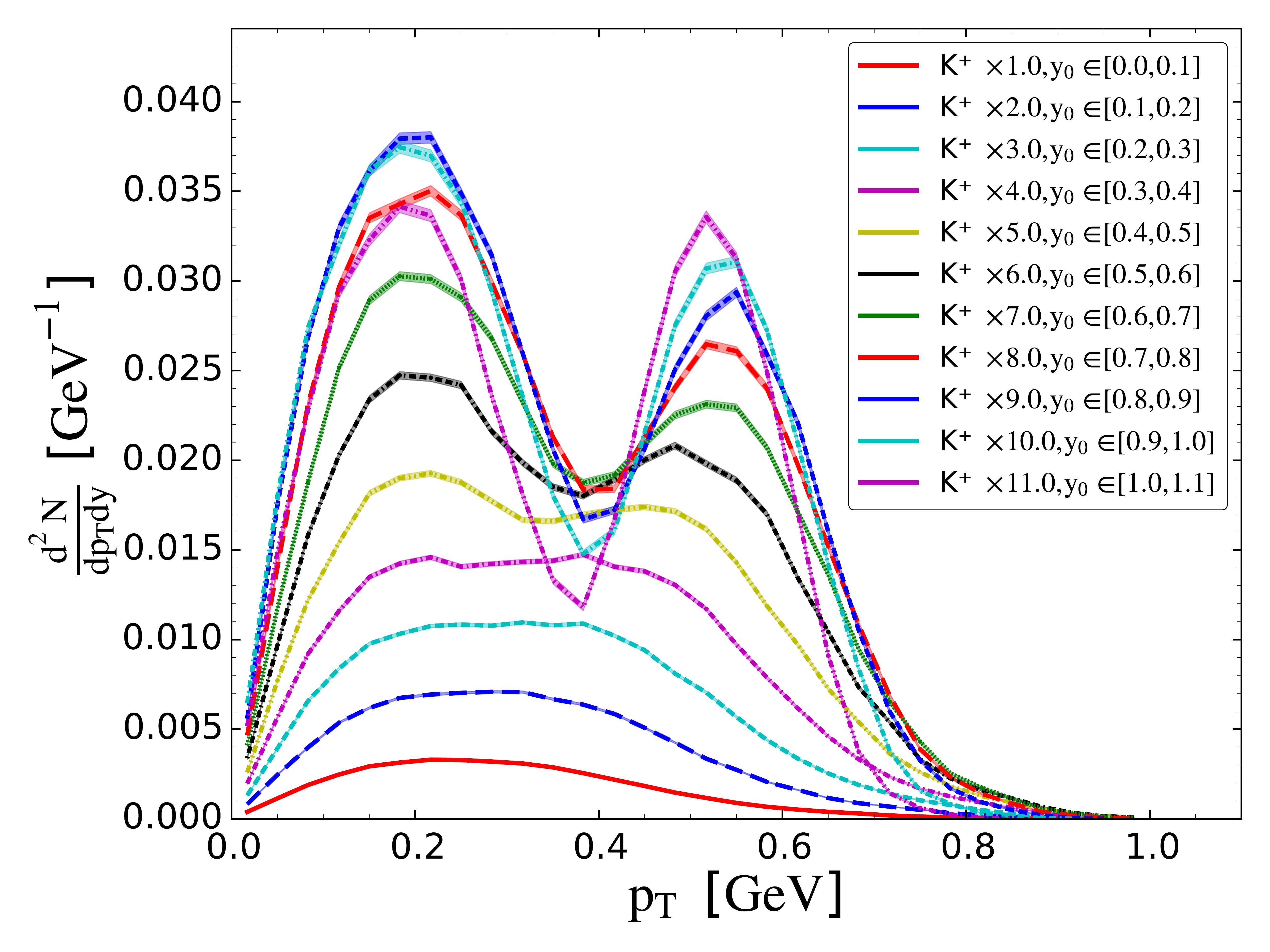}
\includegraphics[height=0.31\textheight]{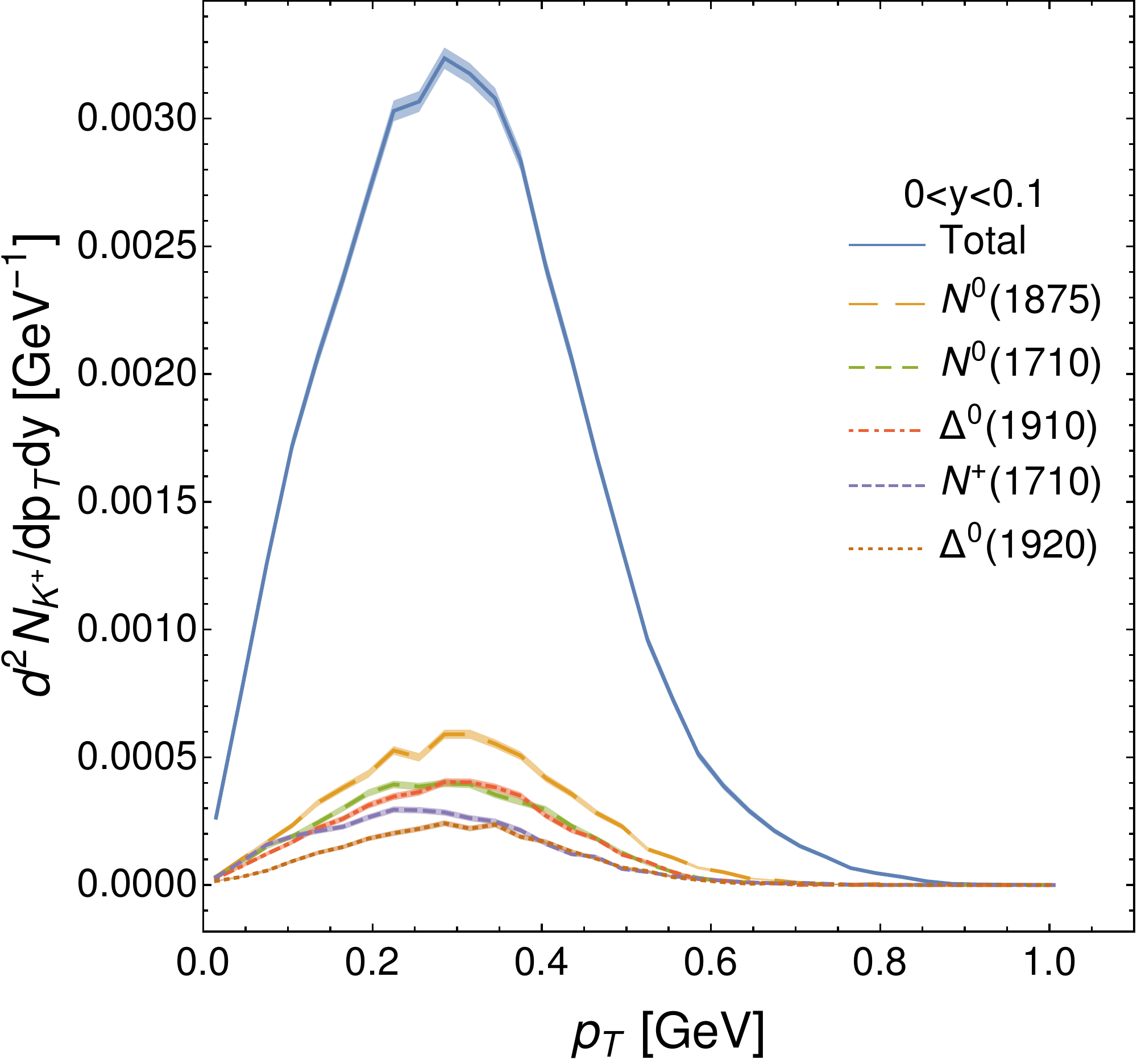}
\includegraphics[height=0.31\textheight]{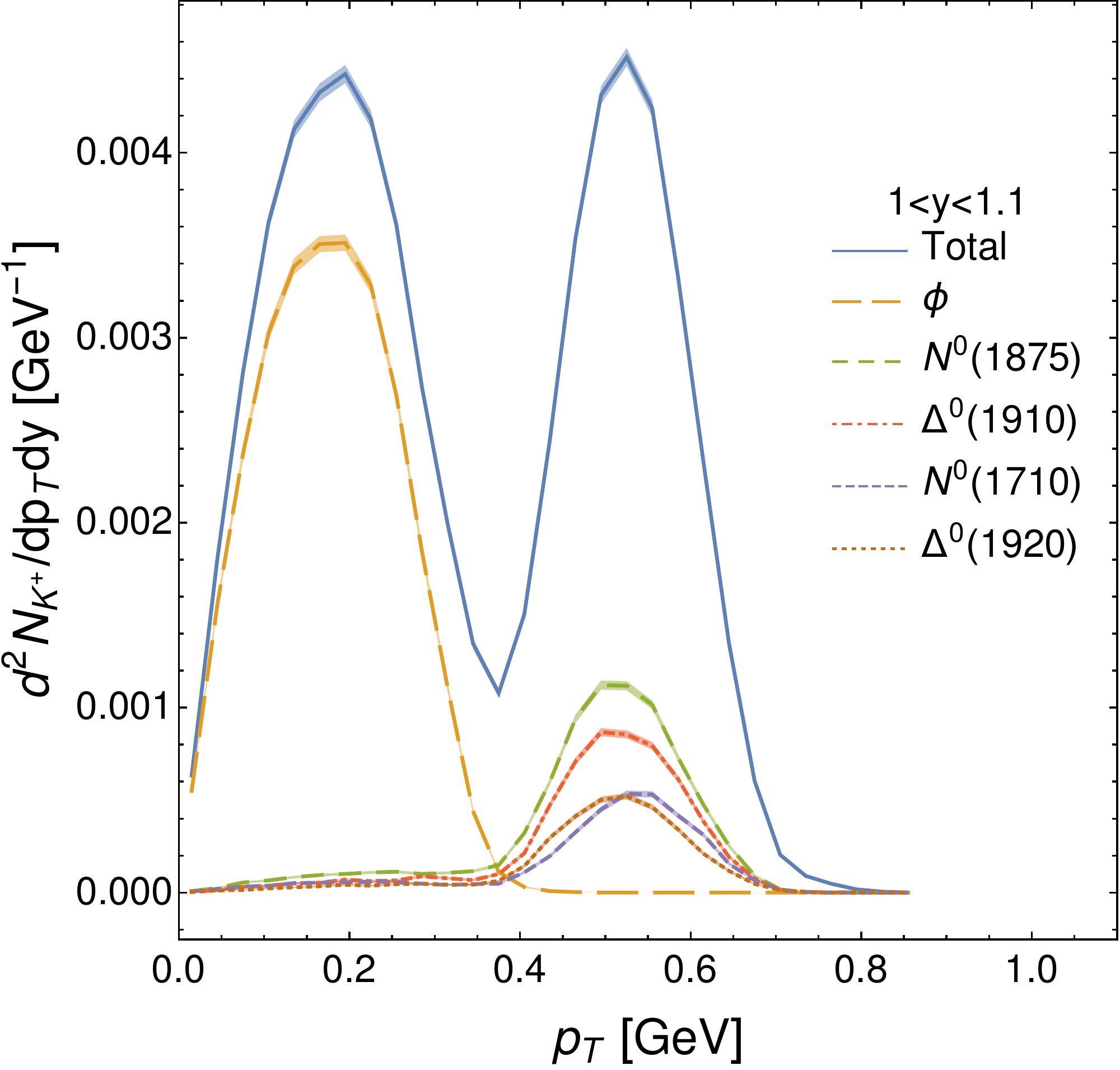}
\caption{
$p_T$ spectra of the average $K^+$s produced in ca.~$620\,000\,000$ $pi^-C$~collisions at $E_\text{kin} = 1.7\,\text{GeV}$ for different rapidities~$y$ (upper plot). For $y\in[0,0.1]$ and $[1,1.1]$, the contributions from the five most important resonances are shown (middle and lower plot).
}
\label{fig:pt_piC}
\end{figure}

At midrapidity, the $p_T$ spectrum of $K^+$ from SMASH~(\cref{fig:pt_piC}) is consistent with a Boltzmann distribution with a temperature of about $87\,\text{MeV}$.
Such a Boltzmann shape of a $p_T$ spectrum is typical for heavy-ion collisions at midrapidity.
In contrast, at higher rapidities an unusual two-peak structure emerges.
Plotting the separate resonance contributions to the $K^+$ spectra in \cref{fig:pt_piC}, we demonstrate that the peak at low~$p_T$ is from $\phi$ decays, while the peak at high~$p_T$ is from $N^*$ and $\Delta^*$ decays.
At midrapidity, only the $(N^*, \Delta^*)$ peak is present.

Two features of the SMASH model might be responsible for the two peak-structure:
\begin{enumerate}
\item In SMASH, high-energy resonances usually decay into only two particles.
      This allows to maintain detailed balance, but may lead to an overstimated $p_T$ of the decay products, because there should more particles in the final state.
      The introduction of more decays with more than two particles in the final state would populate lower transverse momenta.
      On the other hand, such decays of $N^*$ and $\Delta^*$ are rarely measured and their branching ratios are not well constrained.
\item Currently all resonance decays and formations in SMASH are isotropic.
      More realistic angular distributions might move the products of $N^*$ and $\Delta^*$ decays to higher rapidities and lower transverse momentum, assuming the total number of collisions is small enough not to isotropize the fireball.
\end{enumerate}

In any case, the upcoming HADES pion beam data will provide very helpful constraints for the resonance model applied here.

\section{Summary and outlook}
\label{sec:summary}

Strangeness production in heavy-ion collisions at SIS~energies has been analyzed with a hadron-resonance approach.
Elementary cross sections (inclusive and exclusive) have been applied to narrow down the otherwise poorly constrained branching ratios of $N^*$, $\Delta^*$ and hyperon resonances.
The experimental data on cross sections was found to be insufficient to constrain the $\phi$ production, which has been remedied by considering dilepton spectra from proton-niobium collisions.
Without further tuning of the parameters, SMASH has been compared to strangeness production in intermediately sized (Ni-Ni, Ar-KCl) and large (Au-Au) systems.
For the intermediately sized systems, there is a rough agreement with the data, while for large systems the agreement was only good for low participant numbers or high rapidities, hinting at strangeness suppressing in-medium effects missing in SMASH.
Predictions for the upcoming HADES pion-beam results have been shown, demonstrating a high sensitivity to the resonance properties in the model.

The resonance approach discussed here lays the foundation for future studies at higher energies with string fragmentation and an additive quark model, while the resonances are employed for low energies.
It is also planned to look at strangeness production applying local forced thermalization~\cite{Oliinychenko:2016vkg} in conjunction with the resonance approach.
Meanwhile, a hyperon-nucleon potential~\cite{Petschauer:2015nea} based on the qualitative features of a chiral effective theory at the next leading order will also be implemented in the future.
Finally, studies to utilize a Bayesian fit~\cite{Bernhard:2016tnd} instead of the manual tuning of the branching ratios are in preparation.

\section{Acknowledgements}
\label{sec:ack}
The authors thank J.~Steinheimer, M.~Lorenz, J.~Wirth and L.~Fabietti for fruitful discussions. Computational resources have been provided by the Center for Scientific Computing (CSC) at the Goethe-University of Frankfurt and by the Green IT Cube at Gesellschaft für Schwerionenforschung (GSI).
The authors acknowledge funding of a Helmholtz Young Investigator Group VH-NG-822 from the Helmholtz Association and GSI. This work was supported by the Helmholtz International
Center for the Facility for Antiproton and Ion Research (HIC for FAIR) within the
framework of the Landes-Offensive zur Entwicklung Wissenschaftlich-Ökonomischer Exzellenz (LOEWE) program launched by the State of Hesse.
V.~S. and J.~S. acknowledge support by the Helmholtz Graduate School for Hadron and Ion Research (HGS-HIRe).
H.~P. and J.~S. acknowledge support by the Deutsche Forschungsgemeinschaft (DFG) through the grant CRC-TR 211 ``Strong-interaction matter under extreme conditions''.
D.~O. was supported by the U.S. Department of Energy, Office of Science, Office of Nuclear Physics, under contract number DE-AC02-05CH11231 and received support within the framework of the Beam Energy Scan Theory (BEST) Topical Collaboration.

\appendix

\section{$\bar K N$ cross section background}
\label{sec:KbarN_xs}

The $\bar K N$ cross sections have a non-resonant background that has to be parametrized.
In this section, the parametrizations employed by SMASH for these contributions are described.

\begin{figure}
\centering
\includegraphics[width=\linewidth]{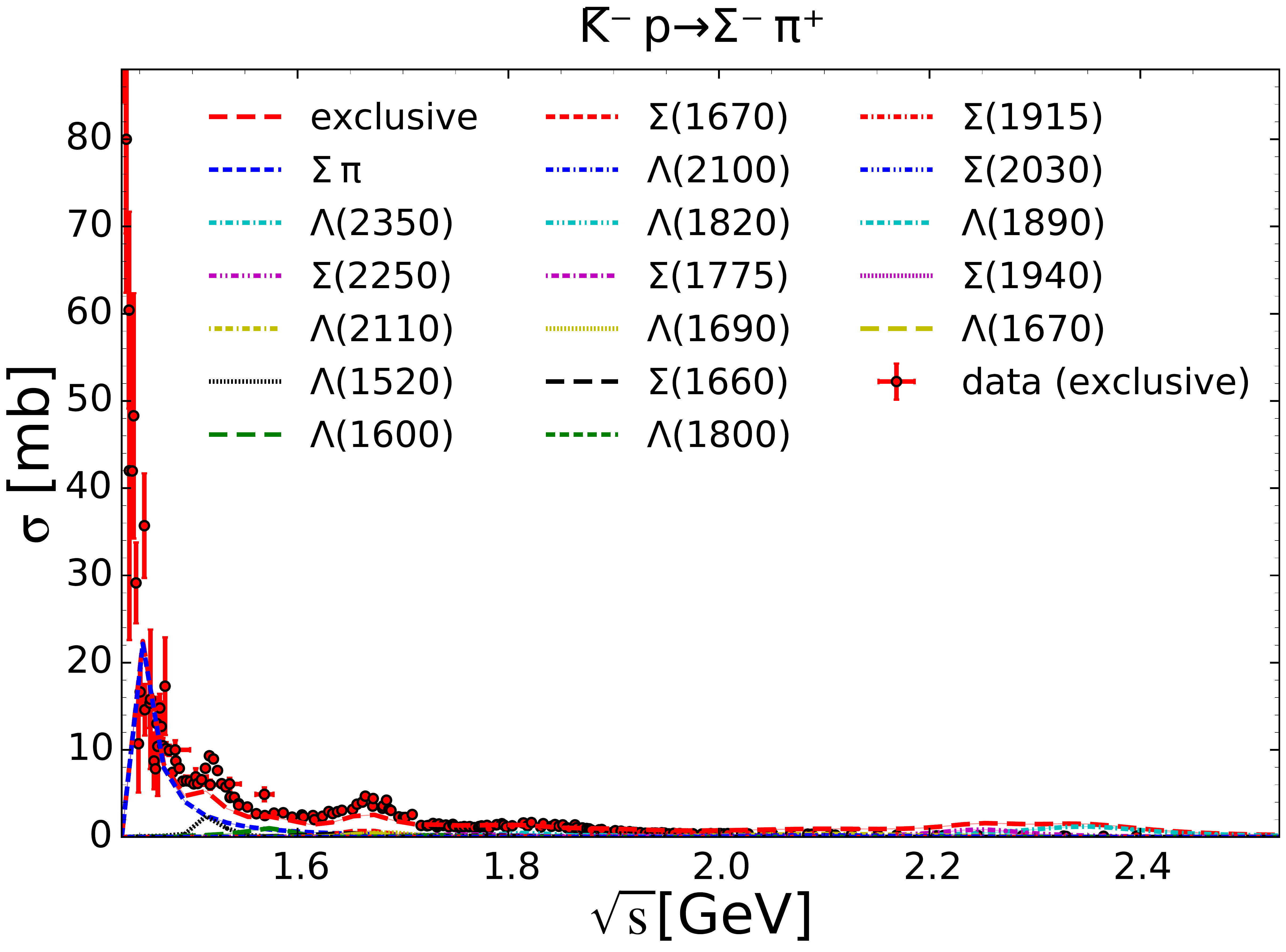}
\caption{$K^- p \to \Sigma^- \pi^+$ cross section from SMASH compared to experimental data~\cite{LaBoer}.}
\label{fig:xs_K-P_Sigma-pi+}
\end{figure}

\begin{figure}
\centering
\includegraphics[width=\linewidth]{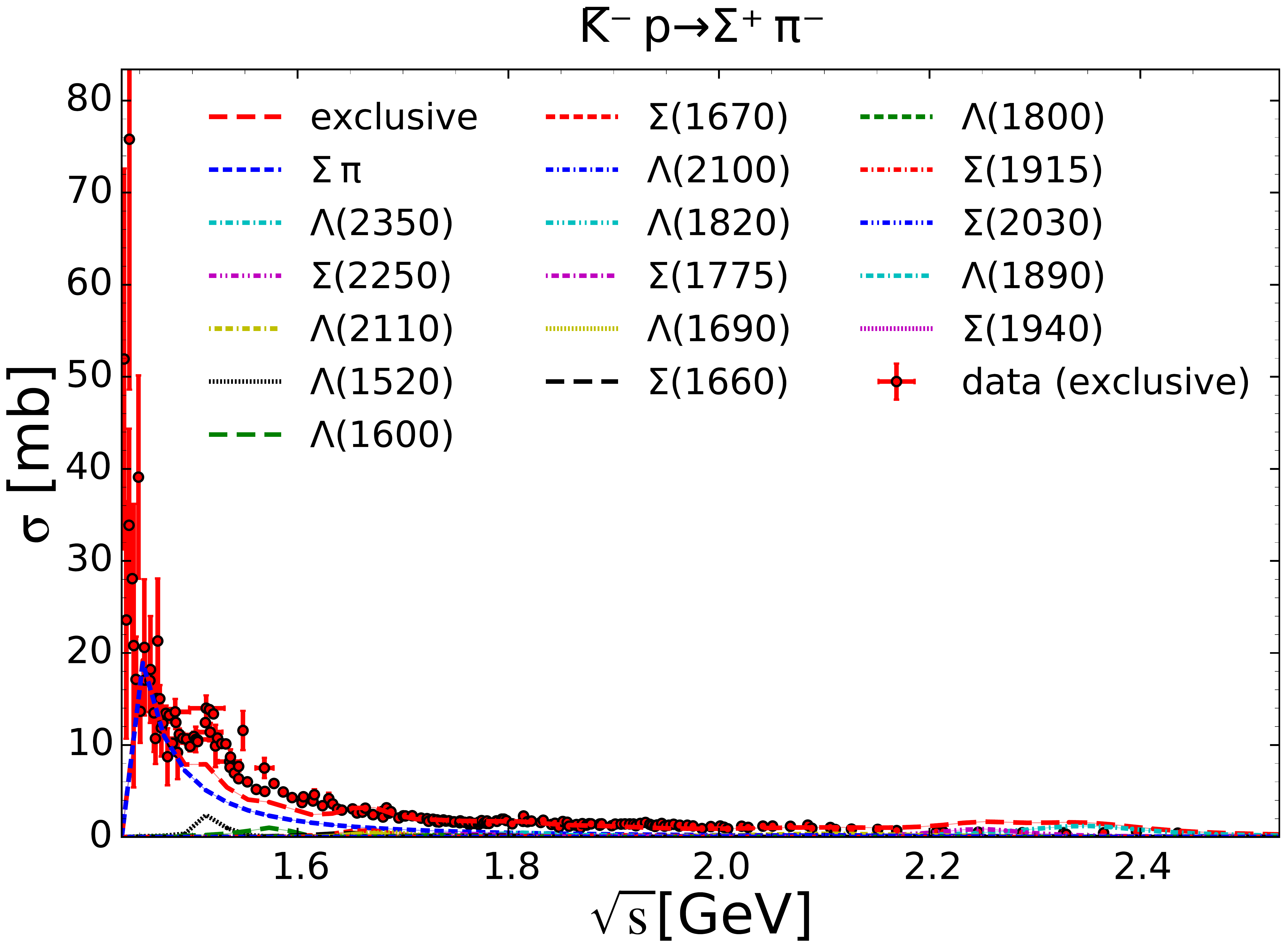}
\caption{$K^- p \to \Sigma^+ \pi^-$ cross section from SMASH compared to experimental data~\cite{LaBoer}.}
\label{fig:xs_K-P_Sigma+pi-}
\end{figure}

The first contribution is an inelastic background diverging towards the threshold.
In this approach, the same parametrization as in UrQMD~\cite{Graef:2014mra} is used.
It is fitted to the exclusive $K^- p \to \Lambda \pi^0, \Sigma^\pm \pi^\mp, \Sigma^0 \pi^0$ cross sections with
\begin{equation}
\sigma_{\bar K N \to \pi Y}(\sqrt s) = \frac{A}{(\sqrt s - B)^2}
\, ,
\label{eqn:s_exchange_param}
\end{equation}
where $A$ and $B$ are free parameters.
To better reproduce the experimental threshold, different parameters than in~\cite{Graef:2014mra} are employed for $K^- p \to \Lambda \pi^0$ and $K^- p \to \Sigma^0 \pi^0$.
For the other two reactions, the same parameters are used (see \cref{tab:s_exchange_param}).
Assuming isospin symmetry and detailed balance, the background parametrization for the backwards reactions $\pi Y \to \bar K N$ can be calculated.
It can be seen in \cref{fig:xs_K-P,fig:xs_K-P_Lambdapi0,fig:xs_K-P_Sigma-pi+,fig:xs_K-P_Sigma+pi-,fig:xs_K-P_Sigma0pi0} that the threshold of the total and exclusive $\bar K N$ cross sections is well described.


\begin{table}
\caption{
Parameters of the strangeness exchange background (\cref{eqn:s_exchange_param}), comparing UrQMD~\cite{Graef:2014mra} with SMASH.
}
\label{tab:s_exchange_param}
\begin{tabular}{clcc}
\toprule
model & \multicolumn{1}{c}{reaction} & $A$ & $B$ \\
\midrule
UrQMD & $K^- p \to \pi^- \Sigma^+$ & 0.0788265 & $1.38841\,\mathrm{GeV}$ \\
      & $K^- p \to \pi^+ \Sigma^-$ & 0.0196741 & $1.42318\,\mathrm{GeV}$ \\
      & $K^- p \to \pi^0 \Sigma^0$ & $0.55 \times 0.0508208$ & $1.38837\,\mathrm{GeV}$ \\
      & $K^- p \to \pi^0 \Lambda$ & $0.45 \times 0.0508208$ & $1.38837\,\mathrm{GeV}$ \\
\midrule
SMASH & $K^- p \to \pi^- \Sigma^+$ & 0.0788265 & $1.38841\,\mathrm{GeV}$ \\
      & $K^- p \to \pi^+ \Sigma^-$ & 0.0196741 & $1.42318\,\mathrm{GeV}$ \\
      & $K^- p \to \pi^0 \Sigma^0$ & 0.0403364 & $1.39830\,\mathrm{GeV}$ \\
      & $K^- p \to \pi^0 \Lambda$ & 0.0593256 & $1.38787\,\mathrm{GeV}$ \\
\bottomrule
\end{tabular}
\end{table}

Resonances are not sufficient to reproduce the elastic $K^- p$ cross section.
Similar to the $pp$ cross section, it is necessary to parametrize the experimental data.
To get rid of the noise, the PDG data~\cite{Patrignani:2016xqp} is smoothed with the LOWESS algorithm~\cite{Cleveland1979,Cleveland1981} and linearly interpolated.
If there is more than one measurement for one energy, the average is taken.
Additionally, the elastic contribution of hyperon resonances ($K^- p \to Y^* \to K^- p$) has to be considered and subtracted from the parametrization.
The result can be seen in \cref{fig:xs_K-P}, where the elastic cross section is perfectly reproduced.

For the charge exchange $K^- p \leftrightarrow \bar K^0 n$ and for $K^- n \to K^- n$, the same parametrization as in GiBUU~\cite{Buss:2011mx} is employed for the non-resonant background.
While this affects kinematics rather than strangeness production, it still has to be considered when tuning the branching ratios to the total $\bar K N$ cross sections (\cref{fig:xs_K-P,fig:xs_K-N}).

To reproduce the total $K^- p$ cross section for $\sqrt s > 2\,\mathrm{GeV}$, channels with more than two final-state particles can be taken into account.
In \cite{Buss:2011mx} this was done by implementing an $\bar K N \to Y^* \pi$ process with constant matrix element for hyperon resonances~$Y^*$.
This contribution is currently not implemented in SMASH, because it is not important for low-energy heavy-ion collisions and does not help to constrain the $Y^*$~branching ratios due to the uncertainty of the matrix element.

\begin{figure}
\centering
\includegraphics[width=\linewidth]{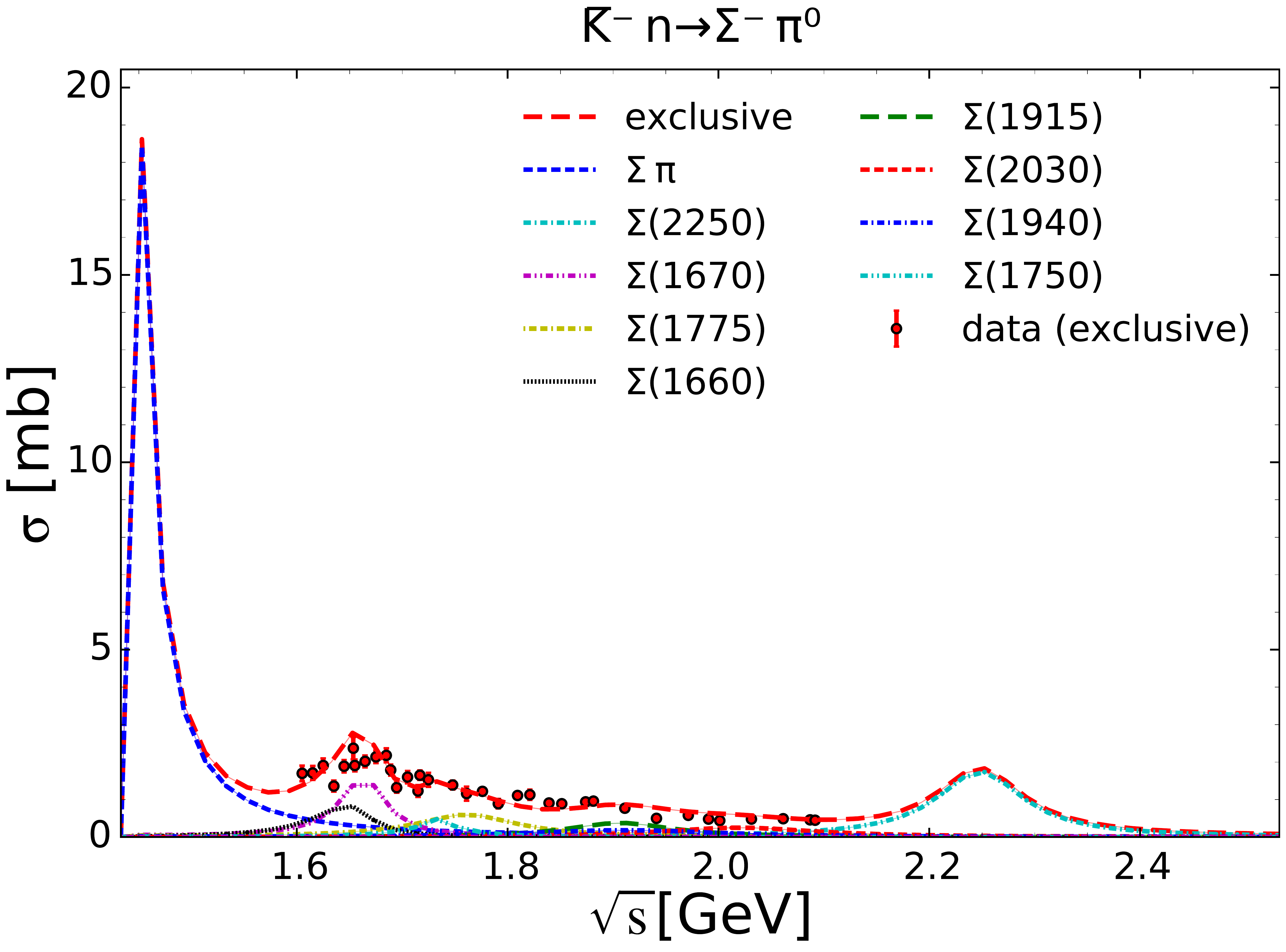}
\caption{$K^- n \to \Sigma^- \pi^0$ cross section from SMASH compared to experimental data~\cite{LaBoer}.}
\label{fig:xs_K-N_Sigma-pi0}
\end{figure}

\section{$KN$ cross section}
\label{sec:KN_xs}

For heavy-ion collision, the $KN$~cross section is important as a mechanism to transfer momentum from the medium to the kaons~\cite{Hartnack:2011cn}.
The kaon multiplicity is not affected, except for the $K^+ n \leftrightarrow K^0 p$ charge exchange.
In that regard, it is important to reproduce the total cross section correctly, but the exclusive cross sections are not as crucial.

Like in GiBUU~\cite{Buss:2011mx}, a parametrization of the experimental data for the elastic $K^+ p$ cross section~\cite{Patrignani:2016xqp} is employed.
Assuming the scattering amplitudes for isospin~$I=1$ are much larger than for $I=0$, the elastic and charge-exchange cross sections are related by the following identity:
\begin{equation}
\sigma_{K^+ n \to K^+ n} = \sigma_{K^+ n \to K^0 p} = \frac14 \sigma_{K^+ p \to K^+ p}
\end{equation}
In GiBUU the factor $\frac12$ is taken instead.
The factor $\frac14$ is derived in \cref{sec:KN_amplitude}.

\begin{figure}
\centering
\includegraphics[width=\linewidth]{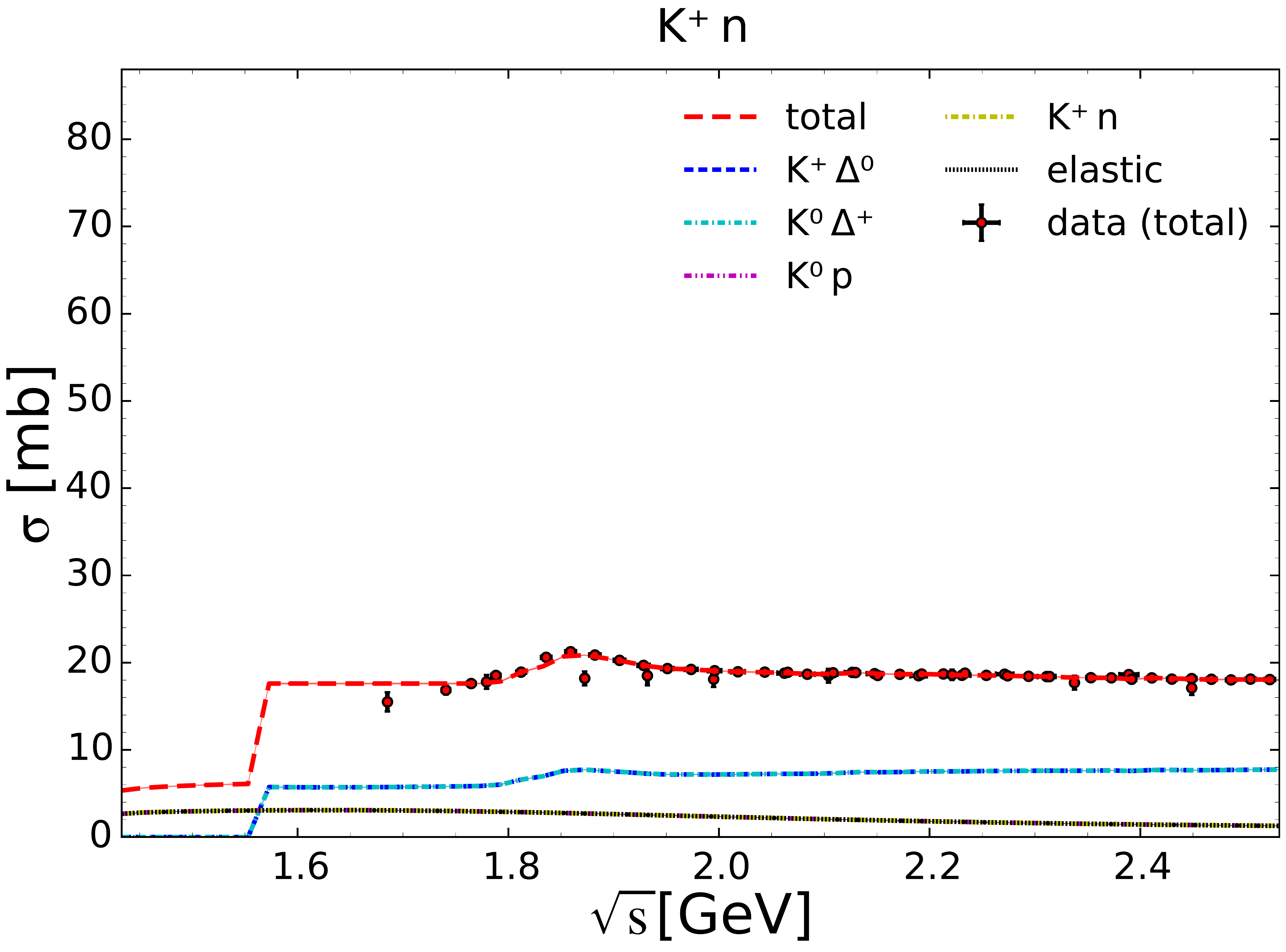}
\caption{$K^+ n$ cross section from SMASH compared to experimental data~\cite{Patrignani:2016xqp}.}
\label{fig:xs_K+N}
\end{figure}

For the inelastic $K^+ N$ cross section, the experimental data is smoothed like for $K^- p$ and the elastic and charge-exchange contributions are subtracted.
(To reproduce the peak at $\sqrt s = 1.87\,\text{GeV}$ in $K^+ n$, the outlier in the experimental data was ignored, see~\cref{fig:xs_K+N}.)
The remaining inelastic cross section is assumed to entirely produce $K \pi N$.
Unlike GiBUU, we assume that this production happens via $K \Delta$, so that the backwards reaction is still possible without having to implement $3 \to 2$ reactions, maintaining detailed balance.
The $K^0 N$ cross sections are derived from the $K^+ N$ cross section by assuming isospin symmetry.

\begin{figure}
\centering
\includegraphics[width=\linewidth]{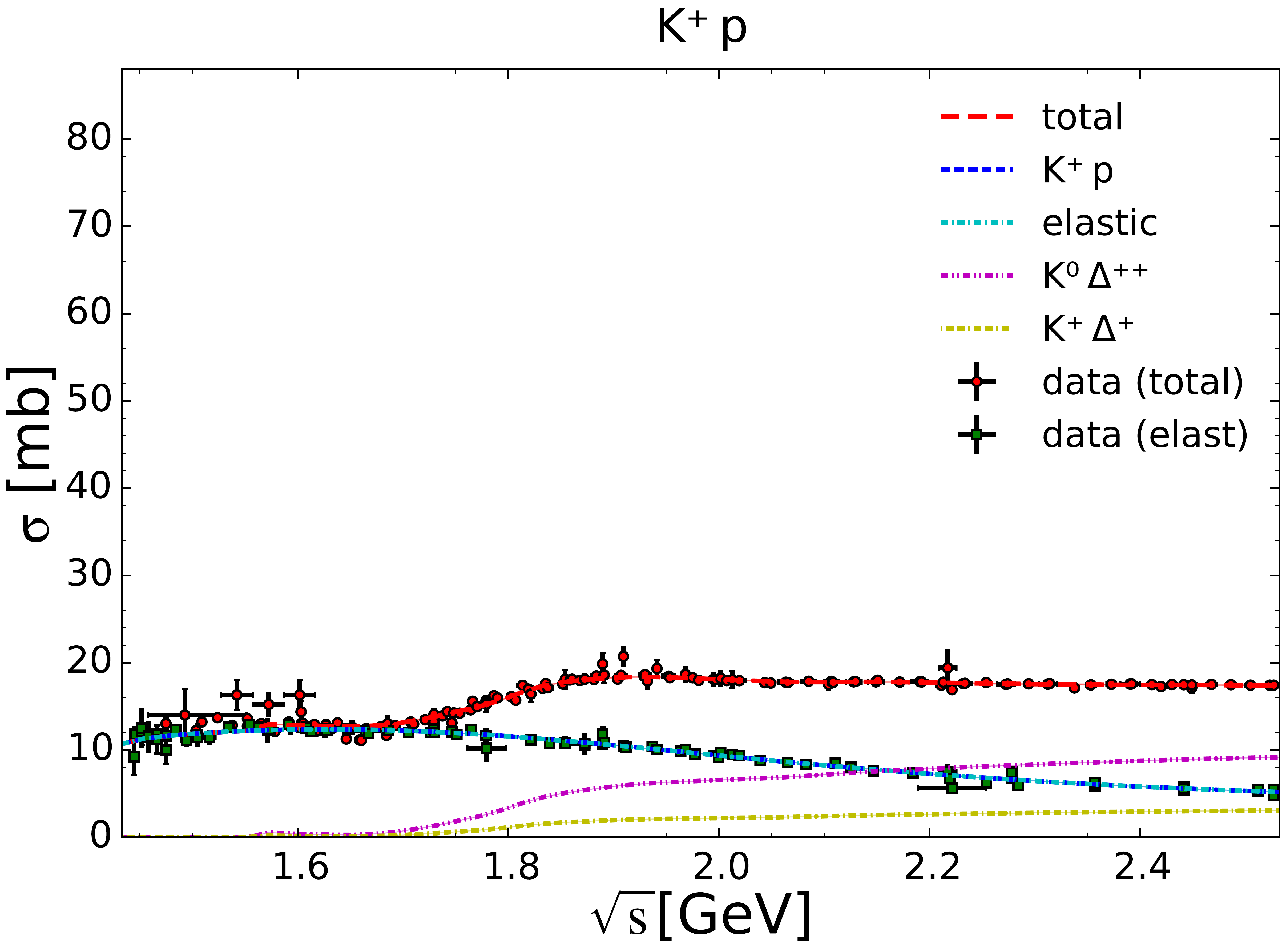}
\caption{$K^+ p$ cross section from SMASH compared to experimental data~\cite{Patrignani:2016xqp}.}
\label{fig:xs_K+P}
\end{figure}

\begin{figure}
\centering
\includegraphics[width=\linewidth]{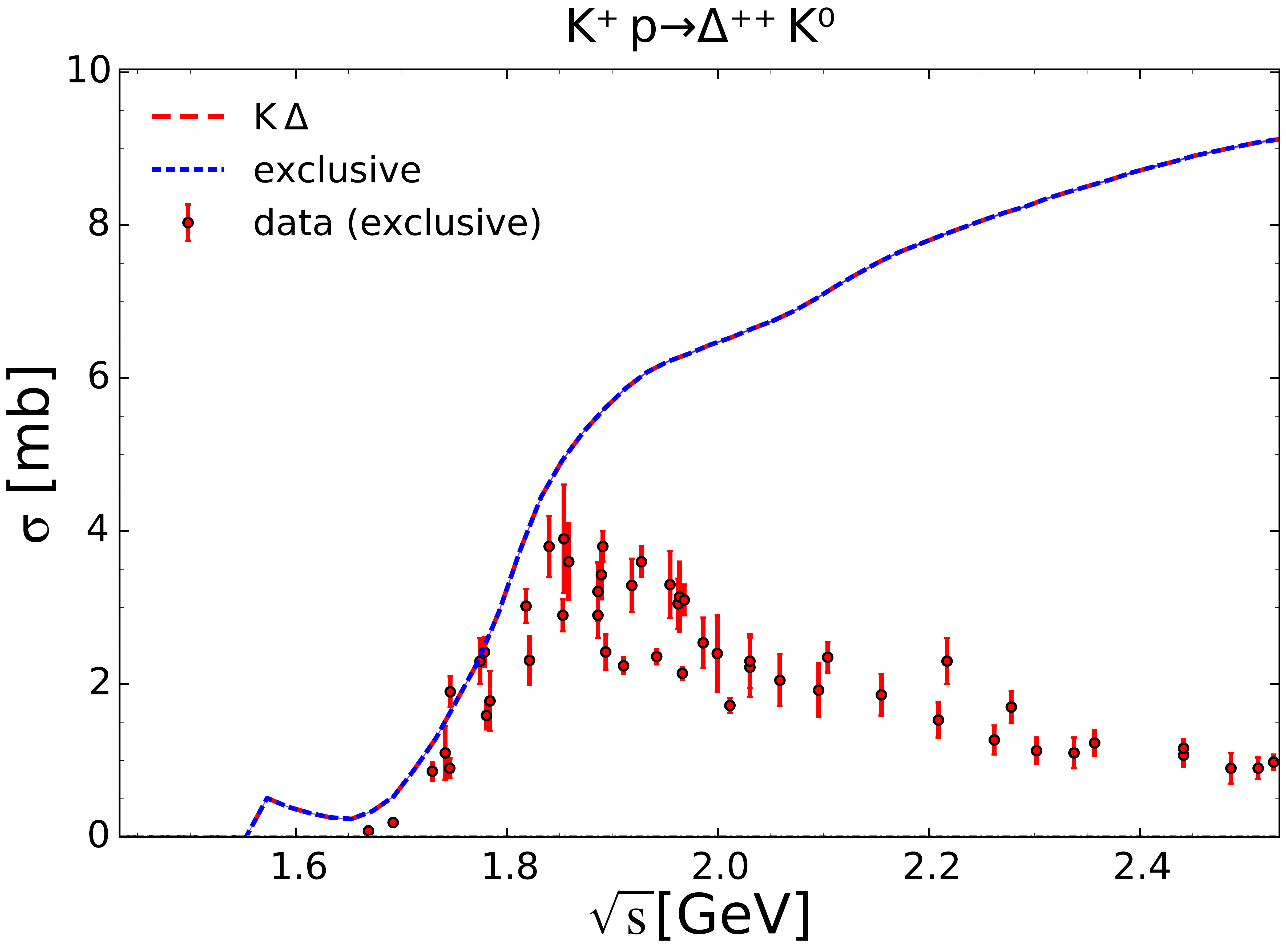}
\caption{
$K^+ p \to \Delta^{++} K^0$ cross section from SMASH compared to experimental data~\cite{LaBoer}.
The parametrization used in SMASH was not designed to reproduce this cross section.
}
\label{fig:xs_K+P_Delta++K0}
\end{figure}

This adequately reproduces the total and elastic $KN$~cross sections (\cref{fig:xs_K+P,fig:xs_K+N}), but it is not designed to reproduce the exclusive cross sections.
They have been measured in experiment~\cite{LaBoer}:
\begin{itemize}
\item $K^+ p \to \Delta^+ K^+,\, \Delta^{++} K^0$
\item $K^+ n \to p \pi^- K+$
\end{itemize}
For $K^+ p \to \Delta^{++} K^0$ (\cref{fig:xs_K+P_Delta++K0}), the data is reproduced for $\sqrt s < 1.85\,\mathrm{GeV}$, but above that energy, the experimental cross section falls off while our parametrization still increases.
The $K^+ p \to \Delta^+ K^+$ parametrization is identical and has the same issues (not shown).
This suggests that reactions with more pions in the final state have to be considered.
Finally, the $K^+ n \to p \pi^- K^+$ cross section is not well reproduced either (not shown).

The observed discrepancies demonstrate that the assumptions about the $KN$~cross section do not work well for the exclusive cross sections.
However, this is not considered important for the systems studied in this work, because the main motivation for the $KN$~cross section is the momentum transfer from the nuclear medium to the kaons, which is mostly affected by the total, not the exclusive cross sections.

\section{Isospin factors for $KN$~scattering}
\label{sec:KN_amplitude}

Considering the reactions $K^+ n \to K^+ n$, $K^+ n \to K^0 p$ and $K^+ p \to K^+ p$, the following eigenstates~$\big|I, I_3\big>$ of the isospin~$I$ and its projection~$I_3$ are relevant:
\begin{align}
\big|K^0 p \big> &= \frac{1}{\sqrt2} \big|1, 0\big> - \frac{1}{\sqrt2} \big|0, 0\big> \\
\big|K^+ n \big> &= \frac{1}{\sqrt2} \big|1, 0\big> + \frac{1}{\sqrt2} \big|0, 0\big> \\
\big|K^+ p \big> &= \big|1, 1\big>
\end{align}
Assuming isospin symmetry, the scattering matrix elements only depend on~$I$.
For the interacting part of the Hamiltonian~$\hat V$, the following scattering amplitudes are obtained:
\begin{align}
\big<K^0 p \big| \hat V \big| K^0 p\big> &= \frac12 M_1 + \frac12 M_0 \\
\big<K^0 p \big| \hat V \big| K^+ n\big> &= \frac12 M_1 - \frac12 M_0 \\
\big<K^+ p \big| \hat V \big| K^+ p\big> &= M_1
\;,
\end{align}
where $M_I = \big<I, I_3\big|\hat V\big|I, I_3\big>$.
Consequently, one obtains for the cross sections assuming $|M_0| \ll |M_1|$:
\begin{align}
\sigma_{K^0 p \to K^0 p} \propto \frac14 |M_1 + M_0|^2 \approx \frac14 |M_1|^2 \\
\sigma_{K^0 p \to K^+ n} \propto \frac14 |M_1 - M_0|^2 \approx \frac14 |M_1|^2 \\
\sigma_{K^+ p \to K^+ p} \propto |M_1|^2
\end{align}
This implies the following relation for the cross sections:
\begin{equation}
\sigma_{K^0 p \to K^0 p} = \sigma_{K^0 p \to K^+ n} = \frac14 \sigma_{K^+ p \to K^+ p}
\end{equation}

\bibliography{inspire,non_inspire}

\begin{thebibliography}{63}
\expandafter\ifx\csname natexlab\endcsname\relax\def\natexlab#1{#1}\fi
\expandafter\ifx\csname bibnamefont\endcsname\relax
  \def\bibnamefont#1{#1}\fi
\expandafter\ifx\csname bibfnamefont\endcsname\relax
  \def\bibfnamefont#1{#1}\fi
\expandafter\ifx\csname citenamefont\endcsname\relax
  \def\citenamefont#1{#1}\fi
\expandafter\ifx\csname url\endcsname\relax
  \def\url#1{\texttt{#1}}\fi
\expandafter\ifx\csname urlprefix\endcsname\relax\def\urlprefix{URL }\fi
\providecommand{\bibinfo}[2]{#2}
\providecommand{\eprint}[2][]{\url{#2}}

\bibitem[{\citenamefont{Blume}(2017)}]{Blume:2017icv}
\bibinfo{author}{\bibfnamefont{C.}~\bibnamefont{Blume}} (\bibinfo{year}{2017}),
  \bibinfo{note}{[EPJ Web Conf.171,03001(2018)]}, \eprint{1710.07508}.

\bibitem[{\citenamefont{Blume and Markert}(2011)}]{Blume:2011sb}
\bibinfo{author}{\bibfnamefont{C.}~\bibnamefont{Blume}} \bibnamefont{and}
  \bibinfo{author}{\bibfnamefont{C.}~\bibnamefont{Markert}},
  \bibinfo{journal}{Prog. Part. Nucl. Phys.} \textbf{\bibinfo{volume}{66}},
  \bibinfo{pages}{834} (\bibinfo{year}{2011}), \eprint{1105.2798}.

\bibitem[{\citenamefont{Adamczewski-Musch
  et~al.}(2018)}]{Adamczewski-Musch:2017rtf}
\bibinfo{author}{\bibfnamefont{J.}~\bibnamefont{Adamczewski-Musch}}
  \bibnamefont{et~al.} (\bibinfo{collaboration}{HADES}),
  \bibinfo{journal}{Phys. Lett. B} \textbf{\bibinfo{volume}{778}},
  \bibinfo{pages}{403} (\bibinfo{year}{2018}), \eprint{1703.08418}.

\bibitem[{\citenamefont{Agakishiev
  et~al.}(2009{\natexlab{a}})}]{Agakishiev:2009rr}
\bibinfo{author}{\bibfnamefont{G.}~\bibnamefont{Agakishiev}}
  \bibnamefont{et~al.} (\bibinfo{collaboration}{HADES}),
  \bibinfo{journal}{Phys. Rev. Lett.} \textbf{\bibinfo{volume}{103}},
  \bibinfo{pages}{132301} (\bibinfo{year}{2009}{\natexlab{a}}),
  \eprint{0907.3582}.

\bibitem[{\citenamefont{Ablyazimov et~al.}(2017)}]{Ablyazimov:2017guv}
\bibinfo{author}{\bibfnamefont{T.}~\bibnamefont{Ablyazimov}}
  \bibnamefont{et~al.} (\bibinfo{collaboration}{CBM}), \bibinfo{journal}{Eur.
  Phys. J. A} \textbf{\bibinfo{volume}{53}}, \bibinfo{pages}{60}
  (\bibinfo{year}{2017}), \eprint{1607.01487}.

\bibitem[{\citenamefont{Kekelidze et~al.}(2012)\citenamefont{Kekelidze,
  Lednicky, Matveev, Meshkov, Sorin, and Trubnikov}}]{Kekelidze:2012zz}
\bibinfo{author}{\bibfnamefont{V.}~\bibnamefont{Kekelidze}},
  \bibinfo{author}{\bibfnamefont{R.}~\bibnamefont{Lednicky}},
  \bibinfo{author}{\bibfnamefont{V.}~\bibnamefont{Matveev}},
  \bibinfo{author}{\bibfnamefont{I.}~\bibnamefont{Meshkov}},
  \bibinfo{author}{\bibfnamefont{A.}~\bibnamefont{Sorin}}, \bibnamefont{and}
  \bibinfo{author}{\bibfnamefont{G.}~\bibnamefont{Trubnikov}},
  \bibinfo{journal}{Phys. Part. Nucl. Lett.} \textbf{\bibinfo{volume}{9}},
  \bibinfo{pages}{313} (\bibinfo{year}{2012}).

\bibitem[{\citenamefont{Sako et~al.}(2014)}]{Sako:2014fha}
\bibinfo{author}{\bibfnamefont{H.}~\bibnamefont{Sako}} \bibnamefont{et~al.},
  \bibinfo{journal}{Nucl. Phys. A} \textbf{\bibinfo{volume}{931}},
  \bibinfo{pages}{1158} (\bibinfo{year}{2014}).

\bibitem[{\citenamefont{Aggarwal et~al.}(2010)}]{Aggarwal:2010cw}
\bibinfo{author}{\bibfnamefont{M.~M.} \bibnamefont{Aggarwal}}
  \bibnamefont{et~al.} (\bibinfo{collaboration}{STAR}) (\bibinfo{year}{2010}),
  \eprint{1007.2613}.

\bibitem[{\citenamefont{Hartnack et~al.}(1998)\citenamefont{Hartnack, Puri,
  Aichelin, Konopka, Bass, Stoecker, and Greiner}}]{Hartnack:1997ez}
\bibinfo{author}{\bibfnamefont{C.}~\bibnamefont{Hartnack}},
  \bibinfo{author}{\bibfnamefont{R.~K.} \bibnamefont{Puri}},
  \bibinfo{author}{\bibfnamefont{J.}~\bibnamefont{Aichelin}},
  \bibinfo{author}{\bibfnamefont{J.}~\bibnamefont{Konopka}},
  \bibinfo{author}{\bibfnamefont{S.~A.} \bibnamefont{Bass}},
  \bibinfo{author}{\bibfnamefont{H.}~\bibnamefont{Stoecker}}, \bibnamefont{and}
  \bibinfo{author}{\bibfnamefont{W.}~\bibnamefont{Greiner}},
  \bibinfo{journal}{Eur. Phys. J. A} \textbf{\bibinfo{volume}{1}},
  \bibinfo{pages}{151} (\bibinfo{year}{1998}), \eprint{nucl-th/9811015}.

\bibitem[{\citenamefont{Bass et~al.}(1998)}]{Bass:1998ca}
\bibinfo{author}{\bibfnamefont{S.~A.} \bibnamefont{Bass}} \bibnamefont{et~al.},
  \bibinfo{journal}{Prog. Part. Nucl. Phys.} \textbf{\bibinfo{volume}{41}},
  \bibinfo{pages}{255} (\bibinfo{year}{1998}), \bibinfo{note}{[Prog. Part.
  Nucl. Phys.41,225(1998)]}, \eprint{nucl-th/9803035}.

\bibitem[{\citenamefont{Cassing and Bratkovskaya}(1999)}]{Cassing:1999es}
\bibinfo{author}{\bibfnamefont{W.}~\bibnamefont{Cassing}} \bibnamefont{and}
  \bibinfo{author}{\bibfnamefont{E.~L.} \bibnamefont{Bratkovskaya}},
  \bibinfo{journal}{Phys. Rept.} \textbf{\bibinfo{volume}{308}},
  \bibinfo{pages}{65} (\bibinfo{year}{1999}).

\bibitem[{\citenamefont{Nara et~al.}(2000)\citenamefont{Nara, Otuka, Ohnishi,
  Niita, and Chiba}}]{Nara:1999dz}
\bibinfo{author}{\bibfnamefont{Y.}~\bibnamefont{Nara}},
  \bibinfo{author}{\bibfnamefont{N.}~\bibnamefont{Otuka}},
  \bibinfo{author}{\bibfnamefont{A.}~\bibnamefont{Ohnishi}},
  \bibinfo{author}{\bibfnamefont{K.}~\bibnamefont{Niita}}, \bibnamefont{and}
  \bibinfo{author}{\bibfnamefont{S.}~\bibnamefont{Chiba}},
  \bibinfo{journal}{Phys. Rev. C} \textbf{\bibinfo{volume}{61}},
  \bibinfo{pages}{024901} (\bibinfo{year}{2000}), \eprint{nucl-th/9904059}.

\bibitem[{\citenamefont{Buss et~al.}(2012)\citenamefont{Buss, Gaitanos,
  Gallmeister, van Hees, Kaskulov, Lalakulich, Larionov, Leitner, Weil, and
  Mosel}}]{Buss:2011mx}
\bibinfo{author}{\bibfnamefont{O.}~\bibnamefont{Buss}},
  \bibinfo{author}{\bibfnamefont{T.}~\bibnamefont{Gaitanos}},
  \bibinfo{author}{\bibfnamefont{K.}~\bibnamefont{Gallmeister}},
  \bibinfo{author}{\bibfnamefont{H.}~\bibnamefont{van Hees}},
  \bibinfo{author}{\bibfnamefont{M.}~\bibnamefont{Kaskulov}},
  \bibinfo{author}{\bibfnamefont{O.}~\bibnamefont{Lalakulich}},
  \bibinfo{author}{\bibfnamefont{A.~B.} \bibnamefont{Larionov}},
  \bibinfo{author}{\bibfnamefont{T.}~\bibnamefont{Leitner}},
  \bibinfo{author}{\bibfnamefont{J.}~\bibnamefont{Weil}}, \bibnamefont{and}
  \bibinfo{author}{\bibfnamefont{U.}~\bibnamefont{Mosel}},
  \bibinfo{journal}{Phys. Rept.} \textbf{\bibinfo{volume}{512}},
  \bibinfo{pages}{1} (\bibinfo{year}{2012}), \eprint{1106.1344}.

\bibitem[{\citenamefont{Zhang et~al.}(2018)}]{Zhang:2017esm}
\bibinfo{author}{\bibfnamefont{Y.-X.} \bibnamefont{Zhang}}
  \bibnamefont{et~al.}, \bibinfo{journal}{Phys. Rev. C}
  \textbf{\bibinfo{volume}{97}}, \bibinfo{pages}{034625}
  (\bibinfo{year}{2018}), \eprint{1711.05950}.

\bibitem[{\citenamefont{Agakishiev
  et~al.}(2014{\natexlab{a}})}]{Agakishiev:2014moo}
\bibinfo{author}{\bibfnamefont{G.}~\bibnamefont{Agakishiev}}
  \bibnamefont{et~al.} (\bibinfo{collaboration}{HADES}),
  \bibinfo{journal}{Phys. Rev. C} \textbf{\bibinfo{volume}{90}},
  \bibinfo{pages}{054906} (\bibinfo{year}{2014}{\natexlab{a}}),
  \eprint{1404.7011}.

\bibitem[{\citenamefont{Hartnack et~al.}(2012)\citenamefont{Hartnack, Oeschler,
  Leifels, Bratkovskaya, and Aichelin}}]{Hartnack:2011cn}
\bibinfo{author}{\bibfnamefont{C.}~\bibnamefont{Hartnack}},
  \bibinfo{author}{\bibfnamefont{H.}~\bibnamefont{Oeschler}},
  \bibinfo{author}{\bibfnamefont{Y.}~\bibnamefont{Leifels}},
  \bibinfo{author}{\bibfnamefont{E.~L.} \bibnamefont{Bratkovskaya}},
  \bibnamefont{and} \bibinfo{author}{\bibfnamefont{J.}~\bibnamefont{Aichelin}},
  \bibinfo{journal}{Phys. Rept.} \textbf{\bibinfo{volume}{510}},
  \bibinfo{pages}{119} (\bibinfo{year}{2012}), \eprint{1106.2083}.

\bibitem[{\citenamefont{Steinheimer and Bleicher}(2016)}]{Steinheimer:2015sha}
\bibinfo{author}{\bibfnamefont{J.}~\bibnamefont{Steinheimer}} \bibnamefont{and}
  \bibinfo{author}{\bibfnamefont{M.}~\bibnamefont{Bleicher}},
  \bibinfo{journal}{J. Phys. G} \textbf{\bibinfo{volume}{43}},
  \bibinfo{pages}{015104} (\bibinfo{year}{2016}), \eprint{1503.07305}.

\bibitem[{\citenamefont{Graef et~al.}(2014)\citenamefont{Graef, Steinheimer,
  Li, and Bleicher}}]{Graef:2014mra}
\bibinfo{author}{\bibfnamefont{G.}~\bibnamefont{Graef}},
  \bibinfo{author}{\bibfnamefont{J.}~\bibnamefont{Steinheimer}},
  \bibinfo{author}{\bibfnamefont{F.}~\bibnamefont{Li}}, \bibnamefont{and}
  \bibinfo{author}{\bibfnamefont{M.}~\bibnamefont{Bleicher}},
  \bibinfo{journal}{Phys. Rev. C} \textbf{\bibinfo{volume}{90}},
  \bibinfo{pages}{064909} (\bibinfo{year}{2014}), \eprint{1409.7954}.

\bibitem[{\citenamefont{Gallmeister et~al.}(2017)\citenamefont{Gallmeister,
  Beitel, and Greiner}}]{Gallmeister:2017ths}
\bibinfo{author}{\bibfnamefont{K.}~\bibnamefont{Gallmeister}},
  \bibinfo{author}{\bibfnamefont{M.}~\bibnamefont{Beitel}}, \bibnamefont{and}
  \bibinfo{author}{\bibfnamefont{C.}~\bibnamefont{Greiner}}
  (\bibinfo{year}{2017}), \eprint{1712.04018}.

\bibitem[{\citenamefont{Karpenko et~al.}(2015)\citenamefont{Karpenko, Huovinen,
  Petersen, and Bleicher}}]{Karpenko:2015xea}
\bibinfo{author}{\bibfnamefont{I.~A.} \bibnamefont{Karpenko}},
  \bibinfo{author}{\bibfnamefont{P.}~\bibnamefont{Huovinen}},
  \bibinfo{author}{\bibfnamefont{H.}~\bibnamefont{Petersen}}, \bibnamefont{and}
  \bibinfo{author}{\bibfnamefont{M.}~\bibnamefont{Bleicher}},
  \bibinfo{journal}{Phys. Rev. C} \textbf{\bibinfo{volume}{91}},
  \bibinfo{pages}{064901} (\bibinfo{year}{2015}), \eprint{1502.01978}.

\bibitem[{\citenamefont{Denicol et~al.}(2018)\citenamefont{Denicol, Gale, Jeon,
  Monnai, Schenke, and Shen}}]{Denicol:2018wdp}
\bibinfo{author}{\bibfnamefont{G.~S.} \bibnamefont{Denicol}},
  \bibinfo{author}{\bibfnamefont{C.}~\bibnamefont{Gale}},
  \bibinfo{author}{\bibfnamefont{S.}~\bibnamefont{Jeon}},
  \bibinfo{author}{\bibfnamefont{A.}~\bibnamefont{Monnai}},
  \bibinfo{author}{\bibfnamefont{B.}~\bibnamefont{Schenke}}, \bibnamefont{and}
  \bibinfo{author}{\bibfnamefont{C.}~\bibnamefont{Shen}}
  (\bibinfo{year}{2018}), \eprint{1804.10557}.

\bibitem[{\citenamefont{Akamatsu et~al.}(2018)\citenamefont{Akamatsu, Asakawa,
  Hirano, Kitazawa, Morita, Murase, Nara, Nonaka, and
  Ohnishi}}]{Akamatsu:2018olk}
\bibinfo{author}{\bibfnamefont{Y.}~\bibnamefont{Akamatsu}},
  \bibinfo{author}{\bibfnamefont{M.}~\bibnamefont{Asakawa}},
  \bibinfo{author}{\bibfnamefont{T.}~\bibnamefont{Hirano}},
  \bibinfo{author}{\bibfnamefont{M.}~\bibnamefont{Kitazawa}},
  \bibinfo{author}{\bibfnamefont{K.}~\bibnamefont{Morita}},
  \bibinfo{author}{\bibfnamefont{K.}~\bibnamefont{Murase}},
  \bibinfo{author}{\bibfnamefont{Y.}~\bibnamefont{Nara}},
  \bibinfo{author}{\bibfnamefont{C.}~\bibnamefont{Nonaka}}, \bibnamefont{and}
  \bibinfo{author}{\bibfnamefont{A.}~\bibnamefont{Ohnishi}},
  \bibinfo{journal}{Phys. Rev. C} \textbf{\bibinfo{volume}{98}},
  \bibinfo{pages}{024909} (\bibinfo{year}{2018}), \eprint{1805.09024}.

\bibitem[{\citenamefont{Weil et~al.}(2016{\natexlab{a}})}]{Weil:2016zrk}
\bibinfo{author}{\bibfnamefont{J.}~\bibnamefont{Weil}} \bibnamefont{et~al.},
  \bibinfo{journal}{Phys. Rev. C} \textbf{\bibinfo{volume}{94}},
  \bibinfo{pages}{054905} (\bibinfo{year}{2016}{\natexlab{a}}),
  \eprint{1606.06642}.

\bibitem[{\citenamefont{Tindall et~al.}(2017)\citenamefont{Tindall,
  Torres-Rincon, Rose, and Petersen}}]{Tindall:2016try}
\bibinfo{author}{\bibfnamefont{J.}~\bibnamefont{Tindall}},
  \bibinfo{author}{\bibfnamefont{J.~M.} \bibnamefont{Torres-Rincon}},
  \bibinfo{author}{\bibfnamefont{J.~B.} \bibnamefont{Rose}}, \bibnamefont{and}
  \bibinfo{author}{\bibfnamefont{H.}~\bibnamefont{Petersen}},
  \bibinfo{journal}{Phys. Lett. B} \textbf{\bibinfo{volume}{770}},
  \bibinfo{pages}{532} (\bibinfo{year}{2017}), \eprint{1612.06436}.

\bibitem[{\citenamefont{Staudenmaier
  et~al.}(2017{\natexlab{a}})\citenamefont{Staudenmaier, Weil, Steinberg,
  Endres, and Petersen}}]{Staudenmaier:2017vtq}
\bibinfo{author}{\bibfnamefont{J.}~\bibnamefont{Staudenmaier}},
  \bibinfo{author}{\bibfnamefont{J.}~\bibnamefont{Weil}},
  \bibinfo{author}{\bibfnamefont{V.}~\bibnamefont{Steinberg}},
  \bibinfo{author}{\bibfnamefont{S.}~\bibnamefont{Endres}}, \bibnamefont{and}
  \bibinfo{author}{\bibfnamefont{H.}~\bibnamefont{Petersen}}
  (\bibinfo{year}{2017}{\natexlab{a}}), \eprint{1711.10297}.

\bibitem[{\citenamefont{Rose et~al.}(2018)\citenamefont{Rose, Torres-Rincon,
  Schäfer, Oliinychenko, and Petersen}}]{Rose:2017bjz}
\bibinfo{author}{\bibfnamefont{J.~B.} \bibnamefont{Rose}},
  \bibinfo{author}{\bibfnamefont{J.~M.} \bibnamefont{Torres-Rincon}},
  \bibinfo{author}{\bibfnamefont{A.}~\bibnamefont{Schäfer}},
  \bibinfo{author}{\bibfnamefont{D.~R.} \bibnamefont{Oliinychenko}},
  \bibnamefont{and} \bibinfo{author}{\bibfnamefont{H.}~\bibnamefont{Petersen}},
  \bibinfo{journal}{Phys. Rev. C} \textbf{\bibinfo{volume}{97}},
  \bibinfo{pages}{055204} (\bibinfo{year}{2018}), \eprint{1709.03826}.

\bibitem[{\citenamefont{Agakishiev
  et~al.}(2014{\natexlab{b}})}]{Agakishiev:2014wqa}
\bibinfo{author}{\bibfnamefont{G.}~\bibnamefont{Agakishiev}}
  \bibnamefont{et~al.}, \bibinfo{journal}{Eur. Phys. J. A}
  \textbf{\bibinfo{volume}{50}}, \bibinfo{pages}{82}
  (\bibinfo{year}{2014}{\natexlab{b}}), \eprint{1403.3054}.

\bibitem[{\citenamefont{Patrignani et~al.}(2016)}]{Patrignani:2016xqp}
\bibinfo{author}{\bibfnamefont{C.}~\bibnamefont{Patrignani}}
  \bibnamefont{et~al.} (\bibinfo{collaboration}{Particle Data Group}),
  \bibinfo{journal}{Chin. Phys. C} \textbf{\bibinfo{volume}{40}},
  \bibinfo{pages}{100001} (\bibinfo{year}{2016}).

\bibitem[{\citenamefont{Manley and Saleski}(1992)}]{Manley:1992yb}
\bibinfo{author}{\bibfnamefont{D.~M.} \bibnamefont{Manley}} \bibnamefont{and}
  \bibinfo{author}{\bibfnamefont{E.~M.} \bibnamefont{Saleski}},
  \bibinfo{journal}{Phys. Rev. D} \textbf{\bibinfo{volume}{45}},
  \bibinfo{pages}{4002} (\bibinfo{year}{1992}).

\bibitem[{\citenamefont{Forster et~al.}(2007)}]{Forster:2007qk}
\bibinfo{author}{\bibfnamefont{A.}~\bibnamefont{Forster}} \bibnamefont{et~al.},
  \bibinfo{journal}{Phys. Rev. C} \textbf{\bibinfo{volume}{75}},
  \bibinfo{pages}{024906} (\bibinfo{year}{2007}), \eprint{nucl-ex/0701014}.

\bibitem[{\citenamefont{Agakishiev
  et~al.}(2009{\natexlab{b}})}]{Agakishiev:2009ar}
\bibinfo{author}{\bibfnamefont{G.}~\bibnamefont{Agakishiev}}
  \bibnamefont{et~al.} (\bibinfo{collaboration}{HADES}),
  \bibinfo{journal}{Phys. Rev. C} \textbf{\bibinfo{volume}{80}},
  \bibinfo{pages}{025209} (\bibinfo{year}{2009}{\natexlab{b}}),
  \eprint{0902.3487}.

\bibitem[{\citenamefont{Münzer et~al.}(2017)}]{Munzer:2017hbl}
\bibinfo{author}{\bibfnamefont{R.}~\bibnamefont{Münzer}} \bibnamefont{et~al.}
  (\bibinfo{year}{2017}), \eprint{1703.01978}.

\bibitem[{\citenamefont{Balewski et~al.}(1998)}]{Balewski:1998pd}
\bibinfo{author}{\bibfnamefont{J.~T.} \bibnamefont{Balewski}}
  \bibnamefont{et~al.}, \bibinfo{journal}{Phys. Lett. B}
  \textbf{\bibinfo{volume}{420}}, \bibinfo{pages}{211} (\bibinfo{year}{1998}),
  \eprint{nucl-ex/9803003}.

\bibitem[{\citenamefont{Sewerin et~al.}(1999)}]{Sewerin:1998ky}
\bibinfo{author}{\bibfnamefont{S.}~\bibnamefont{Sewerin}} \bibnamefont{et~al.},
  \bibinfo{journal}{Phys. Rev. Lett.} \textbf{\bibinfo{volume}{83}},
  \bibinfo{pages}{682} (\bibinfo{year}{1999}), \eprint{nucl-ex/9811004}.

\bibitem[{\citenamefont{Kowina et~al.}(2004)}]{Kowina:2004kr}
\bibinfo{author}{\bibfnamefont{P.}~\bibnamefont{Kowina}} \bibnamefont{et~al.},
  \bibinfo{journal}{Eur. Phys. J. A} \textbf{\bibinfo{volume}{22}},
  \bibinfo{pages}{293} (\bibinfo{year}{2004}), \eprint{nucl-ex/0402008}.

\bibitem[{\citenamefont{Abd El-Samad et~al.}(2010)}]{AbdElSamad:2010tz}
\bibinfo{author}{\bibfnamefont{S.}~\bibnamefont{Abd El-Samad}}
  \bibnamefont{et~al.} (\bibinfo{collaboration}{TOF}), \bibinfo{journal}{Phys.
  Lett. B} \textbf{\bibinfo{volume}{688}}, \bibinfo{pages}{142}
  (\bibinfo{year}{2010}), \eprint{1003.0603}.

\bibitem[{\citenamefont{Abdel-Bary et~al.}(2010)}]{AbdelBary:2010pc}
\bibinfo{author}{\bibfnamefont{M.}~\bibnamefont{Abdel-Bary}}
  \bibnamefont{et~al.} (\bibinfo{collaboration}{COSY-TOF}),
  \bibinfo{journal}{Eur. Phys. J. A} \textbf{\bibinfo{volume}{46}},
  \bibinfo{pages}{27} (\bibinfo{year}{2010}), \bibinfo{note}{[Erratum: Eur.
  Phys. J.A46,435(2010)]}, \eprint{1008.4287}.

\bibitem[{\citenamefont{Bilger et~al.}(1998)}]{Bilger:1998jf}
\bibinfo{author}{\bibfnamefont{R.}~\bibnamefont{Bilger}} \bibnamefont{et~al.},
  \bibinfo{journal}{Phys. Lett. B} \textbf{\bibinfo{volume}{420}},
  \bibinfo{pages}{217} (\bibinfo{year}{1998}).

\bibitem[{\citenamefont{Abdel-Samad et~al.}(2006)}]{AbdelSamad:2006qu}
\bibinfo{author}{\bibfnamefont{S.}~\bibnamefont{Abdel-Samad}}
  \bibnamefont{et~al.} (\bibinfo{collaboration}{COSY-TOF}),
  \bibinfo{journal}{Phys. Lett. B} \textbf{\bibinfo{volume}{632}},
  \bibinfo{pages}{27} (\bibinfo{year}{2006}).

\bibitem[{\citenamefont{Baldini}(1988)}]{LaBoer}
\bibinfo{author}{\bibfnamefont{A.}~\bibnamefont{Baldini}},
  \bibinfo{journal}{Landolt-Börnstein. New Series, 1/12B}
  (\bibinfo{year}{1988}).

\bibitem[{\citenamefont{Valdau et~al.}(2010)}]{Valdau:2010kw}
\bibinfo{author}{\bibfnamefont{{\relax Yu}.}~\bibnamefont{Valdau}}
  \bibnamefont{et~al.}, \bibinfo{journal}{Phys. Rev. C}
  \textbf{\bibinfo{volume}{81}}, \bibinfo{pages}{045208}
  (\bibinfo{year}{2010}), \eprint{1002.3459}.

\bibitem[{\citenamefont{Valdau et~al.}(2007)}]{Valdau:2007re}
\bibinfo{author}{\bibfnamefont{{\relax Yu}.}~\bibnamefont{Valdau}}
  \bibnamefont{et~al.}, \bibinfo{journal}{Phys. Lett. B}
  \textbf{\bibinfo{volume}{652}}, \bibinfo{pages}{245} (\bibinfo{year}{2007}),
  \eprint{nucl-ex/0703044}.

\bibitem[{\citenamefont{Ko}(1983)}]{Ko:1983zp}
\bibinfo{author}{\bibfnamefont{C.~m.} \bibnamefont{Ko}},
  \bibinfo{journal}{Phys. Lett. B} \textbf{\bibinfo{volume}{120}},
  \bibinfo{pages}{294} (\bibinfo{year}{1983}).

\bibitem[{\citenamefont{Wolke}(1998)}]{Wolke_thesis}
\bibinfo{author}{\bibfnamefont{M.}~\bibnamefont{Wolke}}, Ph.D. thesis,
  \bibinfo{school}{University of Bonn} (\bibinfo{year}{1998}).

\bibitem[{\citenamefont{Winter et~al.}(2006)}]{Winter:2006vd}
\bibinfo{author}{\bibfnamefont{P.}~\bibnamefont{Winter}} \bibnamefont{et~al.},
  \bibinfo{journal}{Phys. Lett. B} \textbf{\bibinfo{volume}{635}},
  \bibinfo{pages}{23} (\bibinfo{year}{2006}), \eprint{hep-ex/0602030}.

\bibitem[{\citenamefont{Quentmeier et~al.}(2001)}]{Quentmeier:2001ec}
\bibinfo{author}{\bibfnamefont{C.}~\bibnamefont{Quentmeier}}
  \bibnamefont{et~al.}, \bibinfo{journal}{Phys. Lett. B}
  \textbf{\bibinfo{volume}{515}}, \bibinfo{pages}{276} (\bibinfo{year}{2001}),
  \eprint{nucl-ex/0103001}.

\bibitem[{\citenamefont{Maeda et~al.}(2008)}]{Maeda:2007cy}
\bibinfo{author}{\bibfnamefont{Y.}~\bibnamefont{Maeda}} \bibnamefont{et~al.}
  (\bibinfo{collaboration}{ANKE}), \bibinfo{journal}{Phys. Rev. C}
  \textbf{\bibinfo{volume}{77}}, \bibinfo{pages}{015204}
  (\bibinfo{year}{2008}), \eprint{0710.1755}.

\bibitem[{\citenamefont{Balestra et~al.}(2001)}]{Balestra:2000ex}
\bibinfo{author}{\bibfnamefont{F.}~\bibnamefont{Balestra}} \bibnamefont{et~al.}
  (\bibinfo{collaboration}{DISTO}), \bibinfo{journal}{Phys. Rev. C}
  \textbf{\bibinfo{volume}{63}}, \bibinfo{pages}{024004}
  (\bibinfo{year}{2001}), \eprint{nucl-ex/0011009}.

\bibitem[{\citenamefont{Agakishiev et~al.}(2012{\natexlab{a}})}]{HADES:2011ab}
\bibinfo{author}{\bibfnamefont{G.}~\bibnamefont{Agakishiev}}
  \bibnamefont{et~al.} (\bibinfo{collaboration}{HADES}), \bibinfo{journal}{Eur.
  Phys. J. A} \textbf{\bibinfo{volume}{48}}, \bibinfo{pages}{64}
  (\bibinfo{year}{2012}{\natexlab{a}}), \eprint{1112.3607}.

\bibitem[{\citenamefont{Agakishiev
  et~al.}(2012{\natexlab{b}})}]{Agakishiev:2012vj}
\bibinfo{author}{\bibfnamefont{G.}~\bibnamefont{Agakishiev}}
  \bibnamefont{et~al.} (\bibinfo{collaboration}{HADES}),
  \bibinfo{journal}{Phys. Lett. B} \textbf{\bibinfo{volume}{715}},
  \bibinfo{pages}{304} (\bibinfo{year}{2012}{\natexlab{b}}),
  \eprint{1205.1918}.

\bibitem[{\citenamefont{Staudenmaier
  et~al.}(2017{\natexlab{b}})\citenamefont{Staudenmaier, Weil, and
  Petersen}}]{Staudenmaier:2016hmh}
\bibinfo{author}{\bibfnamefont{J.}~\bibnamefont{Staudenmaier}},
  \bibinfo{author}{\bibfnamefont{J.}~\bibnamefont{Weil}}, \bibnamefont{and}
  \bibinfo{author}{\bibfnamefont{H.}~\bibnamefont{Petersen}},
  \bibinfo{journal}{J. Phys. Conf. Ser.} \textbf{\bibinfo{volume}{832}},
  \bibinfo{pages}{012037} (\bibinfo{year}{2017}{\natexlab{b}}),
  \eprint{1611.09164}.

\bibitem[{\citenamefont{Weil et~al.}(2016{\natexlab{b}})\citenamefont{Weil,
  Staudenmaier, and Petersen}}]{Weil:2016fxr}
\bibinfo{author}{\bibfnamefont{J.}~\bibnamefont{Weil}},
  \bibinfo{author}{\bibfnamefont{J.}~\bibnamefont{Staudenmaier}},
  \bibnamefont{and} \bibinfo{author}{\bibfnamefont{H.}~\bibnamefont{Petersen}},
  \bibinfo{journal}{J. Phys. Conf. Ser.} \textbf{\bibinfo{volume}{742}},
  \bibinfo{pages}{012034} (\bibinfo{year}{2016}{\natexlab{b}}),
  \eprint{1604.07028}.

\bibitem[{\citenamefont{Adamczyk et~al.}(2017)}]{Adamczyk:2017iwn}
\bibinfo{author}{\bibfnamefont{L.}~\bibnamefont{Adamczyk}} \bibnamefont{et~al.}
  (\bibinfo{collaboration}{STAR}), \bibinfo{journal}{Phys. Rev. C}
  \textbf{\bibinfo{volume}{96}}, \bibinfo{pages}{044904}
  (\bibinfo{year}{2017}), \eprint{1701.07065}.

\bibitem[{\citenamefont{Becattini and Manninen}(2009)}]{Becattini:2008ya}
\bibinfo{author}{\bibfnamefont{F.}~\bibnamefont{Becattini}} \bibnamefont{and}
  \bibinfo{author}{\bibfnamefont{J.}~\bibnamefont{Manninen}},
  \bibinfo{journal}{Phys. Lett. B} \textbf{\bibinfo{volume}{673}},
  \bibinfo{pages}{19} (\bibinfo{year}{2009}), \eprint{0811.3766}.

\bibitem[{\citenamefont{Agakishiev et~al.}(2011)}]{Agakishiev:2010rs}
\bibinfo{author}{\bibfnamefont{G.}~\bibnamefont{Agakishiev}}
  \bibnamefont{et~al.} (\bibinfo{collaboration}{HADES}), \bibinfo{journal}{Eur.
  Phys. J. A} \textbf{\bibinfo{volume}{47}}, \bibinfo{pages}{21}
  (\bibinfo{year}{2011}), \eprint{1010.1675}.

\bibitem[{\citenamefont{Adamczewski-Musch
  et~al.}(2017)}]{Adamczewski-Musch:2016vrc}
\bibinfo{author}{\bibfnamefont{J.}~\bibnamefont{Adamczewski-Musch}}
  \bibnamefont{et~al.} (\bibinfo{collaboration}{HADES}),
  \bibinfo{journal}{Phys. Rev. C} \textbf{\bibinfo{volume}{95}},
  \bibinfo{pages}{015207} (\bibinfo{year}{2017}), \eprint{1611.01040}.

\bibitem[{\citenamefont{Lorenz}(2014)}]{Lorenz:2014eja}
\bibinfo{author}{\bibfnamefont{M.}~\bibnamefont{Lorenz}}
  (\bibinfo{collaboration}{HADES}), \bibinfo{journal}{Nucl. Phys. A}
  \textbf{\bibinfo{volume}{931}}, \bibinfo{pages}{785} (\bibinfo{year}{2014}).

\bibitem[{\citenamefont{Mangiarotti et~al.}(2003)}]{Mangiarotti:2002mw}
\bibinfo{author}{\bibfnamefont{A.}~\bibnamefont{Mangiarotti}}
  \bibnamefont{et~al.} (\bibinfo{collaboration}{FOPI}), \bibinfo{journal}{Nucl.
  Phys. A} \textbf{\bibinfo{volume}{714}}, \bibinfo{pages}{89}
  (\bibinfo{year}{2003}), \eprint{nucl-ex/0209012}.

\bibitem[{\citenamefont{Oliinychenko and
  Petersen}(2017)}]{Oliinychenko:2016vkg}
\bibinfo{author}{\bibfnamefont{D.}~\bibnamefont{Oliinychenko}}
  \bibnamefont{and} \bibinfo{author}{\bibfnamefont{H.}~\bibnamefont{Petersen}},
  \bibinfo{journal}{J. Phys. G} \textbf{\bibinfo{volume}{44}},
  \bibinfo{pages}{034001} (\bibinfo{year}{2017}), \eprint{1609.01087}.

\bibitem[{\citenamefont{Petschauer et~al.}(2016)\citenamefont{Petschauer,
  Haidenbauer, Kaiser, Meißner, and Weise}}]{Petschauer:2015nea}
\bibinfo{author}{\bibfnamefont{S.}~\bibnamefont{Petschauer}},
  \bibinfo{author}{\bibfnamefont{J.}~\bibnamefont{Haidenbauer}},
  \bibinfo{author}{\bibfnamefont{N.}~\bibnamefont{Kaiser}},
  \bibinfo{author}{\bibfnamefont{U.-G.} \bibnamefont{Meißner}},
  \bibnamefont{and} \bibinfo{author}{\bibfnamefont{W.}~\bibnamefont{Weise}},
  \bibinfo{journal}{Eur. Phys. J. A} \textbf{\bibinfo{volume}{52}},
  \bibinfo{pages}{15} (\bibinfo{year}{2016}), \eprint{1507.08808}.

\bibitem[{\citenamefont{Bernhard et~al.}(2016)\citenamefont{Bernhard, Moreland,
  Bass, Liu, and Heinz}}]{Bernhard:2016tnd}
\bibinfo{author}{\bibfnamefont{J.~E.} \bibnamefont{Bernhard}},
  \bibinfo{author}{\bibfnamefont{J.~S.} \bibnamefont{Moreland}},
  \bibinfo{author}{\bibfnamefont{S.~A.} \bibnamefont{Bass}},
  \bibinfo{author}{\bibfnamefont{J.}~\bibnamefont{Liu}}, \bibnamefont{and}
  \bibinfo{author}{\bibfnamefont{U.}~\bibnamefont{Heinz}},
  \bibinfo{journal}{Phys. Rev. C} \textbf{\bibinfo{volume}{94}},
  \bibinfo{pages}{024907} (\bibinfo{year}{2016}), \eprint{1605.03954}.

\bibitem[{\citenamefont{Cleveland}(1979)}]{Cleveland1979}
\bibinfo{author}{\bibfnamefont{W.~S.} \bibnamefont{Cleveland}},
  \bibinfo{journal}{Journal of the American statistical association}
  \textbf{\bibinfo{volume}{74}}, \bibinfo{pages}{829} (\bibinfo{year}{1979}).

\bibitem[{\citenamefont{Cleveland}(1981)}]{Cleveland1981}
\bibinfo{author}{\bibfnamefont{W.~S.} \bibnamefont{Cleveland}},
  \bibinfo{journal}{The American Statistician} \textbf{\bibinfo{volume}{35}},
  \bibinfo{pages}{54} (\bibinfo{year}{1981}).

\end{thebibliography}

\end{document}